\pdfoutput=1
\documentclass{article}

\PassOptionsToPackage{numbers,compress}{natbib}
\usepackage[eandd, final]{neurips_2026}

\usepackage[utf8]{inputenc}
\usepackage[T1]{fontenc}
\usepackage[dvipsnames,table]{xcolor}
\usepackage{hyperref}
\usepackage{url}
\usepackage{booktabs}
\usepackage{microtype}
\usepackage{adjustbox}
\usepackage{amsmath}
\usepackage{caption}
\usepackage{enumitem}
\usepackage[capitalize]{cleveref}
\usepackage{algorithm}
\usepackage{algpseudocode}
\usepackage{longtable}
\usepackage{amssymb}
\usepackage{subcaption}
\usepackage{placeins}

\crefname{appendix}{Appx.}{Appx.}
\Crefname{appendix}{Appendix}{Appendices}

\usepackage{arydshln}
\newcommand{\field}[1]{\texttt{\footnotesize #1}\hspace{0.45em}}

\usepackage[skins,breakable]{tcolorbox}
\newcommand{\definition}[1]{    \tcbset{width=\textwidth,boxrule=0.5pt,colback=lightgray!40,arc=3pt,auto outer arc,left=3pt,right=3pt,boxsep=3pt,colframe=black}
    \begin{center}
        \begin{minipage}{\textwidth}
            \begin{tcolorbox}
                \textbf{Definition:} #1
            \end{tcolorbox}
        \end{minipage}
    \end{center}
}

\usepackage{array}

\newcommand{\nDatasetsAll}{1034}
\newcommand{\nDatasetsOpenNeuro}{782}
\newcommand{\nDatasetsMerged}{253}
\newcommand{\nVolumes}{564,885}
\newcommand{\nVolumesShort}{565k}
\newcommand{\nDuplicates}{64,888}
\newcommand{\nDuplicatesShort}{65k}
\newcommand{\nHashesShort}{526k}
\newcommand{\nDuplicatePairs}{118,487}
\newcommand{\nDuplicatePairsShort}{118k}
\newcommand{\nDuplicatePairsAuto}{56,051}
\newcommand{\nDuplicatePairsExact}{44,727}
\newcommand{\nDuplicatePairsManual}{62,436}
\newcommand{\nRemovedCopies}{40,635}
\newcommand{\nUniqueVolumes}{524,250}
\newcommand{\nReduction}{7.2}
\newcommand{\nDatasetsOverlap}{136}
\newcommand{\nMainCluster}{107}
\title{Dataset Collections: Challenges of Large-Scale Data Aggregation in 3D Medical Image Datasets}

\author{%
  \parbox{\dimexpr\textwidth-2\tabcolsep\relax}{\centering
  \mbox{Yannick Kirchhoff\textsuperscript{1,2,3,*}} \quad
  \mbox{Saikat Roy\textsuperscript{1,3,*}} \quad
  \mbox{Elisa Stegmeier\textsuperscript{1,*}} \quad
  \mbox{Hamideh Haghiri\textsuperscript{1,4,*}} \quad
  \mbox{Constantin Ulrich\textsuperscript{1}} \quad
  \mbox{Maximilian Rokuss\textsuperscript{1,5}} \quad
  \mbox{Tassilo Wald\textsuperscript{6}} \quad
  \mbox{Karol Gotkowski\textsuperscript{1,5}} \quad
  \mbox{Benjamin Hamm\textsuperscript{1}} \quad
  \mbox{Raphael Stock\textsuperscript{1,3,5,7}} \quad
  \mbox{Nico Disch\textsuperscript{1,2}} \quad
  \mbox{Michael Baumgartner\textsuperscript{6}} \quad
  \mbox{Dimitrios Bounias\textsuperscript{1,7,8}} \quad
  \mbox{Anand Deshpande\textsuperscript{1,4}} \quad
  \mbox{Stefan Dinkelacker\textsuperscript{1}} \quad
  \mbox{Stefan Dvoretskii\textsuperscript{6}} \quad
  \mbox{Katharina Eckstein\textsuperscript{1,8,9}} \quad
  \mbox{Selen Erkan\textsuperscript{6}} \quad
  \mbox{Jessica Kächele\textsuperscript{1,8,9}} \quad
  \mbox{Kim-Celine Kahl\textsuperscript{6}} \quad
  \mbox{Balint Kovacs\textsuperscript{1,2,7,8}} \quad
  \mbox{Lucas Kulla\textsuperscript{1,4,7,10}} \quad
  \mbox{Moritz Langenberg\textsuperscript{1,3}} \quad
  \mbox{Philipp Schader\textsuperscript{1,3}} \quad
  \mbox{Stephen Schaumann\textsuperscript{6}} \quad
  \mbox{Dasha Trofimova\textsuperscript{6}} \quad
  \mbox{Shuhan Xiao\textsuperscript{6}} \quad
  \mbox{Sebastian Ziegler\textsuperscript{1,5}} \quad
  \mbox{David Zimmerer\textsuperscript{1,5}} \quad
  \mbox{Marco Nolden\textsuperscript{1,4}} \quad
  \mbox{Fabian Isensee\textsuperscript{1,5,7}} \quad
  \mbox{Klaus Maier-Hein\textsuperscript{1,2,3,4,5,7,8,11,12}} \quad
  \mbox{Ralf Floca\textsuperscript{1}} \\[0.8em]
  {\mdseries\small
  \textsuperscript{1}Division of Medical Image Computing, German Cancer Research Center (DKFZ), Heidelberg, Germany \\
  \textsuperscript{2}HIDSS4Health -- Helmholtz Information and Data Science School for Health, Helmholtz Association, Karlsruhe/Heidelberg, Germany \\
  \textsuperscript{3}Faculty of Mathematics and Computer Science, Heidelberg University, Heidelberg, Germany \\
  \textsuperscript{4}Helmholtz Metadata Collaboration (HMC) Hub Health, German Cancer Research Center (DKFZ), Heidelberg, Germany \\
  \textsuperscript{5}Helmholtz Imaging, German Cancer Research Center (DKFZ), Heidelberg, Germany \\
  \textsuperscript{6}Formerly: Division of Medical Image Computing, German Cancer Research Center (DKFZ), Heidelberg, Germany \\
  \textsuperscript{7}National Center for Tumor Diseases (NCT) Heidelberg, a partnership between the German Cancer Research Center (DKFZ) and Heidelberg University Hospital (UKHD), Heidelberg, Germany \\
  \textsuperscript{8}Medical Faculty Heidelberg, Heidelberg University, Heidelberg, Germany \\
  \textsuperscript{9}German Cancer Consortium (DKTK), DKFZ, core center Heidelberg, Heidelberg, Germany \\
  \textsuperscript{10}NCT Digital Oncology Program \\
  \textsuperscript{11}Pattern Analysis and Learning Group, Department of Radiation Oncology, Heidelberg University Hospital, Heidelberg, Germany \\
  \textsuperscript{12}Division of Computing and Mathematical Sciences, Mohamed bin Zayed University of Artificial Intelligence (MBZUAI), Abu Dhabi, United Arab Emirates \\
  \textsuperscript{*}Equal contribution}}
}

\begin{document}

\maketitle

\begin{abstract}
The development of deep learning-based medical AI depends on high-quality annotated data. Yet, dataset creation remains constrained by the need for expert radiologist annotations, limiting curated datasets to relatively small cohorts. To overcome this limitation, recent efforts increasingly aggregate disparate public datasets into large-scale \textit{dataset collections}. However, combining datasets at scale often obscures data provenance, introducing substantial risks of reuse propagation, unintended leakage, bias amplification and distorted data distributions. In this work, we study these problems on a massive collection of \textbf{\nVolumesShort{}} 3D radiological images across \textbf{\nDatasetsAll{} public datasets} and provide practical resources to address them. \textit{Firstly}, we formalize \textit{dataset collections} as a distinct paradigm in medical image analysis and characterize their structural challenges, including reuse propagation, duplication, and provenance fragmentation. \textit{Secondly}, we perform a large-scale empirical study of duplication on this collection and demonstrate extensive exact and near-duplicate reuse across public datasets via perceptual hashing and manual tracing, identifying a lower bound of \textbf{\nDuplicatePairsShort{}} duplicate pairs. We make the perceptual hashes publicly accessible for comparison through our website, providing researchers with a standardized reference for identifying potential overlaps and shared provenance. \textit{Thirdly}, since hashing alone cannot detect all duplicates, we introduce \textit{MAP} (\textbf{M}etadata for \textbf{A}ggregation and \textbf{P}rovenance), a lightweight, portable metadata profile for documenting provenance, known overlaps, and deduplication decisions in dataset collections. Our contributions help \textit{dataset collections} shift from coarse aggregations to reliable and provenance-aware data scaling mechanisms in medical image analysis. A web tool supporting our work is available at \url{https://mic-dct.dkfz.de/}.

\end{abstract}

\flushbottom

\thispagestyle{empty}

\section{Introduction}
Medical image datasets are critical to the training of deep neural networks for tasks such as segmentation \cite{azad2024medical,litjens2017survey}, detection \cite{baumgartner2021nndetection,sobek2024medyolo}, reconstruction \cite{yaqub2022deep,ben2021deep} and registration \cite{fu2020deep,zou2022review}, which form the backbone of pipelines for healthcare AI \cite{xing2017deep,peiffer2020machine,adlung2021machine,huang2021application,appelt2022deep}. In particular, publicly available medical images have been critical to methodological development over the past decades. Deep learning in natural images has long leveraged large-scale datasets to achieve tangible benefits in representation learning. In contrast, in the early 2010s, a publicly available medical image dataset would commonly refer to a monolithic aggregation of medical images uniquely sourced from one or more clinics or hospitals, often with sample sizes in the order of tens to hundreds of samples, small enough for visual inspection. However, with the popularity of large-scale pretraining in natural computer vision \cite{goldblum2023battle}, efforts followed to develop large-scale public datasets for medical image analysis.

One useful approach towards this was the simple aggregation of the aforementioned existing small-scale datasets to enable what we term as \textit{dataset collections} in this work.
Dataset collections have enabled massive data scaling for training deep neural networks in 3D medical image analysis \cite{wald2025openmind,xu2026cads,wald2025revisiting,roy2025mednext}. Training at scale has enabled powerful networks for tasks such as large-scale segmentation \cite{ulrich2023multitalent} and detection \cite{eckstein2025missing}, lesion tracking \cite{rokuss2025lesionlocator,kirchhoff2026exploiting}, prompted segmentation of medical structures \cite{ma2024segment,wang2025sam}, visual question answering \cite{gai20253d}, and grounded 3D medical segmentation \cite{rokuss2026voxtell}.
Yet, despite their strong benefits, dataset collections are heterogeneous datasets with implicit dependencies from a public data ecosystem and should not be treated as independent datasets.

While dataset collections enable scale (\cref{tab:dataset_collections}), they fundamentally alter the statistical structure of training data. In contrast to curated cohorts, collections inherit hidden dependencies across datasets, including duplicated samples, near-duplicates, and multi-hop reuse of underlying sources. These artifacts violate common assumptions of independence between training and evaluation data, inflate effective dataset size, and can lead to overly optimistic conclusions about model performance and scaling behavior (\cref{sec:dataset_collections}). Crucially, such issues are often difficult to detect post hoc, as dataset collections typically lack explicit documentation of data provenance (\cref{sec:data_dedup}). As a result, the reliability of empirical findings derived from these collections becomes increasingly uncertain at scale without explicit tracking of provenance (\cref{sec:metadata}).

\begin{figure}[t]
    \centering
    \includegraphics[
    width=\textwidth,
    trim=0 0 0 10,
    clip]{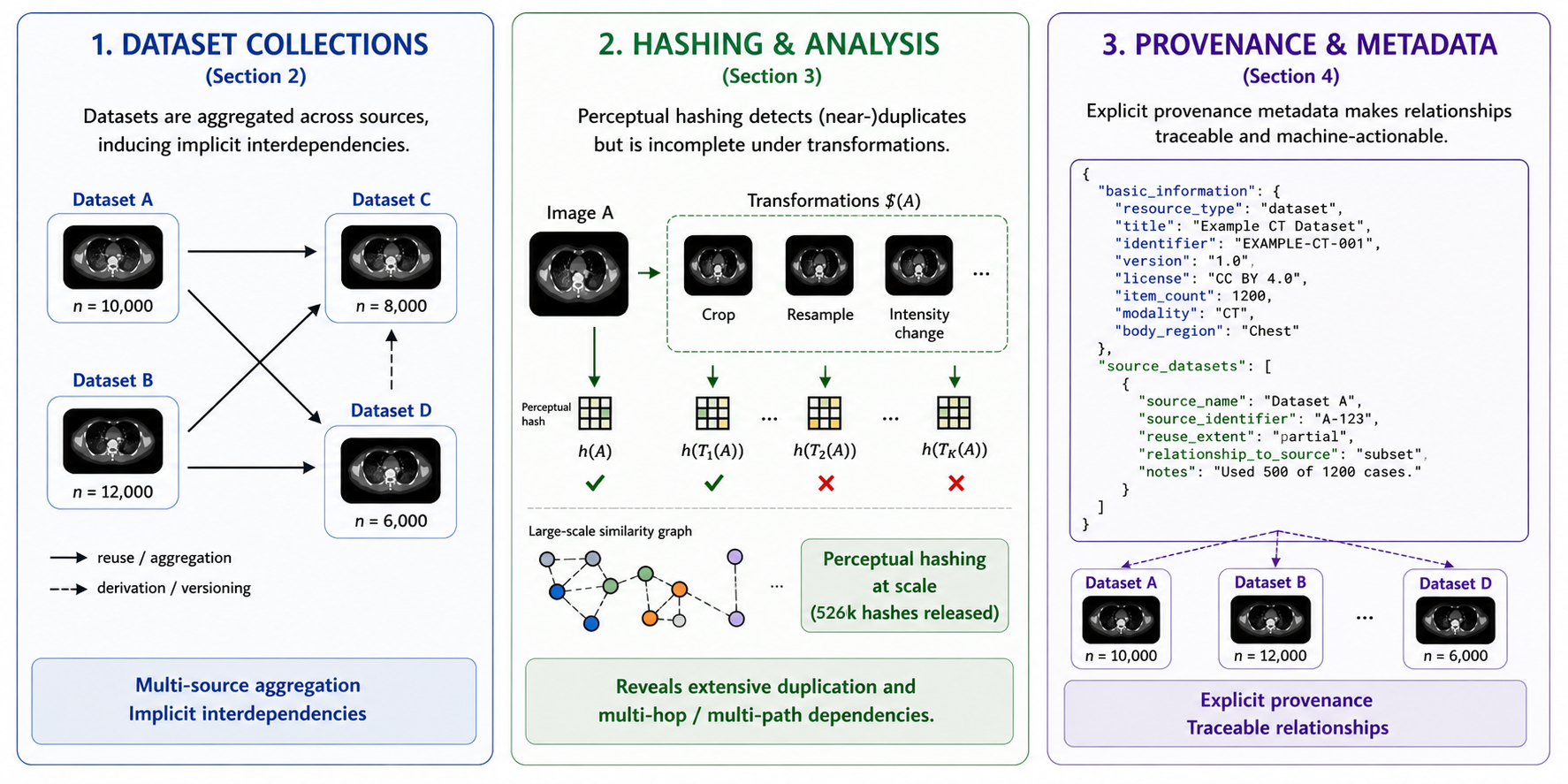}
    \caption{\textbf{Dataset Collections in Medical Image Analysis.} Large-scale medical datasets are increasingly dataset collections formed through the aggregation of images, labels and metadata from multiple sources. However, limited provenance can introduce hidden dependencies and duplicates. In this paper, we (1) highlight the risks of dataset reuse, (2) detect duplicates via perceptual hashing, (3) document provenance through structured metadata profiles, and (4) provide associated tooling through a web-based platform (\cref{apx:webtool}).}
    \label{fig:website-pipeline}
\end{figure}

In this paper, structured as in \cref{fig:website-pipeline}, we introduce dataset collections for medical images, and contrast their challenges against those of individual datasets and multi-center cohorts. We argue that post hoc deduplication is inherently incomplete, and that reliable use of dataset collections requires explicit provenance. Our contributions are as follows:

\begin{enumerate}[leftmargin=*, itemsep=0pt, topsep=0pt]
    \item \textit{Formalizing Dataset Collections.}
    We formalize the concept of dataset collections as large-scale aggregations of existing datasets used to enable training at scale in medical AI. We distinguish them from curated multi-center cohorts and identify fundamental challenges arising from this paradigm, including issues of data provenance, duplication, and implicit dependencies between datasets. We further analyze how these factors impact both model development and evaluation.
    \item \textit{Large-Scale Empirical Analysis via Perceptual Hashing.}
    To systematically study these challenges, we construct a large-scale dataset collection comprising \textit{\nVolumesShort{} 3D medical images}. Using perceptual hashing, we uncover extensive duplication and multi-hop relationships across constituent datasets, revealing a high degree of structural entanglement. To facilitate reproducibility and external validation, we make our hashes available for public comparison.
    \item \textit{Minimal Provenance Metadata and Tooling.}
    We propose a lightweight metadata profile \textit{(MAP)} for documenting provenance in dataset collections, capturing key attributes such as data origin, composition, and processing. To support adoption, we provide a web-based tool that assists dataset authors in constructing such metadata while enabling automated comparisons against our perceptual hash database. This framework offers a practical pathway toward improving traceability and responsible use of dataset collections in medical AI, and can be used both prospectively and retrospectively.
\end{enumerate}

\begin{figure}
    \centering
\includegraphics[width=\linewidth]{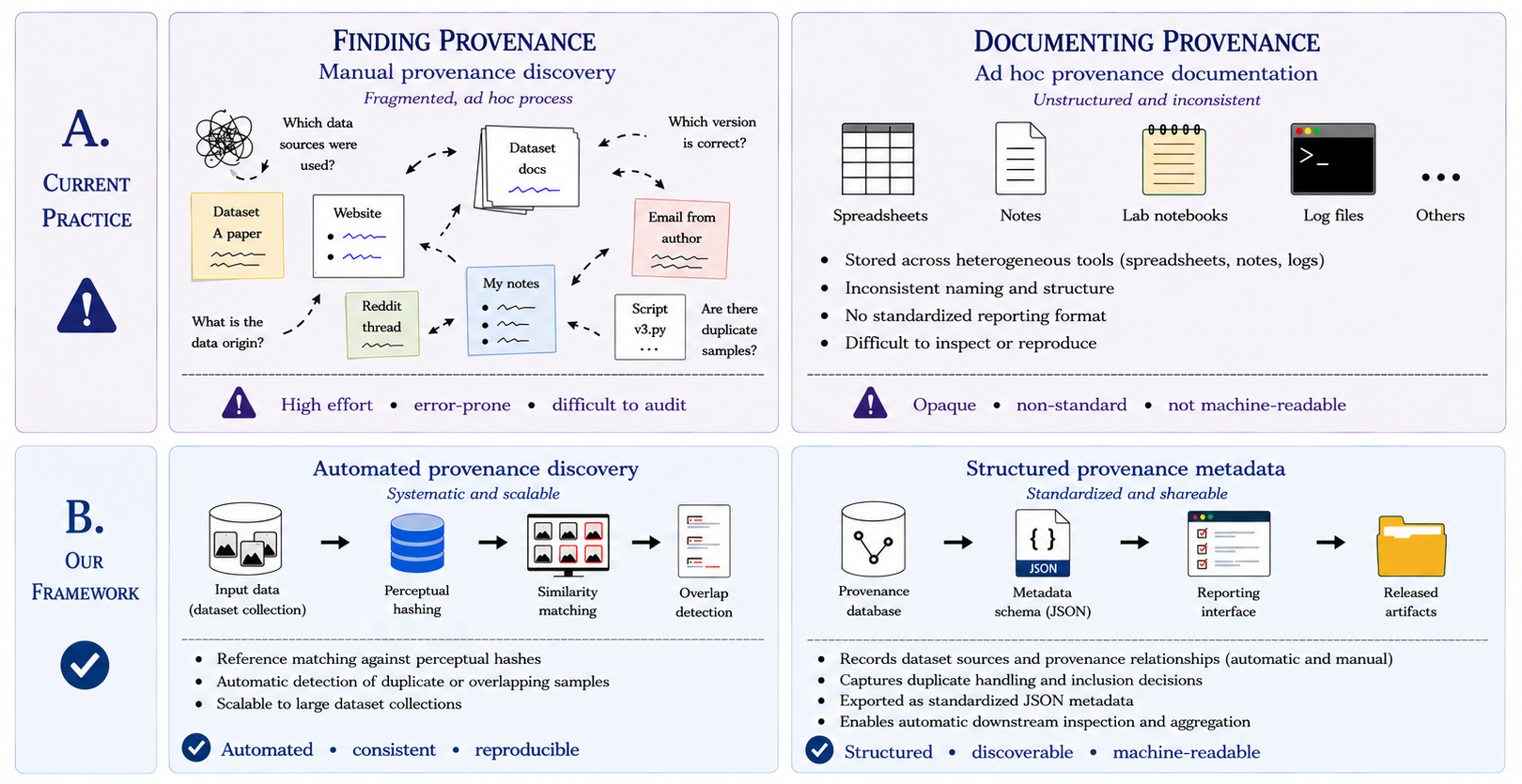}
    \caption{\textbf{Structural Provenance Documentation via Metadata Profiling.} (A) Provenance is currently reconstructed manually from papers, websites, notes and correspondence, and documented ad hoc across heterogeneous tools. (B) We screen dataset collections for overlap by perceptual hashing against a reference hash database (\cref{sec:data_dedup}) and record provenance and deduplication decisions as structured, machine-readable MAP metadata (\cref{sec:metadata}).}
    \label{fig:framework}
\end{figure}

\textbf{Scope and Positioning.}
We do not aim to exhaustively enumerate all duplication in medical imaging data (\cref{fig:umap_viz}) or to list every available dataset collection (\cref{tab:dataset_collections}). Instead, we formalize the \textit{concept} of dataset collections, providing examples to distinguish them from existing use cases, and study their challenges through a large-scale analysis of $\sim$\nVolumesShort{} 3D images. As practical steps, we evaluate perceptual hashing for overlap detection while making the hashes publicly accessible for comparison through our website, and introduce a minimal, portable provenance metadata profile for dataset collections (\cref{fig:framework}).

\section{Dataset Collections: Structure, Provenance, and Challenges} \label{sec:dataset_collections}
\definition{We define \textit{\textbf{Dataset Collections}} in medical image analysis as aggregates of public or non-public medical imaging datasets, which may or may not be uniquely sourced themselves from imaging endpoints (e.g. a clinic or a hospital).}

Dataset collections do not merely represent \textit{one large dataset} made from many small datasets. They can be thought of as aggregations of nodes in the open data ecosystem of medical image analysis (\cref{fig:umap_viz}). The objective of dataset collections in medical image analysis is to enable training at scale, which may be supervised \cite{xu2026cads,bassi2024touchstone}, semi- or self-supervised \cite{MedIA-FLARE21,qu2023abdomenatlas}, followed by evaluation via task-specific fine-tuning \cite{wald2025openmind,cerri2025large,roy2025mednext} or direct inference \cite{bassi2024touchstone,xu2026cads,rokuss2025lesionlocator}. Such collections systematically aggregate data from multiple geographies, demographics, study types (screening vs diagnostic), scanner types, licenses, etc., levying a massive coordination burden on authors. While issues of provenance and deduplication are often addressed on a \textit{per-collection} basis, it is often not done in a manner that can be trivially replicated downstream. This is exacerbated by increasing scale over the last half decade (\cref{tab:dataset_collections}), with collections increasing from 1k images to more than 300k. We discuss points of interest, as well as issues with dataset collections, while acknowledging the correlations between these points.

\textbf{Distinction from Multi-Center Cohorts.}
Dataset collections are aggregations of publicly or privately available datasets, usually constructed to enable scaling of AI training. They share similarities with multi-center datasets and cohorts (e.g., \texttt{PI-CAI} \cite{saha2024artificial}, \texttt{TotalSegmentator} \cite{wasserthal2023totalsegmentator}, \texttt{PANORAMA} \cite{alves2026artificial}), and may even appear to provide broad coverage owing to the use of multiple data sources. But they are distinct in that they typically aggregate pre-existing public datasets or collections, and are primarily intended for scaling AI training, rather than being carefully curated with explicit inclusion criteria, study design, and population definitions for rigorously designed clinical studies as enabled by multi-center cohorts.

\textbf{Data Provenance in Collections.}
Provenance of a data object refers to its origin, modification, and usage over time \cite{fairweather2020non}.
Data provenance is a key component of reliable clinical decision-making \cite{murray2022data} and the development of medical AI \cite{shaw2019artificial,werder2022establishing,ahmed2023data}, the lack of which can lead to cascades of studies built on unreliable models \cite{gibson2026evidence_2}. The FAIR principles emphasize the importance of provenance as part of data stewardship \cite{wilkinson2016fair}, and similarly, the EU Medical Device Regulation (MDR) emphasizes documentation of data used for validation and verification in clinical systems \cite{eu_mdr_2017_745}. More recently, \textit{Chapter III, Section 2, Article 10} of the EU Artificial Intelligence Act \cite{eu_ai_act_2024_article10} codifies requirements for data governance in high-risk AI systems (which may include medical AI), explicitly emphasizing data collection processes and the origin of data.
However, dataset collections often do not actively address provenance, as their primary goal is typically large-scale data aggregation for model training. As a result, the burden of establishing provenance is implicitly shifted to end users, who must infer data origins, preprocessing steps, disease states, and inclusion criteria. With the growing scale and complexity of dataset collections, this process is increasingly infeasible, motivating the need for explicit mechanisms to ensure provenance and support reliable use of collections in AI development.

\begin{table}[t]
\centering
\captionsetup{skip=6pt}
\caption{
\textbf{The Rapid Growth of Dataset Collections in 3D Medical Image Analysis.} Dataset collections have grown massively leveraging disparate public sources, while \textit{standard} datasets are released through a unified curation pipeline at a slower pace. We conduct our analysis on one of the largest collections to date with \textbf{$\sim$\nVolumesShort{}} images from \textbf{\nDatasetsAll{}} individual datasets (\cref{sec:data_dedup}).
}
\setlength{\tabcolsep}{4pt}
\renewcommand{\arraystretch}{1.0}
\begin{adjustbox}{width=0.975\linewidth}
\begin{tabular}{r c c c r c c }
\toprule
\textbf{Dataset} & \textbf{Year} & \textbf{Modality} & \textbf{Category} & \textbf{Scale} & \textbf{Composition} & \textbf{Provenance} \\
\midrule
\rowcolor{yellow!20}\textbf{PI-CAI} \cite{saha2024artificial} & 2023 & MRI & Multi-Center & $\sim$1.5k & 4 Centers & - \\
\rowcolor{yellow!20}\textbf{CT-RATE} \cite{hamamci2026generalist} & 2026 & CT & Single-Center & $\sim$25k & 1 Center & - \\
\midrule
\rowcolor{gray!20} \textbf{Abd.CT-1K} \cite{ma2021abdomenct} & 2020 & CT & Collection& $\sim$1k & 6 Datasets & Limited \\
\textbf{FLARE21} \cite{MedIA-FLARE21} & 2021 & CT & Collection & $<$1k & 2 Centers & Limited \\
\rowcolor{gray!20} \textbf{ULS23} \cite{degrauw2025uls23} & 2023 & CT & Collection& $\sim$10k & 14 Datasets & Limited \\
\textbf{FLARE23} \cite{ma2024automatic} & 2023 & CT & Collection & $\sim$4k & 50 Centers & Limited \\
\rowcolor{gray!20} \textbf{Abd.Atl.1.0} \cite{qu2023abdomenatlas} & 2023 & CT & Collection & $\sim$5k & 20 Datasets & Limited \\
\textbf{FLARE24} \cite{codabench_flare2024} & 2024 & CT & Collection& $\sim$10k & \textit{Unpublished} & Limited \\
\rowcolor{gray!20} \textbf{Abd.Atl.1.1} \cite{li2024abdomenatlas} & 2024 & CT & Collection & $\sim$10k & 88 Centers & Limited \\
\textbf{FLARE25} \cite{flare2025_challenge} & 2025 & CT & Collection & $\sim$10-20k & \textit{Unpublished} & Limited \\
\rowcolor{gray!20} \textbf{CADS} \cite{xu2026cads} & 2025 & CT & Collection & $\sim$22k & 42 Datasets & Filenames \\
\textbf{FOMO60k} \cite{munk2026towards} & 2025 & MRI & Collection& $\sim$60k & 16 Datasets & Limited \\
\rowcolor{gray!20} \textbf{OpenMind} \cite{wald2025openmind} & 2025 & MRI & Collection& $\sim$114k & 800 Datasets & Filenames \\
\textbf{FOMO300k} \cite{cerri2025large} & 2025 & MRI & Collection& $\sim$300k & 37 Datasets & Limited \\
\midrule
\rowcolor{blue!20}\textbf{Our Collection} & 2026 & Mixed & Collection& $\sim$\textbf{\nVolumesShort{}} & \textbf{\nDatasetsAll{} datasets} & \cref{sec:data_dedup} \\
\bottomrule
\end{tabular}
\end{adjustbox}
\label{tab:dataset_collections}
\end{table}

\textbf{The Structural Problem of Repeated Samples.}
A central consequence of limited data provenance in dataset collections is the repeated reuse of datasets and, in many cases, individual data samples. As collections grow through multi-source aggregation, implicit relationships between datasets emerge, often leading to duplicated or near-duplicated samples across sources. This violates the assumption that samples are independently drawn and inflates the effective size of the dataset.

A useful way to understand this phenomenon is to view dataset reuse as a directed graph \(G=(V,E)\), where nodes signify datasets and directed edges encode reuse, containment, or derivation. In this view, dataset collections are not flat unions of independent sources, but reuse graphs in which dependencies propagate along edges and accumulate across paths. The public medical image ecosystem has been driven by substantial effort from individual teams, each with their own data release strategies. Over time, this has led to the emergence of multi-hop relationships between datasets, whether through successive versions of a dataset (e.g., \texttt{BraTS}, which has been reused, reannotated, and updated over a decade \cite{ghaffari2019automated}) or through overlapping datasets released under different names (e.g., \texttt{LiTS} \cite{bilic2023liver} and \texttt{MSD03} \cite{antonelli2022medical}). At current scales, such dependencies are difficult to detect, particularly when datasets are repeatedly aggregated across multiple stages (e.g., \texttt{MSD} \(\rightarrow\) \texttt{AbdomenCT1k} \(\rightarrow\) \texttt{CADS}). As a result, dataset collections may contain fewer unique samples than their nominal size suggests, with direct implications for both model training and evaluation. We highlight two interesting cases below:

\begin{itemize}[leftmargin=*, itemsep=0pt, topsep=0pt]
\item[]
\textbf{Case 1: \textit{Multi-path reuse}.}
A source dataset may be reused along multiple outgoing edges in the reuse graph. For example, both \texttt{FLARE23} \cite{ma2024automatic} and \texttt{DAP-Atlas} \cite{jaus2024unifying} include \texttt{AutoPET} \cite{gatidis2022whole}. This creates a shared upstream dependency, meaning that training on one collection and evaluating on the other may violate independence assumptions unless the overlap is explicitly handled.

\item[] \textbf{Case 2: \textit{Multi-hop dependencies}.}
Dependencies may also propagate across multiple edges. For example, \texttt{AbdomenCT1k} \cite{ma2021abdomenct} aggregates \texttt{KiTS19} \cite{heller2019kits19}, \texttt{LiTS}/\texttt{MSD03-Liver}, \texttt{MSD07-Pancreas}, \texttt{MSD09-Spleen} \cite{antonelli2022medical,bilic2023liver}, and \texttt{NIH Pancreas} \cite{roth2015deeporgan}. However, \texttt{LiTS} also appears in \texttt{CT-ORG} \cite{rister2020ct}, and both \texttt{CT-ORG} and \texttt{AbdomenCT1k} are later included in \texttt{AbdomenAtlas1.1} \cite{li2024abdomenatlas}. This illustrates how overlapping lineage can arise even in carefully curated collections, and why safe inclusion has required and continues to require substantial manual provenance tracking by dataset authors.

\end{itemize}

In our analysis of a dataset collection comprising \textbf{\nVolumesShort{}} 3D volumes (\cref{sec:data_dedup}) using perceptual hashing (\cref{alg:hash_similarity_search}), we detect at least \textbf{\nDuplicatePairsShort{}} duplicate pairs, involving \textbf{\nDuplicatesShort{}} images. This scale of duplication provides empirical evidence that reuse relationships are both widespread and structurally non-trivial. The resulting reuse graph, visualized in \cref{fig:umap_viz}, reveals a dense web of interdependencies across datasets, highlighting how duplication arises not only from direct reuse but also through multi-hop and multi-path relationships. These observations underscore the difficulty of resolving such overlap post hoc and motivate the need for systematic approaches to make dataset provenance explicit.

\textbf{Challenges of Dataset Collections.} Building on the above, we outline key challenges that dataset collections introduce for model training and medical studies.

\begin{itemize}[leftmargin=*]

\item \textbf{Breakdown of Scaling Laws.}
Research into scaling laws assumes that performance, $f(x)$, improves predictably with increasing scale ($x$) in terms of data, model size, and compute, often expressed as a power-law relationship, $f(x)=\alpha x^\beta$ \cite{hestness2017deep,kaplan2020scaling,klug2022scaling}. This has also recently been explored in medical image analysis \cite{ilse2025data,chu2025revisiting,chen2025scaling}. However, there is evidence of different scaling behavior in the absence of independent and identically distributed (i.i.d.) data (e.g., underrepresented concepts in synthetic text-to-image data \cite{fan2024scaling}). Dataset collections with a lack of provenance guarantees can therefore lead to misleading conclusions about data efficiency, model capacity, and generalization in scaling law studies due to duplication. This is particularly problematic if such studies govern the future construction of larger medical image datasets, which is an expensive endeavor.

\item \textbf{Implicit Bias Amplification.}
Bias in AI models for medical image analysis refers to systematic differences in model behavior arising from imbalances or confounding factors in the data (e.g., certain scanners, institutions, or patient demographics) \cite{cross2024bias,hanna2025ethical,hasanzadeh2025bias,drenkow2025detecting,xiao2025bias}. Unmitigated biases can reinforce subgroup performance disparities \cite{stanley2025and} or lead to shortcut learning \cite{boland2024there,lin2024shortcut}. Dataset collections of increasing size with limited provenance increase the likelihood that certain subgroup-specific patterns are overrepresented in the training data, thereby distorting the data distribution and amplifying implicit biases. Such aggregation introduces multiple sources of bias, including selection bias (systematic differences in cohort inclusion criteria), label bias (inconsistencies in annotation protocols or ground truth definitions) in case existing labels are re-used \cite{ulrich2023multitalent}, and prevalence bias (shifts in class distributions across datasets). This effect is further exacerbated when knowledge from such models is propagated downstream (e.g., pseudo-labeling \cite{liang2020we,qu2023abdomenatlas} or zero-shot inference \cite{wasserthal2023totalsegmentator}).

\item \textbf{Overestimated Robustness Measures.} Large curated multi-center cohorts \cite{saha2024artificial,alves2026artificial,stegeman2026designing} often explicitly ensure traceability of data sources and controlled distributional coverage to support studies of algorithmic robustness \cite{paschali2018}. Dataset collections, in contrast, are designed primarily to increase training scale. While measuring robustness at scale is critical to developers and medical AI vendors, caution should be exercised in replacing such cohorts with collections, as apparent dataset diversity in collections does not necessarily correspond to independent or controlled distributions.

\item \textbf{Limited Diversity Leads to Performance Saturation.}
Foundation models in medical image analysis have been enabled by aggressive data scaling \cite{moor2023foundation,khan2025comprehensive}. However, there is an increasing body of research suggesting that data diversity plays a critical role in efficient training \cite{joshi2024data}. This has recently been explored in biomedical image analysis \cite{bosch2026diversity,wang2026data}, with findings indicating that diversity, rather than raw dataset size alone, governs performance gains. Dataset collections with limited provenance run the risk of aggregating data without equivalently increasing diversity. Duplication collapses the effective number of samples in the dataset \cite{cui2019class}, thereby leading to premature performance saturation in models trained on such data, despite continued increases in nominal dataset size.

\item \textbf{Unintended Data Leakage.}
The independence between training and evaluation sets is fundamental to reliable empirical assessment \cite{bishop-2006}. Dataset collections constructed from multiple sources can induce implicit connections between training and evaluation data through shared or duplicated samples. This is particularly problematic when -- \textit{\textbf{A)}} collections involve \textit{multiple} and \textit{multi-hop} aggregations (e.g., Dataset A $\rightarrow$ B $\rightarrow$ C and Dataset A $\rightarrow$ D) as seen in \cref{fig:umap_viz}, where tracking overlaps becomes harder, and \textit{\textbf{B)}} dataset collections increasingly exhaust public datasets, so the pool of publicly available non-overlapping evaluation datasets shrinks \cite{xu2026cads}. As a result, the risk of violating independence assumptions between training and evaluation sets increases, leading to overestimation of robustness, unreliable effect sizes, and misleading comparisons across methods.
Furthermore, standard validation strategies such as cross-validation within dataset collections \cite{isensee2021nnu} become unreliable unless duplication is explicitly controlled, as folds may inadvertently share duplicated samples.

\end{itemize}

\section{Data Deduplication: An Imperfect Solution to an Ill-Posed Problem}
\label{sec:data_dedup}

A natural response to hidden overlap in dataset collections is to attempt deduplication after the collection has already been assembled. In principle, this appears straightforward: if duplicated or reused samples can be detected, they can be removed or at least flagged before model development and evaluation. In practice, however, post hoc deduplication is an imperfect solution to an ill-posed problem. Public medical imaging datasets often lack stable cross-dataset identifiers, explicit lineage information and details about imaging data handling, including transformations like intensity clipping, resampling or cropping.

This ambiguity means that deduplication cannot be reduced to a single well-defined operation. Exact file-level deduplication, using cryptographic checksums such as MD5 or SHA-family hashes for example, is useful when the same file is redistributed without modification. However, these hashes are intentionally sensitive to any change in the input: even a minor alteration in metadata, compression, orientation, voxel spacing, or file encoding can produce a completely different hash value.

In this work, we use perceptual hashing as a lightweight mechanism for large-scale duplicate screening. The method is designed to detect exact and near-exact image-level reuse while tolerating small changes that would break cryptographic hashes. This makes it well-suited as a first-pass tool for dataset collections: it is computationally inexpensive, produces compact representations that can be shared without distributing the underlying medical images, and enables efficient similarity search over hundreds of thousands of volumes. However, its outputs should be interpreted conservatively as candidate evidence of overlap, not as proof of complete lineage reconstruction.

\subsection{Perceptual Hashing}

Perceptual hashing maps an image to a compact representation that preserves coarse visual similarity rather than exact file identity \cite{monga2006perceptual,zauner2010implementation}. This differs fundamentally from cryptographic hashing: cryptographic hashes are designed so that small input changes produce unrelated outputs, a property that is essential for integrity verification, but poorly matched to duplicate detection under benign preprocessing. Perceptual hashes, by contrast, are designed so that visually similar inputs produce identical or nearby hash codes. Similarity can then be measured by comparing hashes, commonly using Hamming distance for binary hash vectors.

Our perceptual hash calculation is inspired by the average-hash family of methods described by Krawetz \cite{krawetz2011looks}, but adapted to volumetric medical images. Given a 3D image $I$, we first resample it to a fixed $16 \times 16 \times 16$ grid, producing $I_r$. We then compute the global median intensity $m$ of the resampled image and encode each voxel according to whether it lies above or below this median. The resulting binary volume is flattened into a fixed-length hash.

This procedure intentionally discards high frequency details and intensity scales, making it robust to transformations typically used in data processing pipelines, such as intensity clipping, while being substantially cheaper than registration-based or learned similarity methods.
To detect duplicates we compute the 4096-dimensional perceptual hashes for all eligible images and combine the resulting binary vectors into a single shared hash matrix. This matrix is indexed with an Annoy approximate nearest-neighbor index with 1024 trees~\cite{bernhardsson_annoy}, enabling an efficient large-scale similarity search across the full dataset collection without exhaustive pairwise comparison. Candidate matches are retrieved from the ANN index and evaluated by Hamming distance, after which duplicate and near-duplicate image pairs are defined using configurable distance thresholds. The resulting duplicate graph is then summarized at dataset level and used to generate filtered edge tables, statistics, and stricter low-distance analyses. The hash similarity search is also summarized in \cref{alg:hash_similarity_search} (\cref{apx:percetual hashing}).

The main advantage of this approach is its simplicity. It does not require access to patient identifiers, DICOM headers, scanner metadata, or original acquisition records. It can be computed locally, shared as compact binary signatures, and compared against a reference database without exposing the original medical images. This makes perceptual hashing useful as a lightweight, privacy-conscious screening tool for dataset collections. Importantly, the method is more tolerant than cryptographic hashing: two images that differ only by minor preprocessing may still produce hashes with small Hamming distance, whereas their cryptographic hashes would be unrelated.

\subsection{Results}\label{sec:dedup_results}

\begin{figure}[ht]
    \centering
    \includegraphics[width=\textwidth]{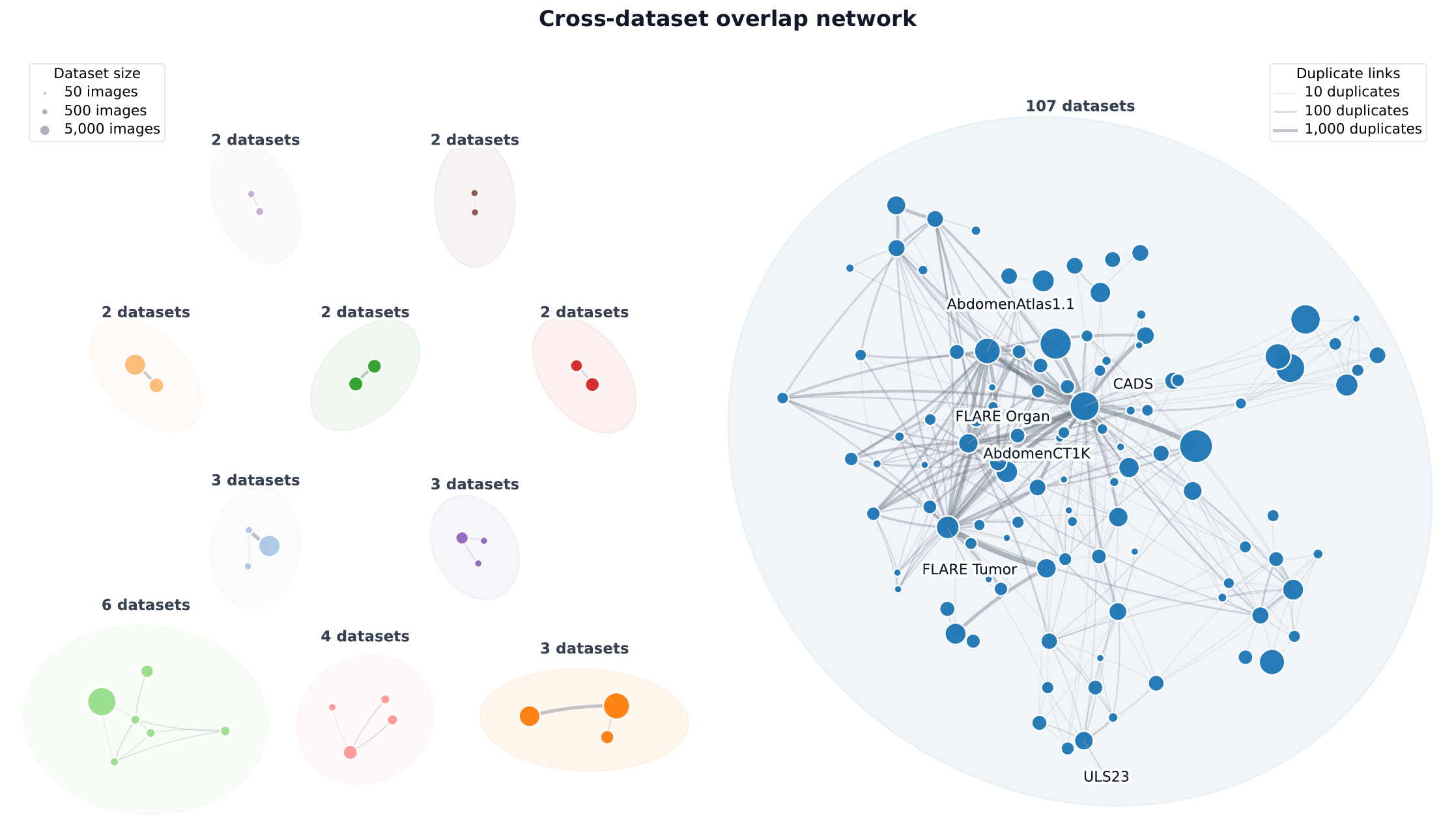}
    \caption{\textbf{Cross-dataset overlap network revealed by exact and near-duplicate 3D medical images.} Each node represents a public dataset or dataset collection, and edges connect datasets that share at least one image pair with perceptual Hamming distance $\leq 100$ or identified through dataset-specific deduplication and manual overlap tracing. Node size indicates the number of images in a dataset, edge size indicates the number of duplicate links associated with a dataset, and shaded regions denote connected components in the overlap graph. The largest component contains \nMainCluster{} datasets, including widely reused datasets and collections such as \texttt{AMOS}, \texttt{KiTS}, and \texttt{AbdomenAtlas}, illustrating that public dataset collections are not independent flat unions but part of a densely connected reuse ecosystem. Because perceptual hashing detects visual identity rather than causal lineage, edges should be interpreted as evidence of image-level overlap rather than complete provenance relationships.}
    \label{fig:umap_viz}
\end{figure}

We collected a total of \nDatasetsAll{} datasets from public sources, including \texttt{TCIA}, \texttt{OpenNeuro}, \texttt{Huggingface}, \texttt{grand-challenge} and \texttt{Kaggle}, collectively comprising a total of \nVolumes{} 3D volumes. Of these, \nDatasetsOpenNeuro{} datasets come from \texttt{OpenNeuro}; for the dataset-level analysis, we treat them as a single dataset, which gives \nDatasetsMerged{} datasets. A detailed overview of all datasets can be found in \cref{apx:collection}. Out of these, we identified a total of \nDuplicatePairs{} duplicate pairs. Perceptual hashing (Hamming distance $\leq 100$) found \nDuplicatePairsAuto{} of them, including \nDuplicatePairsExact{} exact duplicate pairs (Hamming distance 0). Another \nDuplicatePairsManual{} duplicate pairs were identified using dataset-specific modifications of the deduplication algorithm, such as intensity clipping, and manual tracing of dataset overlaps. Overall, \nDuplicates{} volumes have at least one duplicate; deduplication removes \nRemovedCopies{} duplicated copies and leaves \nUniqueVolumes{} unique volumes, a reduction in effective dataset size of \nReduction{}\%.

\cref{fig:umap_viz} visualizes the connections between public datasets and dataset collections. In total, we find at least partial overlaps for \nDatasetsOverlap{} of the \nDatasetsMerged{} datasets. We identify multiple clusters, with the main cluster comprising a total of \nMainCluster{} datasets, including popular datasets like \texttt{AMOS}~\cite{ji2022amos}, \texttt{KiTS}~\cite{heller2019kits19,heller2023kits21} and \texttt{LiTS}~\cite{bilic2023liver} and collections such as \texttt{AbdomenCT-1K}~\cite{ma2021abdomenct}, \texttt{FLARE}~\cite{ma2024automatic} and \texttt{AbdomenAtlas}~\cite{li2024abdomenatlas,bassi2024touchstone}. The remaining smaller clusters mainly correspond to multiple iterations of challenges, reuse of data in different challenge tasks, and the OpenNeuro ecosystem.

\subsection{Validation \& Limitations}
\label{sec:vallim}

\textbf{Manual validation.}\label{sec:manval}
We manually validated the duplicate detection on 5{,}000 image pairs, using stratified pairwise distance-based random sampling: 1{,}000 pairs were sampled randomly below the threshold (Hamming distance $< 100$), which the pipeline flags as duplicates, and 4{,}000 pairs above it (Hamming distance $> 100$), which it does not flag. Each pair consists of an image and its top-1 candidate from the ANN search, and each pair was labeled manually as duplicate or non-duplicate. Of the flagged pairs, 989 were true duplicates (positive predictive value, PPV, 0.99); of the unflagged pairs, 3{,}026 were indeed no duplicates (negative predictive value, NPV, 0.76), while 974 were duplicates missed at this threshold, many of them different reconstructions or contrast phases of the same scan (\cref{apx:manval}). These findings suggest that, at the chosen threshold, perceptual hashing yields precise duplicate candidates but has limited recall, likely underestimating the full extent of provenance-related duplication, in particular for less similar pairs. We deliberately chose this low threshold so that detected duplicates are reliable, which makes the number of detected duplicates a lower bound on the true number of duplicates. Depending on the use case, the threshold can be adapted to balance false negatives and false positives. More details can be found in \cref{apx:manval}.

\textbf{Looking Beyond Similarity.}
Perceptual hashing \cite{samanta2021analysis,zhou2025survey} is a simple and scalable heuristic for detecting image-level duplication, but it has important limitations. \textit{Firstly}, it is not invariant to transformations such as rotations, flips, cropping, padding, or changes in field of view. \textit{Secondly}, the resulting hash depends on the image loading and preprocessing pipeline, including format conversion, orientation handling, and resampling. As a consequence, the same underlying scan may yield different hashes after seemingly minor processing changes. \textit{Thirdly}, perceptual hashing is primarily sensitive to exact and near-exact duplicates and is not designed to identify semantically related but non-identical images.
Formally, let $A$ denote an image and $\mathcal{T}(\cdot)$ a simple transformation such as cropping, resampling, or intensity normalization. Then,
$
\exists \ \mathcal{T} \ \text{s.t.} \ \mathrm{pHash}(\mathcal{T}(A)) \neq \mathrm{pHash}(A),
$
even though $\mathcal{T}(A)$ and $A$ may originate from the same underlying acquisition. We provide a more detailed analysis of the effect of transformations on hash similarity in \cref{apx:hashlimits}.
\textit{Finally}, image similarity alone does not establish provenance. Matches indicate potential overlap or reuse, but do not explain how images were derived, transformed, or redistributed across datasets. Deduplication can answer whether samples look similar, but not whether they are causally related, making it a lossy proxy for tracing data reuse. As a result, deduplication is inherently incomplete and cannot explain how data propagates across datasets, motivating the need for explicit provenance.

\section{MAP: A Metadata Profile for Dataset Collections}
\label{sec:metadata}
Public medical imaging datasets rarely provide complete lineage information. As discussed in \cref{sec:dataset_collections}, large-scale collection curation substantially increases the burden on curators to track dataset provenance in a reusable and consistent manner. In practice, a curator may know the repository name, citation, or DOI of a dataset, but not the full history of how it was assembled. Datasets may represent direct institutional releases, subsets of prior datasets, relabeled versions, or aggregates of multiple public sources \cite{jimenez2024copycats}. As collections are constructed through successive aggregation, the connection to the original point of acquisition is often weakened or lost. Consequently, deduplication is typically performed without reliable knowledge of which data points were reused, transformed, or republished. This leads to two key challenges: (i) duplicate detection becomes technically difficult in the absence of stable identifiers or consistent metadata, and (ii) even when overlaps are detected, the results are often incomplete and difficult to interpret.

To address this, we introduce \textbf{MAP (Metadata for Aggregation and Provenance)}, a lightweight metadata profile for dataset collections, together with a web-based tool to support its practical use. MAP is designed to provide a simple, machine-readable representation of dataset provenance, reuse, and deduplication decisions, without imposing the overhead of comprehensive metadata standards. In contrast to narrative documentation frameworks such as \textit{Datasheets for Datasets} \cite{gebru2021datasheets}, which emphasize descriptive transparency, MAP focuses on structured provenance and reuse relationships, enabling scalable analysis of dependencies across dataset collections.

\begin{figure}[h]
    \centering
    \includegraphics[
    width=0.95\linewidth,
    trim=15 25 15 15,
    clip]{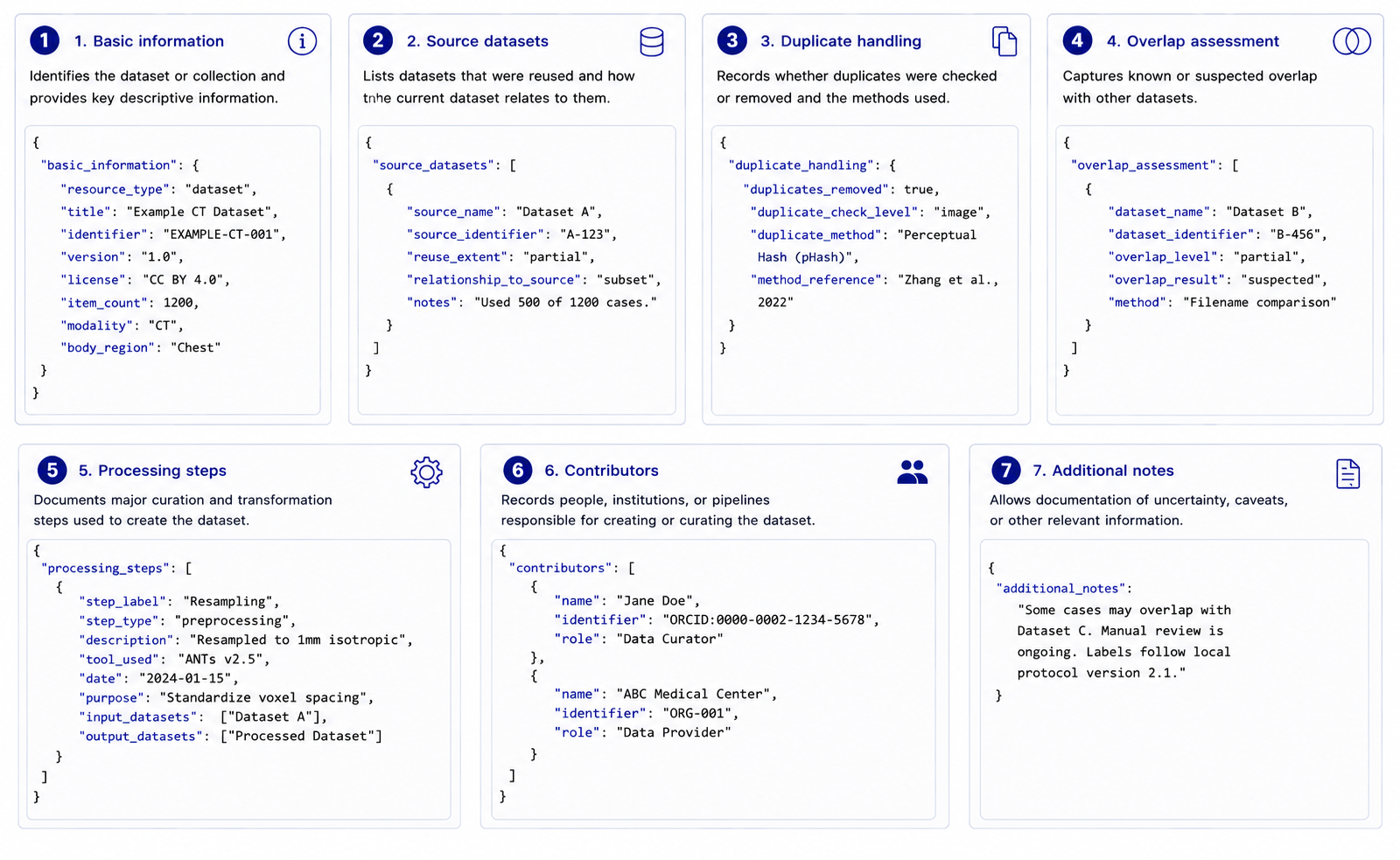}
    \caption{\textbf{The MAP Profile can be fragmented into 7 constituent areas} (\cref{apx:MAP}). Notably, the \textit{Source Dataset} section encodes provenance, while \textit{Duplicate Handling} makes the efforts of the authors transparent regarding their work in handling repeated samples. Finally, \textit{Preprocessing Steps} clarifies the modifications made to the source images in the collection for downstream reference.}
    \label{fig:metadata_snippets}
\end{figure}

\textbf{Rationale and Structure.}
Dataset collections are evolving compositions of datasets with implicit and often undocumented reuse relationships. In the absence of explicit provenance, these relationships remain latent, making it difficult to determine whether datasets are independent, partially overlapping, or derived from one another. This ambiguity directly undermines the interpretation of training scale, evaluation protocols, and reported generalization. Therefore, instead of proposing a comprehensive metadata standard such as \cite{suri2026roadmap}, we ask a more constrained question -- \textit{What is the minimal information required to make reuse, overlap, and lineage explicit at scale}?

As illustrated in \cref{fig:metadata_snippets}, the resulting MAP profile is a structured, machine-readable representation of dataset-level provenance, designed around five principles: (i) \textit{minimal}, capturing only information that affects reuse interpretation; (ii) \textit{simple}, enabling lightweight creation and export as JSON; (iii) \textit{structure-agnostic}, independent of dataset organization; (iv) \textit{traceability-first}, prioritizing lineage and reuse relationships; and (v) \textit{adoptable}, requiring minimal overhead for dataset curators and researchers.
Concretely, MAP encodes provenance at two levels. At the \textit{collection level}, it records identifiers, constituent datasets, overlap assessments, and deduplication practices. At the \textit{dataset level}, it records dataset identity, reuse type (full, partial, subset), derivation relationships, key transformations, and licensing and citation metadata. This structured representation exposes dependencies that are otherwise implicit, enabling systematic analysis of dataset reuse and supporting more reliable utilization of large-scale dataset collections, both existing and in the future (\cref{apx:prospect}). A detailed description of fields and associated data types is available in \cref{apx:MAP}.

\textbf{Implementation and Portability.}
The MAP profile is accompanied by a web-based platform (\cref{apx:webtool}) for analyzing and documenting dataset collections and their deduplication processes. The platform provides visibility into existing collections, including their constituent datasets, detected overlaps, and applied deduplication strategies, and enables similarity-based inspection through a reference database of $\sim$\nHashesShort{} perceptual hashes derived from public datasets from \cref{sec:data_dedup}. This already allows users to compare their own data against a standardized baseline to identify potential overlap and shared provenance. Within this workflow, MAP serves as a lightweight, machine-readable disclosure layer: users can further populate the profile via a guided interface and export it as structured JSON for publication alongside collections. The profile is also portable, capturing provenance with fields that can be readily integrated into existing metadata frameworks (e.g., VIDS \cite{muthu2026vids}).

\section{Conclusion}
We formalize dataset collections as a distinct concept in medical image analysis and show that large-scale aggregation introduces challenges through repeated reuse and hidden dependencies. Using perceptual hashing, we reveal widespread reuse relationships across public datasets and make $\sim$\nHashesShort{} hashes publicly accessible for comparison through our website. Finally, we introduce MAP, a lightweight metadata profile for provenance in dataset collections, to support their reliable use in large-scale medical AI.

\begin{ack}
\paragraph{Funding.}
This work was supported by the Helmholtz Association under the Helmholtz Foundation Model Initiative (HFMI), project ``The Human Radiome Project'' (THRP) (Y.K., C.U., K.G.).
This work was supported by the Helmholtz Association under the joint research program ``HIDSS4Health -- Helmholtz Information and Data Science School for Health'' (Y.K., N.D., B.K.).
This work was supported by HELMHOLTZ IMAGING, a platform of the Helmholtz Information \& Data Science Incubator (M.R., T.W., K.G., R.S., S.E., K.-C.K., D.T., S.Z., D.Z., F.I.).
This work was supported by HELMHOLTZ METADATA COLLABORATION (HMC), a platform of the Helmholtz Information \& Data Science Incubator (H.H., A.D., S. Dvoretskii, L.K., M.N.).
Funded by the Deutsche Forschungsgemeinschaft (DFG, German Research Foundation) -- Project number 402688427 (Y.K.).
Funded by the Deutsche Forschungsgemeinschaft (DFG, German Research Foundation) -- Project number 410981386 (M.B.).
This work was supported by the Medical Informatics Initiative project ``PrivateAIM'' (01ZZ2316M) (E.S.).
This work was supported by NUM 3.0 (01KX2524) (B.H., P.S.).
This work was supported by NUM 2.0 (01KX2121) (B.H.).
This work was supported by the NCT Heidelberg Image Data Analysis Infrastructure (IDAI) (R.S., D.B., B.K., F.I.).
This work was supported by the German Cancer Consortium (DKTK, Strategic Initiative Joint Imaging Platform) (K.E., J.K.).
This work was supported by the European Union project EUCAIM (101100633) (K.E.).
This work was supported by the Google PhD Fellowship Program (M.R.).

\paragraph{Dataset acknowledgments.}
Data used in this work were obtained from The Cancer Imaging Archive (TCIA) \cite{clark2013tcia}, supported by the National Cancer Institute.
Data used in this publication were generated by the National Cancer Institute's Cancer Moonshot Biobank.
Data used in this publication were generated by the National Cancer Institute Clinical Proteomic Tumor Analysis Consortium (CPTAC).
The results shown here are in whole or part based upon data generated by the TCGA Research Network: \url{http://cancergenome.nih.gov/}.
The authors acknowledge the National Cancer Institute and the Foundation for the National Institutes of Health, and their critical role in the creation of the free publicly available LIDC/IDRI Database used in this study.
The COVID-19-AR collection has been funded in whole or in part with federal funds from the National Center for Advancing Translational Sciences UL1 TR003107 and the National Cancer Institute, Contract No. 75N91019D00024, Subcontract 20X023F.
The QIN-PROSTATE-Repeatability data were collected and shared with support from award U01 CA151261, PI Fiona Fennessy.
The QIN-SARCOMA data were collected and shared with support from award U01 CA154602, PI Wei Huang.
Data used in this work were obtained from the Human Tumor Atlas Network (HTAN) \cite{rozenblattrosen2020human}; collection of the HTAN-HMS and HTAN-OHSU images was supported by NCI grants U2CCA233262, U2CCA233280 and U24CA233243.
The imaging and associated clinical data downloaded from MIDRC (The Medical Imaging and Data Resource Center) and used for research in this publication was made possible by the National Institute of Biomedical Imaging and Bioengineering (NIBIB) of the National Institutes of Health under contract 75N92020D00021 and through The Advanced Research Projects Agency for Health (ARPA-H). The views and conclusions contained in this document are those of the authors and should not be interpreted as representing the official policies, either expressed or implied, of the U.S. Government.
We thank Istanbul Medipol University Mega Hospital for providing the CT-RATE data, and we acknowledge the providers of the AbdomenAtlas dataset.
Visible Human Project images courtesy of the U.S. National Library of Medicine.
The MRISegmentator-Abdomen data were provided by the NIH Clinical Center (\url{https://nihcc.box.com/s/zbrocs18g9ctrl0gb4n3hq7dlotpuwg5}) \cite{zhuang2025mrisegmenter}.
Data were provided in part by OASIS-1: Cross-Sectional: Principal Investigators: D. Marcus, R. Buckner, J. Csernansky, J. Morris; P50 AG05681, P01 AG03991, P01 AG026276, R01 AG021910, P20 MH071616, U24 RR021382 \cite{marcus2007open}.
We acknowledge the RibFrac project teams \cite{yang2024ribfrac,jin2020deep}.
Data used in this publication were obtained as part of the Brain Tumor Segmentation (BraTS) Challenge project through Synapse ID: syn53708249.
The SegTHOR database is endorsed by the Centre Henri Becquerel (CHB), Rouen, France.
The VALDO Task 2 data include data of the ALFA Study. We acknowledge the funding sources of the VALDO data: the Wellcome Trust (082464/Z/07/Z), the British Heart Foundation (SP/07/001/23603, PG/08/103, PG/12/29/29497 and CS/13/1/30327), Erasmus MC University Medical Center, the Erasmus University Rotterdam, the Netherlands Organization for Scientific Research (NWO) Grant 918-46-615, the Netherlands Organization for Health Research and Development (ZonMW), the Research Institute for Disease in the Elderly (RIDE), the European Union Seventh Framework Programme (FP7/2007--2013) under grant agreement No.~601055, VPHDARE@IT, and the Dutch Technology Foundation STW.
Data were provided in part by the Brain Genomics Superstruct Project of Harvard University and the Massachusetts General Hospital, (Principal Investigators: Randy Buckner, Joshua Roffman, and Jordan Smoller), with support from the Center for Brain Science Neuroinformatics Research Group, the Athinoula A. Martinos Center for Biomedical Imaging, and the Center for Human Genetic Research. 20 individual investigators at Harvard and MGH generously contributed data to the overall project.
The OpenBHB data further include the IXI dataset (\url{https://brain-development.org/ixi-dataset/}) and data from ABIDE I and II; we acknowledge the funding sources of the contributing sites listed on the ABIDE pages (\url{https://fcon_1000.projects.nitrc.org/indi/abide/}).
This data was obtained from the OpenNeuro database \cite{markiewicz2021openneuro}. Its accession numbers are ds000007, ds000030, ds000031, ds000052, ds000101, ds000102, ds000117, ds000140, ds000148, ds000164, ds000168, ds000203, ds000204, ds000205, ds000213, ds000214, ds000216, ds000217, ds000218, ds000219, ds000220, ds000221, ds000222, ds000223, ds000224, ds000229, ds000231, ds000232, ds000234, ds000235, ds000236, ds000237, ds000239, ds000241, ds000243, ds000244, ds000245, ds000246, ds000247, ds000253, ds000254, ds001226, ds001357, ds001415, ds001761, ds001784, ds002080, ds002158, ds002232, ds002550, ds002634, ds002643, ds002898, ds003020, ds003382, ds003424, ds003823, ds003872, ds003900, ds003974, ds003988, ds004068, ds004285, ds004386, ds004627, ds004910, ds004926 and ds005003.
The OpenNeuro datasets ds001420 and ds001421 originate from the Cimbi database.
We thank the providers of the OpenNeuro datasets ds002422 (Ya. Panikratova, A. Tomyshev, E. Pechenkova and R. Vlasova), ds003469 (E. V. Pechenkova, Y. R. Panikratova, M. A. Fomina, A. D. Rumshiskaya, D. A. Bazhenova, L. A. Makovskaya, I. S. Lebedeva and V. E. Sinitsyn), ds003763 (O. R. Dobrushina, L. A. Dobrynina, G. A. Arina, M. V. Gubanova, A. V. Belopasova, E. S. Novikova, A. D. Suslina, E. I. Kremneva, E. V. Pechenkova and M. V. Krotenkova), ds003082 (J. Cote, J.-P. R. Falet, V. Tarka, Z.-E. Martinez-Moreno and E. de Villers-Sidani) and ds005279 (H. T. Wei, F. B. Faisal, T. Beck, C. Shao and J. A. Meltzer).
The OpenNeuro dataset ds005234 is distributed under the CC BY license.
We acknowledge the Canada First Research Excellence Fund ``BrainsCAN'' grant to Western University for the OpenNeuro datasets ds003433, ds003434, ds003438, ds003439, ds003440, ds003441, ds003442, ds003443, ds003444, ds003445 and ds003446 and cite \url{www.newbi4fmri.com}.
The OpenNeuro dataset ds001415 is a product of dissertation research at the University of Oregon funded in part by NSF DDRI 1634086 and the Lewis Family Endowment.
We acknowledge the core facility MAFIL supported by the Czech-BioImaging large RI project (LM2018129 and LM2023050 funded by MEYS CR), part of the Euro-BioImaging (\url{www.eurobioimaging.eu}) ALM and Medical Imaging Node (Brno, CZ), for their support with obtaining scientific data (OpenNeuro dataset ds004499).
The QTAB project resource was produced as a result of i) the goodwill and contribution of 422 twin/triplet participants and their parents, ii) funding from the National Health and Medical Research Council, Australia (APP1078756) and the Queensland Brain Institute, University of Queensland, iii) access to several key resources, including the Centre for Advanced Imaging, the Human Studies Unit, Institute of Molecular Bioscience, and the Queensland Cyber Infrastructure Foundation, at the University of Queensland, local and national twin registries at the QIMR Berghofer Medical Research Institute and Twin Research Australia, as well as the many assessments made available by researchers worldwide, and iv) was established with the purpose of promoting the conduct of health-related research in adolescence.

\paragraph{Competing interests.}
The authors declare no competing interests.

\end{ack}

\newpage
\bibliographystyle{unsrtnat}
\bibliography{references}

@article{adlung2021machine,
  title={Machine learning in clinical decision making},
  author={Adlung, Lorenz and Cohen, Yotam and Mor, Uria and Elinav, Eran},
  journal={Med},
  volume={2},
  number={6},
  pages={642--665},
  year={2021},
  publisher={Elsevier},
  doi = {10.1016/j.medj.2021.04.006},
}

@article{ahmed2023data,
  title={Data provenance in healthcare: approaches, challenges, and future directions},
  author={Ahmed, Mansoor and Dar, Amil Rohani and Helfert, Markus and Khan, Abid and Kim, Jungsuk},
  journal={Sensors},
  volume={23},
  number={14},
  pages={6495},
  year={2023},
  publisher={MDPI},
  doi = {10.3390/s23146495},
}

@article{alves2026artificial,
  title={Artificial intelligence and radiologists in pancreatic cancer detection using standard of care {CT} scans ({PANORAMA}): an international, paired, non-inferiority, confirmatory, observational study},
  author={Alves, Natalia and Schuurmans, Megan and Rutkowski, Dawid and Saha, Anindo and Vendittelli, Pierpaolo and Obuchowski, Nancy and Liedenbaum, Marjolein H and others},
  journal={The Lancet Oncology},
  volume={27},
  number={1},
  pages={116--124},
  year={2026},
  publisher={Elsevier},
  doi = {10.1016/S1470-2045(25)00567-4},
}

@article{antonelli2022medical,
  title={The medical segmentation decathlon},
  author={Antonelli, Michela and Reinke, Annika and Bakas, Spyridon and Farahani, Keyvan and Kopp-Schneider, Annette and Landman, Bennett A and Litjens, Geert and others},
  journal={Nature communications},
  volume={13},
  number={1},
  pages={4128},
  year={2022},
  publisher={Nature Publishing Group UK London},
  doi = {10.1038/s41467-022-30695-9},
}

@article{appelt2022deep,
  title={Deep learning for radiotherapy outcome prediction using dose data--a review},
  author={Appelt, AL and Elhaminia, B and Gooya, A and Gilbert, A and Nix, M},
  journal={Clinical Oncology},
  volume={34},
  number={2},
  pages={e87--e96},
  year={2022},
  publisher={Elsevier},
  doi = {10.1016/j.clon.2021.12.002},
}

@article{azad2024medical,
  title={Medical image segmentation review: The success of {U-Net}},
  author={Azad, Reza and Aghdam, Ehsan Khodapanah and Rauland, Amelie and Jia, Yiwei and Avval, Atlas Haddadi and Bozorgpour, Afshin and Karimijafarbigloo, Sanaz and others},
  journal={IEEE Transactions on Pattern Analysis and Machine Intelligence},
  volume={46},
  number={12},
  pages={10076--10095},
  year={2024},
  publisher={IEEE},
  doi = {10.1109/tpami.2024.3435571},
}

@article{bassi2024touchstone,
  title={Touchstone benchmark: Are we on the right way for evaluating {AI} algorithms for medical segmentation?},
  author={Bassi, Pedro R and Li, Wenxuan and Tang, Yucheng and Isensee, Fabian and Wang, Zifu and Chen, Jieneng and Chou, Yu-Cheng and others},
  journal={Advances in Neural Information Processing Systems},
  volume={37},
  pages={15184--15201},
  year={2024},
  doi = {10.52202/079017-0485},
}

@inproceedings{baumgartner2021nndetection,
  title={{nnDetection}: a self-configuring method for medical object detection},
  author={Baumgartner, Michael and J{\"a}ger, Paul F and Isensee, Fabian and Maier-Hein, Klaus H},
  booktitle={International conference on medical image computing and computer-assisted intervention},
  pages={530--539},
  year={2021},
  organization={Springer},
  doi = {10.1007/978-3-030-87240-3_51},
}

@article{ben2021deep,
  title={Deep learning for biomedical image reconstruction: A survey},
  author={Ben Yedder, Hanene and Cardoen, Ben and Hamarneh, Ghassan},
  journal={Artificial intelligence review},
  volume={54},
  number={1},
  pages={215--251},
  year={2021},
  publisher={Springer},
  doi = {10.1007/s10462-020-09861-2},
}

@article{bernard2018deep,
  title   = {Deep Learning Techniques for Automatic {MRI} Cardiac Multi-Structures Segmentation and Diagnosis: Is the Problem Solved?},
  author  = {Bernard, Olivier and Lalande, Alain and Zotti, Clement and others},
  journal = {IEEE Transactions on Medical Imaging},
  volume  = {37},
  number  = {11},
  pages   = {2514--2525},
  year    = {2018},
  doi     = {10.1109/TMI.2018.2837502},
  url     = {https://doi.org/10.1109/TMI.2018.2837502}
}

@misc{bernhardsson_annoy,
  title        = {{Annoy: Approximate Nearest Neighbors in C++/Python optimized for memory usage and loading/saving to disk}},
  author       = {Erik Bernhardsson},
  year         = {2013},
  note         = {Accessed: 2026-05-06},
  url = {https://github.com/spotify/annoy},
}

@article{bilic2023liver,
  title={The liver tumor segmentation benchmark (lits)},
  author={Bilic, Patrick and Christ, Patrick and Li, Hongwei Bran and Vorontsov, Eugene and Ben-Cohen, Avi and Kaissis, Georgios and Szeskin, Adi and others},
  journal={Medical image analysis},
  volume={84},
  pages={102680},
  year={2023},
  publisher={Elsevier}
}

@book{bishop-2006,
	author = {Bishop, Christopher M.},
	title = {{Pattern Recognition and Machine Learning}},
	year = {2006},
  publisher = {Springer}
}

@inproceedings{boland2024there,
  title={There are no shortcuts to anywhere worth going: identifying shortcuts in deep learning models for medical image analysis},
  author={Boland, Christopher and Goatman, Keith A and Tsaftaris, Sotirios A and Dahdouh, Sonia},
  booktitle={Medical Imaging with Deep Learning},
  year={2024}
}

@article{bosch2026diversity,
  title={Diversity over scale: Whole-slide image variety enables H\&E foundation model training with fewer patches},
  author={Bosch, Christoph and Wong, John KL and Paulikat, Martin and Zapukhlyak, Myroslav and Arora, Bharti and Aichm{\"u}ller-Ratnaparkhe, Manasi and Baumann, Jens and others},
  journal={Journal of Pathology Informatics},
  pages={100648},
  year={2026},
  publisher={Elsevier}
}

@article{cerri2025large,
  title={A large-scale heterogeneous {3D} magnetic resonance brain imaging dataset for self-supervised learning},
  author={Cerri, Stefano and Munk, Asbj{\o}rn and Ambsdorf, Jakob and Machnio, Julia and Llambias, Sebastian N{\o}rgaard and Nersesjan, Vardan and Krag, Christian Hedeager and others},
  journal={Scientific Data},
  year={2026},
  volume = {13},
  pages = {1351},
  doi = {10.1038/s41597-026-07688-0}
}

@inproceedings{chen2025scaling,
  title={Scaling tumor segmentation: Best lessons from real and synthetic data},
  author={Chen, Qi and Zhou, Xinze and Liu, Chen and Chen, Hao and Li, Wenxuan and Jiang, Zekun and Huang, Ziyan and others},
  booktitle={Proceedings of the IEEE/CVF International Conference on Computer Vision},
  pages={24001--24013},
  year={2025},
  doi = {10.1109/iccv51701.2025.02225},
}

@article{chu2025revisiting,
  title={Revisiting Data Scaling Law for Medical Segmentation},
  author={Chu, Yuetan and Han, Zhongyi and Luo, Gongning and Gao, Xin},
  journal={arXiv preprint arXiv:2511.13883},
  year={2025}
}

@article{clark2013tcia,
  title={The Cancer Imaging Archive ({TCIA}): Maintaining and Operating a Public Information Repository},
  author={Clark, Kenneth and Vendt, Bruce and Smith, Kirk and Freymann, John and Kirby, Justin and Koppel, Paul and Moore, Stephen and others},
  journal={Journal of Digital Imaging},
  volume={26},
  number={6},
  pages={1045--1057},
  year={2013},
  publisher={Springer},
  doi={10.1007/s10278-013-9622-7}
}

@misc{codabench_flare2024,
  title        = {{{FLARE24} Challenge}},
  author       = {{CodaBench}},
  year         = {2024},
  note         = {Accessed: 2026-05-06},
  url = {https://www.codabench.org/competitions/2319/},
}

@article{cross2024bias,
  title={Bias in medical {AI}: implications for clinical decision-making},
  author={Cross, James L and Choma, Michael A and Onofrey, John A},
  journal={PLOS digital health},
  volume={3},
  number={11},
  pages={e0000651},
  year={2024},
  publisher={Public Library of Science San Francisco, CA USA},
  doi = {10.1371/journal.pdig.0000651},
}

@inproceedings{cui2019class,
  title={Class-balanced loss based on effective number of samples},
  author={Cui, Yin and Jia, Menglin and Lin, Tsung-Yi and Song, Yang and Belongie, Serge},
  booktitle={Proceedings of the IEEE/CVF conference on computer vision and pattern recognition},
  pages={9268--9277},
  year={2019},
  doi = {10.1109/cvpr.2019.00949},
}

@article{degrauw2025uls23,
  author    = {de Grauw, M.J.J. and Scholten, E.Th. and Smit, E.J. and Rutten, M.J.C.M. and Prokop, M. and van Ginneken, B. and Hering, A.},
  title     = {{The ULS23 challenge: A baseline model and benchmark dataset for 3D universal lesion segmentation in computed tomography}},
  journal   = {Medical Image Analysis},
  publisher = {Elsevier BV},
  year      = {2025},
  volume    = {102},
  pages     = {103525},
  doi       = {10.1016/j.media.2025.103525},
  url       = {https://doi.org/10.1016/j.media.2025.103525}
}

@article{drenkow2025detecting,
  title={Detecting dataset bias in medical {AI}: A generalized and modality-agnostic auditing framework},
  author={Drenkow, Nathan and Pavlak, Mitchell and Harrigian, Keith and Zirikly, Ayah and Subbaswamy, Adarsh and Farhangi, Mohammad Mehdi and Petrick, Nicholas and others},
  journal={arXiv preprint arXiv:2503.09969},
  year={2025}
}

@inproceedings{eckstein2025missing,
  title={The Missing Piece: A Case for Pre-training in {3D} Medical Object Detection},
  author={Eckstein, Katharina and Ulrich, Constantin and Baumgartner, Michael and K{\"a}chele, Jessica and Bounias, Dimitrios and Wald, Tassilo and Floca, Ralf and others},
  booktitle={International Conference on Medical Image Computing and Computer-Assisted Intervention},
  pages={615--626},
  year={2025},
  organization={Springer},
  doi = {10.1007/978-3-032-04965-0_58},
}

@misc{eu_ai_act_2024_article10,
  title        = {{Regulation ({EU}) 2024/1689 of the European Parliament and of the Council laying down harmonised rules on artificial intelligence (Artificial Intelligence Act)}},
  author       = {{European Union}},
  year         = {2024},
  note         = {Article 10: Data and data governance},
  url          = {https://eur-lex.europa.eu/eli/reg/2024/1689/oj}
}

@misc{eu_mdr_2017_745,
  title        = {{Regulation ({EU}) 2017/745 of the European Parliament and of the Council on medical devices}},
  author       = {{European Union}},
  year         = {2017},
  note         = {Official Journal of the European Union, L 117},
  url          = {https://eur-lex.europa.eu/eli/reg/2017/745/oj}
}

@inproceedings{fairweather2020non,
  title={Non-repudiable provenance for clinical decision support systems},
  author={Fairweather, Elliot and Wittner, Rudolf and Chapman, Martin and Holub, Petr and Curcin, Vasa},
  booktitle={International Provenance and Annotation Workshop},
  pages={165--182},
  year={2020},
  organization={Springer},
  doi = {10.1007/978-3-030-80960-7_10},
}

@inproceedings{fan2024scaling,
  title={Scaling laws of synthetic images for model training... for now},
  author={Fan, Lijie and Chen, Kaifeng and Krishnan, Dilip and Katabi, Dina and Isola, Phillip and Tian, Yonglong},
  booktitle={Proceedings of the IEEE/CVF Conference on Computer Vision and Pattern Recognition},
  pages={7382--7392},
  year={2024},
  doi = {10.1109/cvpr52733.2024.00705},
}

@misc{flare2025_challenge,
  title        = {{{FLARE} 2025 Challenge}},
  author       = {{MICCAI 2025}},
  year         = {2025},
  note         = {Accessed: 2026-05-06},
  url = {https://conferences.miccai.org/2025/en/FLARE-2025-Challenge.html},
}

@article{fu2020deep,
  title={Deep learning in medical image registration: a review},
  author={Fu, Yabo and Lei, Yang and Wang, Tonghe and Curran, Walter J and Liu, Tian and Yang, Xiaofeng},
  journal={Physics in Medicine \& Biology},
  volume={65},
  number={20},
  pages={20TR01},
  year={2020},
  publisher={IOP Publishing},
  doi = {10.1088/1361-6560/ab843e},
}

@inproceedings{gai20253d,
  title={{3D-RAD}: A comprehensive {3D} radiology {Med-VQA} dataset with multi-temporal analysis and diverse diagnostic tasks},
  author={Gai, Xiaotang and Liu, Jiaxiang and Li, Yichen and Meng, Zijie and Wu, Jian and Liu, Zuozhu},
  year={2025},
  booktitle = {Advances in Neural Information Processing Systems},
  volume = {38},
  pages = {188177--188218},
  doi = {10.52202/085713-5656}
}

@article{gatidis2022whole,
  title={A whole-body {FDG-PET}/{CT} dataset with manually annotated tumor lesions},
  author={Gatidis, Sergios and Hepp, Tobias and Fr{\"u}h, Marcel and La Foug{\`e}re, Christian and Nikolaou, Konstantin and Pfannenberg, Christina and Sch{\"o}lkopf, Bernhard and others},
  journal={Scientific Data},
  volume={9},
  number={1},
  pages={601},
  year={2022},
  publisher={Nature Publishing Group UK London},
  doi = {10.1038/s41597-022-01718-3},
}

@article{gebru2021datasheets,
  title={Datasheets for datasets},
  author={Gebru, Timnit and Morgenstern, Jamie and Vecchione, Briana and Vaughan, Jennifer Wortman and Wallach, Hanna and Iii, Hal Daum{\'e} and Crawford, Kate},
  journal={Communications of the ACM},
  volume={64},
  number={12},
  pages={86--92},
  year={2021},
  publisher={ACM New York, NY, USA},
  doi = {10.1145/3458723},
}

@article{ghaffari2019automated,
  title={Automated brain tumor segmentation using multimodal brain scans: a survey based on models submitted to the {BraTS} 2012--2018 challenges},
  author={Ghaffari, Mina and Sowmya, Arcot and Oliver, Ruth},
  journal={IEEE reviews in biomedical engineering},
  volume={13},
  pages={156--168},
  year={2019},
  publisher={IEEE},
  doi = {10.1109/rbme.2019.2946868},
}

@article{gibson2026evidence_2,
  title = {{Evidence of unreliable data and poor data provenance in clinical prediction model research and clinical practice}},
  author = {Gibson, Alexander D and White, Nicole M and Collins, Gary S and Barnett, Adrian G},
  year = {2026},
  journal = {BMC Medicine},
  volume = {24},
  number = {1},
  pages = {386},
  doi = {10.1186/s12916-026-04981-y},
}

@article{goldblum2023battle,
  title={Battle of the backbones: A large-scale comparison of pretrained models across computer vision tasks},
  author={Goldblum, Micah and Souri, Hossein and Ni, Renkun and Shu, Manli and Prabhu, Viraj and Somepalli, Gowthami and Chattopadhyay, Prithvijit and others},
  journal={Advances in Neural Information Processing Systems},
  volume={36},
  pages={29343--29371},
  year={2023},
  doi = {10.52202/075280-1277},
}

@article{hamamci2026generalist,
      title={Generalist foundation models from a multimodal dataset for {3D} computed tomography},
      author={Hamamci, Ibrahim Ethem and Er, Sezgin and Wang, Chenyu and Almas, Furkan and Simsek, Ayse Gulnihan and Esirgun, Sevval Nil and Dogan, Irem and others},
      journal={Nature Biomedical Engineering},
      volume={10},
      number={8},
      pages={1610--1628},
      year={2026},
      publisher={Nature Publishing Group UK London},
      doi={10.1038/s41551-025-01599-y}
}

@article{hanna2025ethical,
  title={Ethical and bias considerations in artificial intelligence/machine learning},
  author={Hanna, Matthew G and Pantanowitz, Liron and Jackson, Brian and Palmer, Octavia and Visweswaran, Shyam and Pantanowitz, Joshua and Deebajah, Mustafa and others},
  journal={Modern Pathology},
  volume={38},
  number={3},
  pages={100686},
  year={2025},
  publisher={Elsevier},
  doi = {10.1016/j.modpat.2024.100686},
}

@article{hasanzadeh2025bias,
  title={Bias recognition and mitigation strategies in artificial intelligence healthcare applications},
  author={Hasanzadeh, Fereshteh and Josephson, Colin B and Waters, Gabriella and Adedinsewo, Demilade and Azizi, Zahra and White, James A},
  journal={NPJ Digital Medicine},
  volume={8},
  number={1},
  pages={154},
  year={2025},
  publisher={Nature Publishing Group UK London},
  doi = {10.1038/s41746-025-01503-7},
}

@article{heller2019kits19,
  title={The {KiTS19} challenge data: 300 kidney tumor cases with clinical context, {CT} semantic segmentations, and surgical outcomes},
  author={Heller, Nicholas and Sathianathen, Niranjan and Kalapara, Arveen and Walczak, Edward and Moore, Keenan and Kaluzniak, Heather and Rosenberg, Joel and others},
  journal={arXiv preprint arXiv:1904.00445},
  year={2019},
  doi = {10.48550/arXiv.1904.00445},
}

@article{heller2021state,
  author    = {Heller, Nicholas and Isensee, Fabian and Maier-Hein, Klaus H. and Hou, Xiaoshuai and Xie, Chunmei and Li, Fengyi and Nan, Yang and others},
  title     = {The state of the art in kidney and kidney tumor segmentation in contrast-enhanced {CT} imaging: Results of the {KiTS19} challenge},
  journal   = {Medical Image Analysis},
  year      = {2021},
  volume    = {67},
  pages     = {101821},
  doi       = {10.1016/j.media.2020.101821}
}

@misc{heller2023kits21,
  title = {{The {KiTS21} Challenge: Automatic segmentation of kidneys, renal tumors, and renal cysts in corticomedullary-phase {CT}}}, 
  author = {Nicholas Heller and Fabian Isensee and Dasha Trofimova and Resha Tejpaul and Zhongchen Zhao and Huai Chen and Lisheng Wang and others},
  year = {2023},
  eprint = {2307.01984},
  archivePrefix = {arXiv},
  primaryClass = {cs.CV},
  howpublished = {arXiv preprint arXiv:2307.01984, \doi{10.48550/arXiv.2307.01984}},
  doi = {10.48550/arXiv.2307.01984},
}

@article{hestness2017deep,
  title={Deep learning scaling is predictable, empirically},
  author={Hestness, Joel and Narang, Sharan and Ardalani, Newsha and Diamos, Gregory and Jun, Heewoo and Kianinejad, Hassan and Patwary, Md Mostofa Ali and others},
  journal={arXiv preprint arXiv:1712.00409},
  year={2017}
}

@article{huang2021application,
  title={The application and development of deep learning in radiotherapy: a systematic review},
  author={Huang, Danju and Bai, Han and Wang, Li and Hou, Yu and Li, Lan and Xia, Yaoxiong and Yan, Zhirui and others},
  journal={Technology in Cancer Research \& Treatment},
  volume={20},
  pages={15330338211016386},
  year={2021},
  publisher={SAGE Publications Sage CA: Los Angeles, CA},
  doi = {10.1177/15330338211016386},
}

@article{ilse2025data,
  title={Data Scaling Laws for Radiology Foundation Models},
  author={Ilse, Maximilian and Sharma, Harshita and Schwaighofer, Anton and Bond-Taylor, Sam and P{\'e}rez-Garc{\'\i}a, Fernando and Melnichenko, Olesya and Sykes, Anne-Marie G and others},
  journal={arXiv preprint arXiv:2509.12818},
  year={2025}
}

@book{imran2026multiclass,
  editor = {Imran, Muhammad and Krebs, Jonathan R. and Cooper, Michol A. and Ma, Jun and Zhou, Yuyin and Shao, Wei},
  title = {Multi-class Segmentation of the Aorta: {AortaSeg} 2024 Challenge, Held in Conjunction with {MICCAI} 2024, Virtual Event, October 24, 2024, Proceedings},
  series = {Lecture Notes in Computer Science},
  publisher = {Springer Nature Switzerland},
  year = {2026},
  volume = {16399},
  doi = {10.1007/978-3-032-14246-7},
  url = {https://doi.org/10.1007/978-3-032-14246-7}
}

@article{imran2026multiclass_2,
  title = {{Multi-class segmentation of aortic branches and zones in computed tomography angiography: The AortaSeg24 challenge}},
  author = {Imran, Muhammad and Krebs, Jonathan R. and Sivaraman, Vishal Balaji and Zhang, Teng and Kumar, Amarjeet and Ueland, Walker R. and Fassler, Michael J. and others},
  year = {2026},
  journal = {Medical Image Analysis},
  volume = {113},
  pages = {104188},
  doi = {10.1016/j.media.2026.104188},
}

@article{isensee2021nnu,
  title={{nnU-Net}: a self-configuring method for deep learning-based biomedical image segmentation},
  author={Isensee, Fabian and Jaeger, Paul F and Kohl, Simon AA and Petersen, Jens and Maier-Hein, Klaus H},
  journal={Nature methods},
  volume={18},
  number={2},
  pages={203--211},
  year={2021},
  publisher={Nature Publishing Group},
  doi = {10.1038/s41592-020-01008-z},
}

@inproceedings{jaus2024unifying,
  title = {{Towards Unifying Anatomy Segmentation: Automated Generation of a Full-Body CT Dataset}},
  author = {Jaus, Alexander and Seibold, Constantin and Hermann, Kelsey and Shahamiri, Negar and Walter, Alexandra and Giske, Kristina and Haubold, Johannes and others},
  year = {2024},
  booktitle = {2024 IEEE International Conference on Image Processing (ICIP)},
  pages = {41--47},
  publisher = {IEEE},
  doi = {10.1109/ICIP51287.2024.10647307},
}

@article{ji2022amos,
  title={{AMOS}: A large-scale abdominal multi-organ benchmark for versatile medical image segmentation},
  author={Ji, Yuanfeng and Bai, Haotian and Ge, Chongjian and Yang, Jie and Zhu, Ye and Zhang, Ruimao and Li, Zhen and others},
  journal={Advances in neural information processing systems},
  volume={35},
  pages={36722--36732},
  year={2022},
  doi = {10.52202/068431-2661},
}

@article{jimenez2024copycats,
  title={Copycats: the many lives of a publicly available medical imaging dataset},
  author={Jim{\'e}nez-S{\'a}nchez, Amelia and Avlona, Natalia-Rozalia and Juodelyte, Dovile and Sourget, Th{\'e}o and Vang-Larsen, Caroline and Rogers, Anna and Zaj{\k{a}}c, Hubert and others},
  journal={Advances in Neural Information Processing Systems},
  volume={37},
  pages={113383--113404},
  year={2024}
}

@article{jin2020deep,
  author = {Jin, Liang and Yang, Jiancheng and Kuang, Kaiming and Ni, Bingbing and Gao, Yiyi and Sun, Yingli and Gao, Pan and others},
  title = {Deep-learning-assisted detection and segmentation of rib fractures from {CT} scans: Development and validation of {FracNet}},
  journal = {eBioMedicine},
  year = {2020},
  volume = {62},
  pages = {103106},
  doi = {10.1016/j.ebiom.2020.103106},
  url = {https://doi.org/10.1016/j.ebiom.2020.103106}
}

@inproceedings{joshi2024data,
  title={Data-efficient contrastive language-image pretraining: Prioritizing data quality over quantity},
  author={Joshi, Siddharth and Jain, Arnav and Payani, Ali and Mirzasoleiman, Baharan},
  booktitle={International Conference on Artificial Intelligence and Statistics},
  pages={1000--1008},
  year={2024},
  organization={PMLR}
}

@article{kaplan2020scaling,
  title={Scaling laws for neural language models},
  author={Kaplan, Jared and McCandlish, Sam and Henighan, Tom and Brown, Tom B and Chess, Benjamin and Child, Rewon and Gray, Scott and others},
  journal={arXiv preprint arXiv:2001.08361},
  year={2020}
}

@article{khan2025comprehensive,
  title={A comprehensive survey of foundation models in medicine},
  author={Khan, Wasif and Leem, Seowung and See, Kyle B and Wong, Joshua K and Zhang, Shaoting and Fang, Ruogu},
  journal={IEEE Reviews in Biomedical Engineering},
  year={2025},
  publisher={IEEE},
  doi = {10.1109/rbme.2025.3531360},
}

@misc{kinahan2019data_hnscc,
  author    = {Kinahan, Paul and Muzi, Mark and Bialecki, Brian and Coombs, Laura},
  title     = {{{Data from the ACRIN 6685 Trial HNSCC-FDG-PET/CT} [Data set]}},
  publisher = {The Cancer Imaging Archive},
  year      = {2019},
  doi       = {10.7937/K9/TCIA.2016.JQEJZZNG},
  howpublished = {The Cancer Imaging Archive, \doi{10.7937/K9/TCIA.2016.JQEJZZNG}},
}

@inproceedings{kirchhoff2026exploiting,
  title={Exploiting Longitudinal Context in Clinician-Verified Interactive Lesion Tracking},
  author={Kirchhoff, Yannick and Rokuss, Maximilian and Mertens, Daniel Philipp and F{\"u}ller, David and Hamm, Benjamin and Schreyer, Andreas and Ritter, Oliver and others},
  booktitle={International Conference on Medical Image Computing and Computer-Assisted Intervention},
  pages={259--269},
  year={2026},
  organization={Springer},
  doi = {10.1007/978-3-032-38172-9_25},
}

@inproceedings{klug2022scaling,
  title={Scaling laws for deep learning based image reconstruction},
  author={Klug, Tobit and Heckel, Reinhard},
  year={2023},
  booktitle = {International Conference on Learning Representations (ICLR)}
}

@misc{krawetz2011looks,
  author = {Krawetz, Neal},
  title = {{Looks Like It}},
  year = {2011},
  howpublished = {The Hacker Factor Blog},
  note = {Accessed 2026-05-05}
}

@article{li2024abdomenatlas,
  title={Abdomenatlas: A large-scale, detailed-annotated, \& multi-center dataset for efficient transfer learning and open algorithmic benchmarking},
  author={Li, Wenxuan and Qu, Chongyu and Chen, Xiaoxi and Bassi, Pedro RAS and Shi, Yijia and Lai, Yuxiang and Yu, Qian and others},
  journal={Medical Image Analysis},
  volume={97},
  pages={103285},
  year={2024},
  publisher={Elsevier},
  doi = {10.1016/j.media.2024.103285},
}

@inproceedings{li2024well,
  title={How Well Do Supervised {3D} Models Transfer to Medical Imaging Tasks?},
  author={Li, Wenxuan and Yuille, Alan and Zhou, Zongwei},
  booktitle={International Conference on Learning Representations},
  volume={2024},
  pages={16389--16404},
  year={2024}
}

@inproceedings{liang2020we,
  title={Do we really need to access the source data? source hypothesis transfer for unsupervised domain adaptation},
  author={Liang, Jian and Hu, Dapeng and Feng, Jiashi},
  booktitle={International conference on machine learning},
  pages={6028--6039},
  year={2020},
  organization={PMLR}
}

@inproceedings{lin2024shortcut,
  title={Shortcut learning in medical image segmentation},
  author={Lin, Manxi and Weng, Nina and Mikolaj, Kamil and Bashir, Zahra and Svendsen, Morten BS and Tolsgaard, Martin G and Christensen, Anders N and others},
  booktitle={International Conference on Medical Image Computing and Computer-Assisted Intervention},
  pages={623--633},
  year={2024},
  organization={Springer},
  doi = {10.1007/978-3-031-72111-3_59},
}

@article{litjens2017survey,
  title={A survey on deep learning in medical image analysis},
  author={Litjens, Geert and Kooi, Thijs and Bejnordi, Babak Ehteshami and Setio, Arnaud Arindra Adiyoso and Ciompi, Francesco and Ghafoorian, Mohsen and Van Der Laak, Jeroen Awm and others},
  journal={Medical image analysis},
  volume={42},
  pages={60--88},
  year={2017},
  publisher={Elsevier},
  doi = {10.1016/j.media.2017.07.005},
}

@article{ma2021abdomenct,
  title={{AbdomenCT-1K}: Is abdominal organ segmentation a solved problem?},
  author={Ma, Jun and Zhang, Yao and Gu, Song and Zhu, Cheng and Ge, Cheng and Zhang, Yichi and An, Xingle and others},
  journal={IEEE Transactions on Pattern Analysis and Machine Intelligence},
  volume={44},
  number={10},
  pages={6695--6714},
  year={2021},
  publisher={IEEE},
  doi = {10.1109/tpami.2021.3100536},
  url = {https://doi.org/10.1109/TPAMI.2021.3100536},
}

@article{ma2024automatic,
  title={Automatic organ and pan-cancer segmentation in abdomen {CT}: the {FLARE} 2023 challenge},
  author={Ma, Jun and Zhang, Yao and Gu, Song and Ge, Cheng and Wang, Ershuai and Zhou, Qin and Huang, Ziyan and others},
  journal={arXiv preprint arXiv:2408.12534},
  year={2024},
  doi = {10.48550/arXiv.2408.12534},
}

@article{ma2024segment,
  title={Segment anything in medical images},
  author={Ma, Jun and He, Yuting and Li, Feifei and Han, Lin and You, Chenyu and Wang, Bo},
  journal={Nature communications},
  volume={15},
  number={1},
  pages={654},
  year={2024},
  publisher={Nature Publishing Group UK London},
  doi = {10.1038/s41467-024-44824-z},
}

@article{marcus2007open,
  title   = {Open Access Series of Imaging Studies ({OASIS}): Cross-sectional {MRI} Data in Young, Middle Aged, Nondemented, and Demented Older Adults},
  author  = {Marcus, Daniel S. and Wang, Tracy H. and Parker, Jamie and Csernansky, John G. and Morris, John C. and Buckner, Randy L.},
  journal = {Journal of Cognitive Neuroscience},
  volume  = {19},
  number  = {9},
  pages   = {1498--1507},
  year    = {2007},
  doi     = {10.1162/jocn.2007.19.9.1498}
}

@article{markiewicz2021openneuro,
  title   = {{The OpenNeuro} Resource for Sharing of Neuroscience Data},
  author  = {Markiewicz, Christopher J. and Gorgolewski, Krzysztof J. and Feingold, Franklin and Blair, Ross and Halchenko, Yaroslav O. and Miller, Elizabeth and Hardcastle, Nicholas and others},
  journal = {eLife},
  volume  = {10},
  pages   = {e71774},
  year    = {2021},
  doi     = {10.7554/eLife.71774}
}

@article{MedIA-FLARE21,
title = {Fast and {Low-GPU-memory} abdomen {CT} organ segmentation: The {FLARE} challenge},
author = {Jun Ma and Yao Zhang and Song Gu and Xingle An and Zhihe Wang and Cheng Ge and Congcong Wang and others},
journal = {Medical Image Analysis},
pages = {102616},
volume = {82},
year =  {2022},
  doi = {10.1016/j.media.2022.102616},
}

@article{monga2006perceptual,
  title={Perceptual image hashing via feature points: performance evaluation and tradeoffs},
  author={Monga, Vishal and Evans, Brian L},
  journal={IEEE transactions on Image Processing},
  volume={15},
  number={11},
  pages={3452--3465},
  year={2006},
  publisher={IEEE},
  doi = {10.1109/tip.2006.881948},
}

@article{moor2023foundation,
  title={Foundation models for generalist medical artificial intelligence},
  author={Moor, Michael and Banerjee, Oishi and Abad, Zahra Shakeri Hossein and Krumholz, Harlan M and Leskovec, Jure and Topol, Eric J and Rajpurkar, Pranav},
  journal={Nature},
  volume={616},
  number={7956},
  pages={259--265},
  year={2023},
  publisher={Nature Publishing Group UK London},
  doi = {10.1038/s41586-023-05881-4},
}

@article{munk2026towards,
  title={Towards Brain {MRI} Foundation Models for the Clinic: Findings from the {FOMO25} Challenge},
  author={Munk, Asbj{\o}rn and Cerri, Stefano and Nersesjan, Vardan and Krag, Christian Hedeager and Ambsdorf, Jakob and Garc{\'\i}a, Pablo Rocamora and Machnio, Julia and others},
  journal={arXiv preprint arXiv:2604.11679},
  year={2026}
}

@article{murray2022data,
  title={Data provenance and integrity of health-care systems data for clinical trials},
  author={Murray, Macey L and Love, Sharon B and Carpenter, James R and Hartley, Suzanne and Landray, Martin J and Mafham, Marion and Parmar, Mahesh KB and others},
  journal={The Lancet Digital Health},
  volume={4},
  number={8},
  pages={e567--e568},
  year={2022},
  publisher={Elsevier},
  doi = {10.1016/s2589-7500(22)00122-4},
}

@article{muthu2026vids,
  title={{VIDS}: A Verified Imaging Dataset Standard for Medical {AI}},
  author={Muthu, Joan S and Shalen, John},
  journal={arXiv preprint arXiv:2604.17525},
  year={2026}
}

@misc{openneuro_ds004529_modeling_face_recognition_in_the_pr,
  author = {Nestor Zaragoza-Jimenez and Hauke Niehaus and Ina Thome and Christoph Vogelbacher and Gabriele Ende and Inge Kamp-Becker and Dominik Endres and others},
  title = {{Modeling Face Recognition in the Predictive Coding Framework: 
A Combined Computational Modeling and Functional Imaging Study - Stage 2}},
  publisher = {OpenNeuro},
  year = {2023},
  doi = {10.18112/openneuro.ds004529.v1.1.1},
  url = {https://openneuro.org/datasets/ds004529},
  version = {1.1.1},
  howpublished = {OpenNeuro, \doi{10.18112/openneuro.ds004529.v1.1.1}},
}

@InProceedings{paschali2018,
author="Paschali, Magdalini
and Conjeti, Sailesh
and Navarro, Fernando
and Navab, Nassir",
editor="Frangi, Alejandro F.
and Schnabel, Julia A.
and Davatzikos, Christos
and Alberola-L{\'o}pez, Carlos
and Fichtinger, Gabor",
title="Generalizability vs. Robustness: Investigating Medical Imaging Networks Using Adversarial Examples",
booktitle="Medical Image Computing and Computer Assisted Intervention -- MICCAI 2018",
year="2018",
publisher="Springer International Publishing",
address="Cham",
pages="493--501",
isbn="978-3-030-00928-1",
  doi = {10.1007/978-3-030-00928-1_56},
}

@article{peiffer2020machine,
  title={Machine learning for clinical decision support in infectious diseases: a narrative review of current applications},
  author={Peiffer-Smadja, Nathan and Rawson, Timothy Miles and Ahmad, Raheelah and Buchard, Albert and Georgiou, P and Lescure, F-X and Birgand, Gabriel and others},
  journal={Clinical Microbiology and Infection},
  volume={26},
  number={5},
  pages={584--595},
  year={2020},
  publisher={Elsevier},
  doi = {10.1016/j.cmi.2019.09.009},
}

@article{qu2023abdomenatlas,
  title={{AbdomenAtlas-8K}: Annotating 8,000 {CT} volumes for multi-organ segmentation in three weeks},
  author={Qu, Chongyu and Zhang, Tiezheng and Qiao, Hualin and Tang, Yucheng and Yuille, Alan L and Zhou, Zongwei and others},
  journal={Advances in Neural Information Processing Systems},
  volume={36},
  pages={36620--36636},
  year={2023},
  doi = {10.52202/075280-1592},
}

@article{quinton2023tumour,
  title   = {A Tumour and Liver Automatic Segmentation ({ATLAS}) Dataset on Contrast-Enhanced Magnetic Resonance Imaging for Hepatocellular Carcinoma},
  author  = {Quinton, F{\'e}lix and Popoff, Romain and Presles, Beno{\^\i}t and Leclerc, Sarah and Meriaudeau, Fabrice and Nodari, Guillaume and Lopez, Olivier and others},
  journal = {Data},
  volume  = {8},
  number  = {5},
  pages   = {79},
  year    = {2023},
  doi     = {10.3390/data8050079}
}

@article{rister2020ct,
  title={{CT-ORG}, a new dataset for multiple organ segmentation in computed tomography},
  author={Rister, Blaine and Yi, Darvin and Shivakumar, Kaushik and Nobashi, Tomomi and Rubin, Daniel L},
  journal={Scientific Data},
  volume={7},
  number={1},
  pages={381},
  year={2020},
  publisher={Nature Publishing Group UK London},
  doi = {10.1038/s41597-020-00715-8},
}

@inproceedings{rokuss2025lesionlocator,
  title={{LesionLocator}: Zero-shot universal tumor segmentation and tracking in {3D} whole-body imaging},
  author={Rokuss, Maximilian and Kirchhoff, Yannick and Akbal, Seval and Kovacs, Balint and Roy, Saikat and Ulrich, Constantin and Wald, Tassilo and others},
  booktitle={Proceedings of the Computer Vision and Pattern Recognition Conference},
  pages={30872--30885},
  year={2025},
  doi = {10.1109/cvpr52734.2025.02875},
}

@inproceedings{rokuss2026voxtell,
  title={{VoxTell}: Free-text promptable universal {3D} medical image segmentation},
  author={Rokuss, Maximilian and Langenberg, Moritz and Kirchhoff, Yannick and Isensee, Fabian and Hamm, Benjamin and Ulrich, Constantin and Regnery, Sebastian and others},
  booktitle={Proceedings of the IEEE/CVF conference on computer vision and pattern recognition},
  pages={37538--37557},
  year={2026}
}

@inproceedings{roth2015deeporgan,
  title={{DeepOrgan}: Multi-level deep convolutional networks for automated pancreas segmentation},
  author={Roth, Holger R and Lu, Le and Farag, Amal and Shin, Hoo-Chang and Liu, Jiamin and Turkbey, Evrim B and Summers, Ronald M},
  booktitle={International conference on medical image computing and computer-assisted intervention},
  pages={556--564},
  year={2015},
  organization={Springer},
  doi = {10.1007/978-3-319-24553-9_68},
}

@misc{roth2016data,
  author = {Roth, Holger and Farag, Amal and Turkbey, Evrim B. and Lu, Le and Liu, Jiamin and Summers, Ronald M.},
  title = {{Data From Pancreas-CT (Version 2) [Data set]}},
  publisher = {The Cancer Imaging Archive},
  year = {2016},
  doi = {10.7937/K9/TCIA.2016.tNB1kqBU},
  howpublished = {The Cancer Imaging Archive, \doi{10.7937/K9/TCIA.2016.tNB1kqBU}},
}

@article{roy2025mednext,
  title={{MedNeXt-v2}: Scaling {3D} {ConvNeXts} for Large-Scale Supervised Representation Learning in Medical Image Segmentation},
  author={Roy, Saikat and Kirchhoff, Yannick and Ulrich, Constantin and Rokuss, Maximillian and Wald, Tassilo and Isensee, Fabian and Maier-Hein, Klaus},
  journal={arXiv preprint arXiv:2512.17774},
  year={2025}
}

@article{rozenblattrosen2020human,
  title   = {The {Human Tumor Atlas Network}: Charting Tumor Transitions across Space and Time at Single-Cell Resolution},
  author  = {Rozenblatt-Rosen, Orit and Regev, Aviv and Oberdoerffer, Philipp and Nawy, Tal and Hupalowska, Anna and Rood, Jennifer E. and Ashenberg, Orr and others},
  journal = {Cell},
  volume  = {181},
  number  = {2},
  pages   = {236--249},
  year    = {2020},
  doi     = {10.1016/j.cell.2020.03.053}
}

@misc{rozenfeld2023annotations_2,
  title = {{Annotations for ACRIN-HNSCC-FDG-PET/CT Collection (ACRIN 6685-Tumor-Annotations) (Version 2) [Data set]}},
  author = {Rozenfeld, Michael and Jordan, Petr},
  year = {2023},
  publisher = {The Cancer Imaging Archive},
  doi = {10.7937/jvgc-aq36},
  howpublished = {The Cancer Imaging Archive, \doi{10.7937/jvgc-aq36}},
}

@article{saha2024artificial,
  title={Artificial intelligence and radiologists in prostate cancer detection on {MRI} ({PI-CAI}): an international, paired, non-inferiority, confirmatory study},
  author={Saha, Anindo and Bosma, Joeran S and Twilt, Jasper J and Van Ginneken, Bram and Bjartell, Anders and Padhani, Anwar R and Bonekamp, David and others},
  journal={The Lancet Oncology},
  volume={25},
  number={7},
  pages={879--887},
  year={2024},
  publisher={Elsevier}
}

@article{samanta2021analysis,
  title={Analysis of perceptual hashing algorithms in image manipulation detection},
  author={Samanta, Priyanka and Jain, Shweta},
  journal={Procedia Computer Science},
  volume={185},
  pages={203--212},
  year={2021},
  publisher={Elsevier},
  doi = {10.1016/j.procs.2021.05.021},
}

@article{shaw2019artificial,
  title={Artificial intelligence and the implementation challenge},
  author={Shaw, James and Rudzicz, Frank and Jamieson, Trevor and Goldfarb, Avi},
  journal={Journal of medical Internet research},
  volume={21},
  number={7},
  pages={e13659},
  year={2019},
  publisher={JMIR Publications Toronto, Canada},
  doi = {10.2196/13659},
}

@article{simpson2019large,
  author = {Simpson, Amber L. and Antonelli, Michela and Bakas, Spyridon and Bilello, Michel and Farahani, Keyvan and van Ginneken, Bram and Kopp-Schneider, Annette and others},
  title = {A large annotated medical image dataset for the development and evaluation of segmentation algorithms},
  journal = {arXiv preprint arXiv:1902.09063},
  year = {2019},
  doi = {10.48550/arXiv.1902.09063},
  url = {https://doi.org/10.48550/arXiv.1902.09063}
}

@article{sobek2024medyolo,
  title={{MedYOLO}: a medical image object detection framework},
  author={Sobek, Joseph and Medina Inojosa, Jose R and Medina Inojosa, Betsy J and Rassoulinejad-Mousavi, SM and Conte, Gian Marco and Lopez-Jimenez, Francisco and Erickson, Bradley J},
  journal={Journal of imaging informatics in medicine},
  volume={37},
  number={6},
  pages={3208--3216},
  year={2024},
  publisher={Springer},
  doi = {10.1007/s10278-024-01138-2},
}

@article{stanley2025and,
  title={Where, why, and how is bias learned in medical image analysis models? a study of bias encoding within convolutional networks using synthetic data},
  author={Stanley, Emma AM and Souza, Raissa and Wilms, Matthias and Forkert, Nils D},
  journal={EBioMedicine},
  volume={111},
  year={2025},
  publisher={Elsevier},
  doi = {10.1016/j.ebiom.2024.105501},
}

@article{stegeman2026designing,
  title={Designing {UNICORN}: a Unified Benchmark for Imaging in Computational Pathology, Radiology, and Natural Language},
  author={Stegeman, Michelle and Philipp, Lena and van der Graaf, Fennie and D'Amato, Marina and Grisi, Cl{\'e}ment and Builtjes, Luc and Bosma, Joeran S and others},
  journal={arXiv preprint arXiv:2603.02790},
  year={2026}
}

@article{suri2026roadmap,
  title={{ROADMAP}: an ontology of medical {AI} models and datasets},
  author={Suri, Abhinav and Takahashi, Marcelo Straus and Retson, Tara and Gonzales, Ricardo A and Park, Seong Ho and Kahn Jr, Charles E},
  journal={Radiology: Artificial Intelligence},
  pages={e260069},
  year={2026},
  publisher={Radiological Society of North America},
  doi = {10.1148/ryai.260069},
}

@inproceedings{ulrich2023multitalent,
  title={{MultiTalent}: A multi-dataset approach to medical image segmentation},
  author={Ulrich, Constantin and Isensee, Fabian and Wald, Tassilo and Zenk, Maximilian and Baumgartner, Michael and Maier-Hein, Klaus H},
  booktitle={International Conference on Medical Image Computing and Computer-Assisted Intervention},
  pages={648--658},
  year={2023},
  organization={Springer},
  doi = {10.1007/978-3-031-43898-1_62},
}

@inproceedings{wald2025openmind,
  title={An {OpenMind} for {3D} medical vision self-supervised learning},
  author={Wald, Tassilo and Ulrich, Constantin and Suprijadi, Jonathan and Ziegler, Sebastian and Nohel, Michal and Peretzke, Robin and Kohler, Gregor and others},
  booktitle={Proceedings of the IEEE/CVF International Conference on Computer Vision},
  pages={23839--23879},
  year={2025},
  doi = {10.1109/ICCV51701.2025.02213},
}

@inproceedings{wald2025revisiting,
  title={Revisiting {MAE} pre-training for {3D} medical image segmentation},
  author={Wald, Tassilo and Ulrich, Constantin and Lukyanenko, Stanislav and Goncharov, Andrei and Paderno, Alberto and Miller, Maximilian and Maerkisch, Leander and others},
  booktitle={Proceedings of the Computer Vision and Pattern Recognition Conference},
  pages={5186--5196},
  year={2025},
  doi = {10.1109/cvpr52734.2025.00489},
}

@article{wang2025sam,
  title={{SAM-Med3D}: a vision foundation model for general-purpose segmentation on volumetric medical images},
  author={Wang, Haoyu and Guo, Sizheng and Ye, Jin and Deng, Zhongying and Cheng, Junlong and Li, Tianbin and Chen, Jianpin and others},
  journal={IEEE Transactions on Neural Networks and Learning Systems},
  year={2025},
  publisher={IEEE},
  doi = {10.1109/tnnls.2025.3586694},
}

@article{wang2026data,
  title={A data-and compute-efficient chest X-ray foundation model beyond aggressive scaling},
  author={Wang, Chong and Zhang, Yabin and Gao, Yunhe and Varma, Maya and Mottez, Clemence and Patsatzi, Faidra and Liu, Jiaming and others},
  journal={arXiv preprint arXiv:2602.22843},
  year={2026}
}

@article{wasserthal2023totalsegmentator,
  title={{TotalSegmentator}: robust segmentation of 104 anatomic structures in {CT} images},
  author={Wasserthal, Jakob and Breit, Hanns-Christian and Meyer, Manfred T and Pradella, Maurice and Hinck, Daniel and Sauter, Alexander W and Heye, Tobias and others},
  journal={Radiology: Artificial Intelligence},
  volume={5},
  number={5},
  pages={e230024},
  year={2023},
  publisher={Radiological Society of North America},
  doi = {10.1148/ryai.230024},
}

@article{werder2022establishing,
  title={Establishing data provenance for responsible artificial intelligence systems},
  author={Werder, Karl and Ramesh, Balasubramaniam and Zhang, Rongen},
  journal={ACM Transactions on Management Information Systems (TMIS)},
  volume={13},
  number={2},
  pages={1--23},
  year={2022},
  publisher={ACM New York, NY},
  doi = {10.1145/3503488},
}

@article{wilkinson2016fair,
  title={The {FAIR} Guiding Principles for scientific data management and stewardship},
  author={Wilkinson, Mark D and Dumontier, Michel and Aalbersberg, IJsbrand Jan and Appleton, Gabrielle and Axton, Myles and Baak, Arie and Blomberg, Niklas and others},
  journal={Scientific data},
  volume={3},
  number={1},
  pages={1--9},
  year={2016},
  publisher={Nature Publishing Group},
  doi = {10.1038/sdata.2016.18},
}

@misc{xiao2025bias,
  title= {{Bias in Large {AI} Models for Medicine and Healthcare: Survey and Challenges}},
  author={Xiao, Ying and Chen, Zhenpeng and Huang, Jen-tse and Chen, Wenting and Liu, Yepang and Li, Kezhi and Mousavi, Mohammad Reza and others},
  year={2025},
  howpublished = {Preprints.org, \doi{10.20944/preprints202511.1838.v1}},
  doi = {10.20944/preprints202511.1838.v1}
}

@article{xing2017deep,
  title={Deep learning in microscopy image analysis: A survey},
  author={Xing, Fuyong and Xie, Yuanpu and Su, Hai and Liu, Fujun and Yang, Lin},
  journal={IEEE transactions on neural networks and learning systems},
  volume={29},
  number={10},
  pages={4550--4568},
  year={2017},
  publisher={IEEE},
  doi = {10.1109/tnnls.2017.2766168},
}

@article{xu2026cads,
  title = {{CADS: a comprehensive anatomical dataset and segmentation for whole-body anatomy in computed tomography}},
  author = {Xu, Murong and Navarro, Fernando and Fritsak, Maksym and Hamamci, Ibrahim Ethem and Shit, Suprosanna and Wittmann, Bastian and Er, Sezgin and others},
  year = {2026},
  journal = {Nature Communications},
  doi = {10.1038/s41467-026-78732-1},
}

@article{yang2024ribfrac,
  title={Deep rib fracture instance segmentation and classification from {CT} on the {RibFrac} challenge},
  author={Yang, Jiancheng and Shi, Rui and Jin, Liang and Huang, Xiaoyang and Kuang, Kaiming and Wei, Donglai and Gu, Shixuan and others},
  journal={IEEE Transactions on Medical Imaging},
  year={2025},
  publisher={IEEE},
  doi = {10.1109/TMI.2025.3565514},
}

@article{yaqub2022deep,
  title={Deep learning-based image reconstruction for different medical imaging modalities},
  author={Yaqub, Muhammad and Jinchao, Feng and Arshid, Kaleem and Ahmed, Shahzad and Zhang, Wenqian and Nawaz, Muhammad Zubair and Mahmood, Tariq},
  journal={Computational and Mathematical Methods in Medicine},
  volume={2022},
  number={1},
  pages={8750648},
  year={2022},
  publisher={Wiley Online Library},
  doi = {10.1155/2022/8750648},
}

@mastersthesis{zauner2010implementation,
  author = {Zauner, Christoph},
  title = {Implementation and Benchmarking of Perceptual Image Hash Functions},
  school = {Upper Austria University of Applied Sciences, Hagenberg Campus},
  year = {2010}
}

@article{zhou2025survey,
  title={A survey of perceptual hashing for multimedia},
  author={Zhou, Yuanding and Li, Xinran and Xiong, Cheng and Yao, Heng and Qin, Chuan},
  journal={ACM Transactions on Multimedia Computing, Communications and Applications},
  volume={21},
  number={7},
  pages={1--28},
  year={2025},
  publisher={ACM New York, NY},
  doi = {10.1145/3727880},
}

@article{zhuang2025mrisegmenter,
  title   = {{MRISegmenter}: A Fully Accurate and Robust Automated Multiorgan and Structure Segmentation Tool for {T1}-weighted Abdominal {MRI}},
  author  = {Zhuang, Yan and Mathai, Tejas Sudharshan and Mukherjee, Pritam and Khoury, Brandon and Kim, Boah and Hou, Benjamin and Rabbee, Nusrat and others},
  journal = {Radiology},
  volume  = {315},
  number  = {1},
  pages   = {e241979},
  year    = {2025},
  doi     = {10.1148/radiol.241979}
}

@article{zou2022review,
  title={A review of deep learning-based deformable medical image registration},
  author={Zou, Jing and Gao, Bingchen and Song, Youyi and Qin, Jing},
  journal={Frontiers in Oncology},
  volume={12},
  pages={1047215},
  year={2022},
  publisher={Frontiers Media SA},
  doi = {10.3389/fonc.2022.1047215},
}

\newpage
\appendix

\section{Web-based Platform to Support Hashing Analysis And \textit{MAP}}
\label{apx:webtool}
To support practical adoption of MAP and large-scale provenance analysis, we provide a web-based platform accompanying this work. The platform provides three primary functionalities. \textit{Firstly}, users can explore the analyzed dataset collection and inspect its structure, overlaps, and provenance relationships. \textit{Secondly}, users can download a lightweight script to compute perceptual hashes locally on their own datasets and upload the resulting hashes for comparison against our reference database of public medical imaging datasets, enabling privacy-conscious duplicate screening without sharing the underlying medical images. \textit{Thirdly}, users can generate MAP profiles for their own dataset collections through a guided interface, producing portable structured metadata documenting provenance, dataset composition, and deduplication practices. Together, these tools provide a practical workflow for scalable provenance analysis and transparent documentation of dataset collections.

The platform is publicly available at \url{https://mic-dct.dkfz.de/}.

\section{Perceptual Hashing}
The Perceptual hashing is explained in detail in \cref{alg:hash_similarity_search}.

\label{apx:percetual hashing}
\begin{algorithm}[ht]
\caption{Perceptual hash similarity search with ANN indexing}
\label{alg:hash_similarity_search}
\begin{algorithmic}[1]
\Require Query images $\mathcal{Q}$, reference images $\mathcal{R}$, hash size $s$, ANN trees $T$, distance threshold $\tau$
\Ensure Candidate similarity edges $\mathcal{E}$

\Function{PerceptualHash}{$I, s$}
    \State $I_r \leftarrow \mathrm{Resample}(I; s,s,s)$ \Comment{Resample to fixed $s\times s\times s$ grid}
    \State $m \leftarrow \mathrm{median}(I_r)$ \Comment{Compute global intensity median}
    \State \Return $\mathrm{flatten}\!\left(\mathbf{1}_{\{I_r > m\}}\right)$ \Comment{Median-thresholded binary embedding}
\EndFunction

\State $\mathcal{H}_R \leftarrow \emptyset$ \Comment{Initialize reference hash set}
\For{each image $I_r \in \mathcal{R}$}
    \State $\mathcal{H}_R \leftarrow \mathcal{H}_R \cup \{\Call{PerceptualHash}{I_r, s}\}$ \Comment{Encode reference images as hashes}
\EndFor

\State $\mathcal{A} \leftarrow \mathrm{BuildAnnoyIndex}(\mathcal{H}_R; T)$ \Comment{Build ANN index with $T$ random projection trees}
\State $\mathcal{E} \leftarrow \emptyset$ \Comment{Initialize set of matched pairs}

\For{each image $I_q \in \mathcal{Q}$}
    \State $h_q \leftarrow \Call{PerceptualHash}{I_q, s}$ \Comment{Compute query hash}
    \State $\mathcal{N}_q \leftarrow \mathrm{QueryANN}(\mathcal{A}, h_q)$ \Comment{Retrieve candidate neighbors}
    \For{each $(h_r, I_r) \in \mathcal{N}_q$}
        \State $d \leftarrow \mathrm{Hamming}(h_q, h_r)$ \Comment{Compute exact Hamming distance}
        \If{$d \leq \tau$}
            \State $\mathcal{E} \leftarrow \mathcal{E} \cup \{(I_q, I_r, d)\}$ \Comment{Retain pair if within threshold}
        \EndIf
    \EndFor
\EndFor

\State \Return $\mathcal{E}$ \Comment{Return thresholded similarity graph}
\end{algorithmic}
\end{algorithm}

\textbf{Evaluation of the approximate nearest-neighbor search.}
Annoy returns approximate nearest neighbors and may therefore miss duplicates that an exhaustive search would find. To quantify this, we compared the ANN search of our pipeline (\cref{alg:hash_similarity_search}) with an exhaustive Hamming search on a random sample of 50,000 of the images analyzed in this paper. For each sampled image, the exhaustive search determines the closest image from another dataset; 5,552 of the sampled images have such an image at a Hamming distance of at most 100, 4,909 of them with an identical hash. For 5,539 of these 5,552 images (99.8\%), the ANN search also finds the closest image from another dataset.

\section{Manual validation of perceptual hashing}
\label{apx:manval}

\begin{figure}[ht]
    \centering
    \includegraphics[width=.8\textwidth]{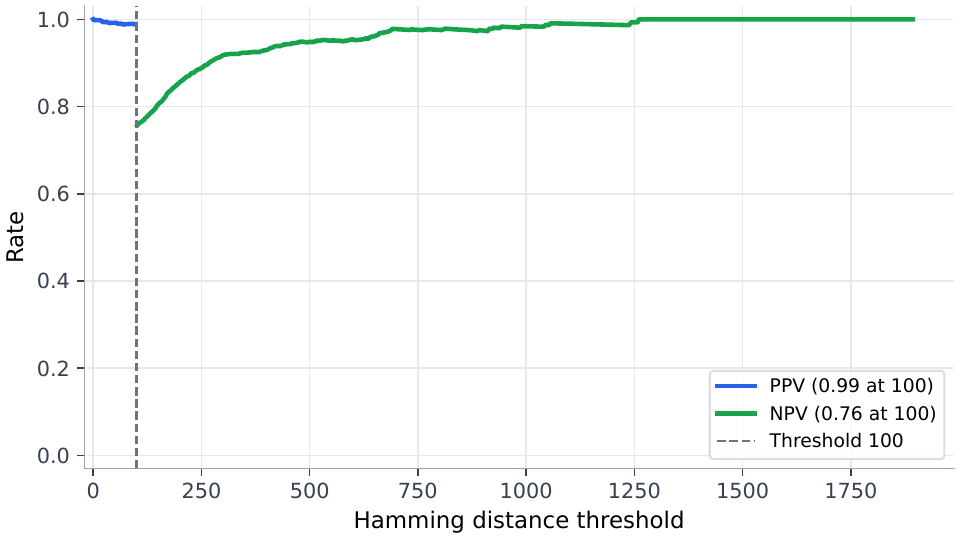}
    \caption{\textbf{Manual validation of the duplicate detection on 5{,}000 image pairs.} The pairs were sampled separately below (1{,}000) and above (4{,}000) the Hamming distance threshold of 100 used in this work and labeled manually; all values are computed on this stratified sample and are conditional on the top-1 candidate of the ANN search. Shown are the PPV for thresholds below 100 and the NPV for thresholds of 100 and above, each estimated within one sampling stratum.}
    \label{fig:ppvnpv}
\end{figure}

\begin{figure}[ht]
    \centering

    \begin{subfigure}[t]{0.8\textwidth}
        \centering
        \includegraphics[width=0.48\linewidth]{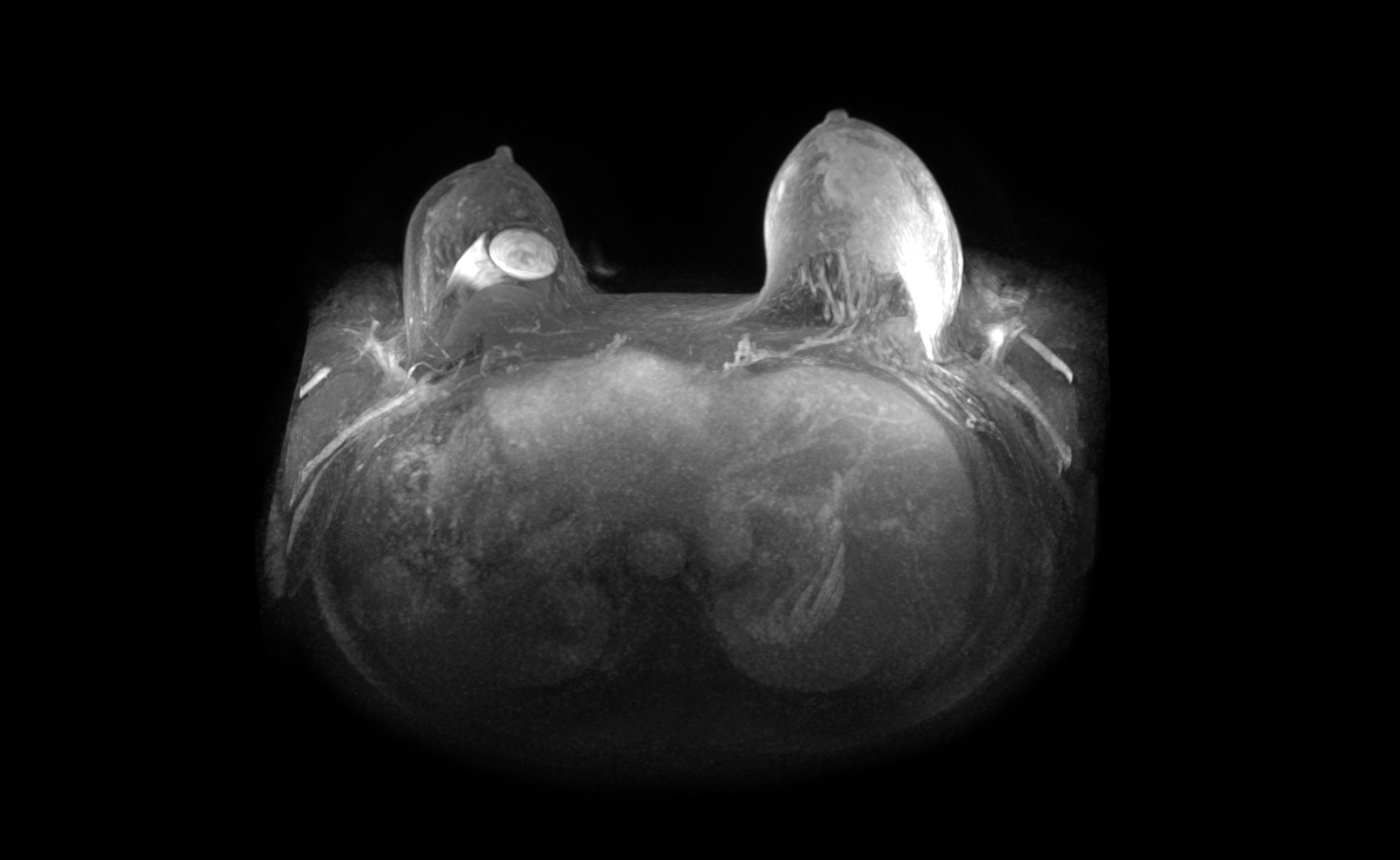}
        \hfill
        \includegraphics[width=0.48\linewidth]{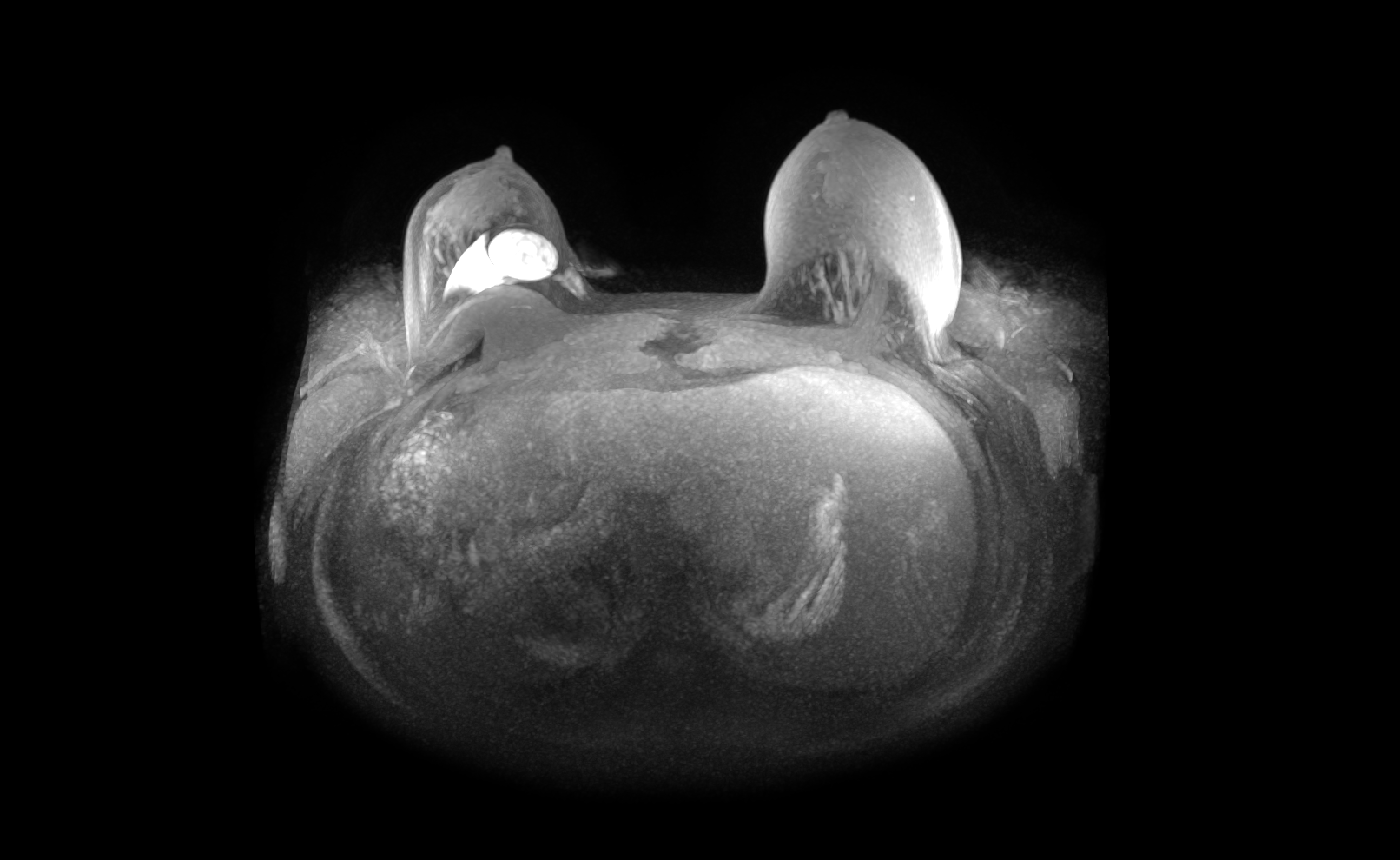}
        \caption{Hash distance: 214}
    \end{subfigure}
    \hfill
    \begin{subfigure}[t]{0.8\textwidth}
        \centering
        \includegraphics[width=0.48\linewidth]{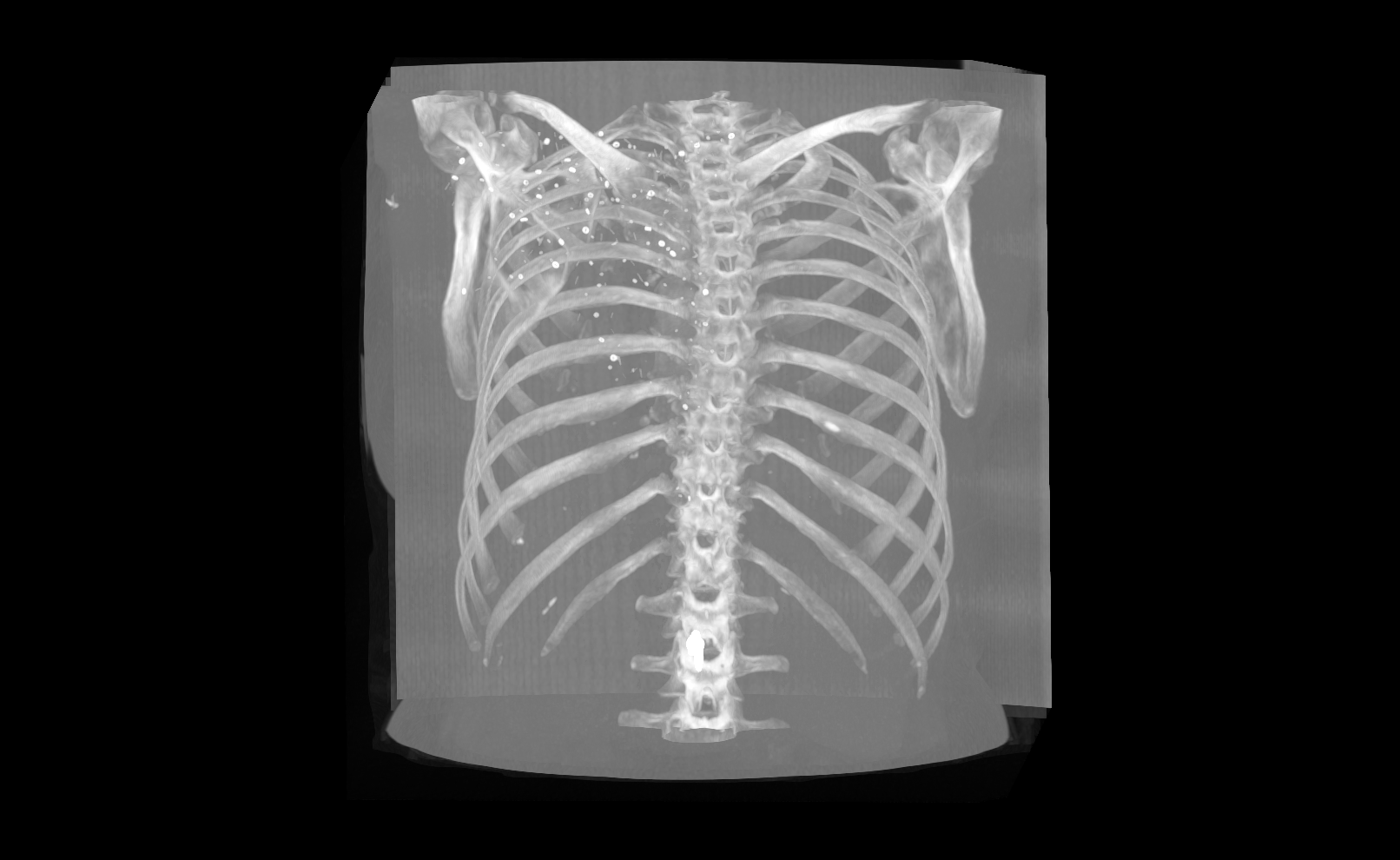}
        \hfill
        \includegraphics[width=0.48\linewidth]{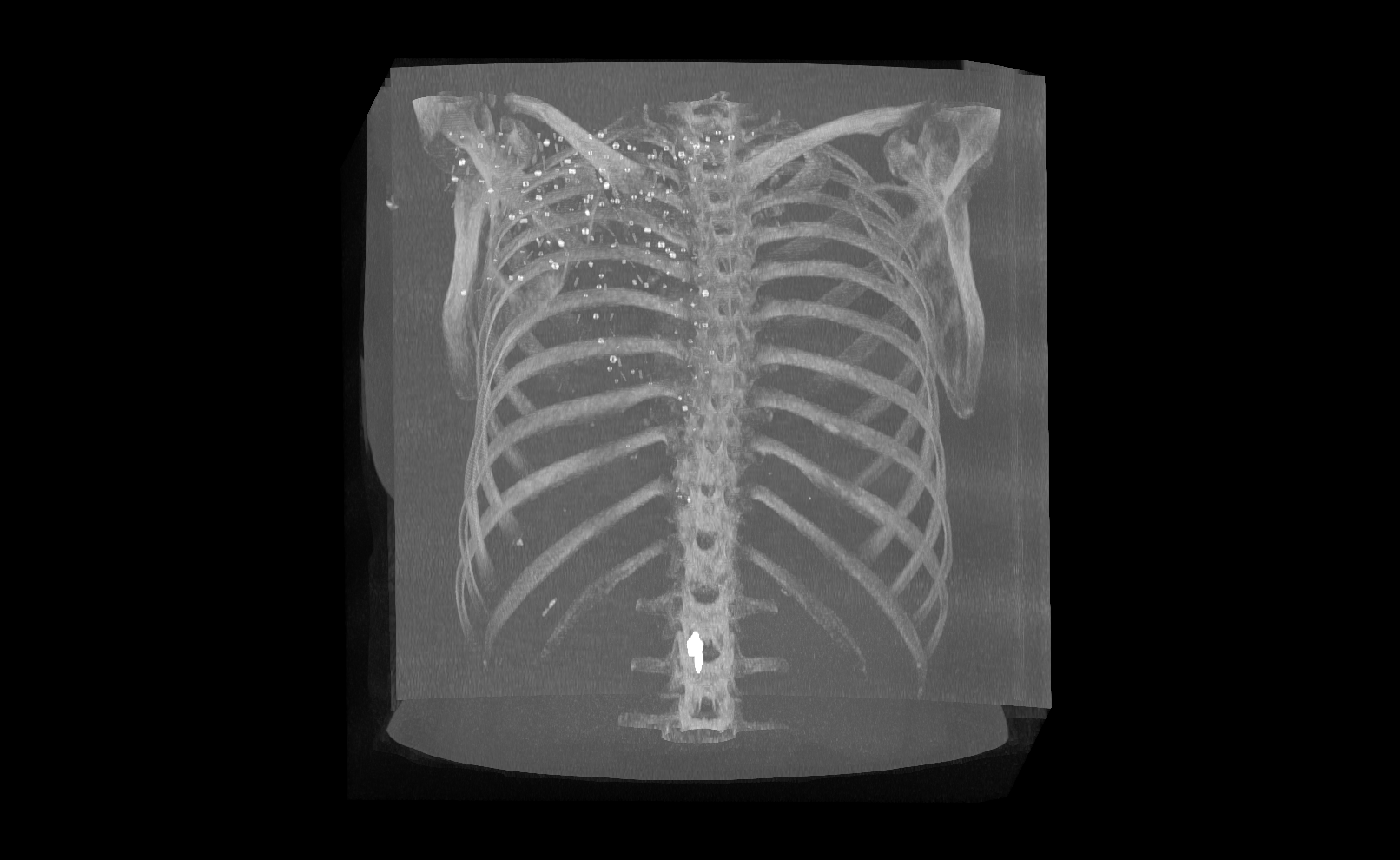}
        \caption{Hash distance: 116}
    \end{subfigure}
    \hfill
    \begin{subfigure}[t]{0.8\textwidth}
        \centering
        \includegraphics[width=0.48\linewidth]{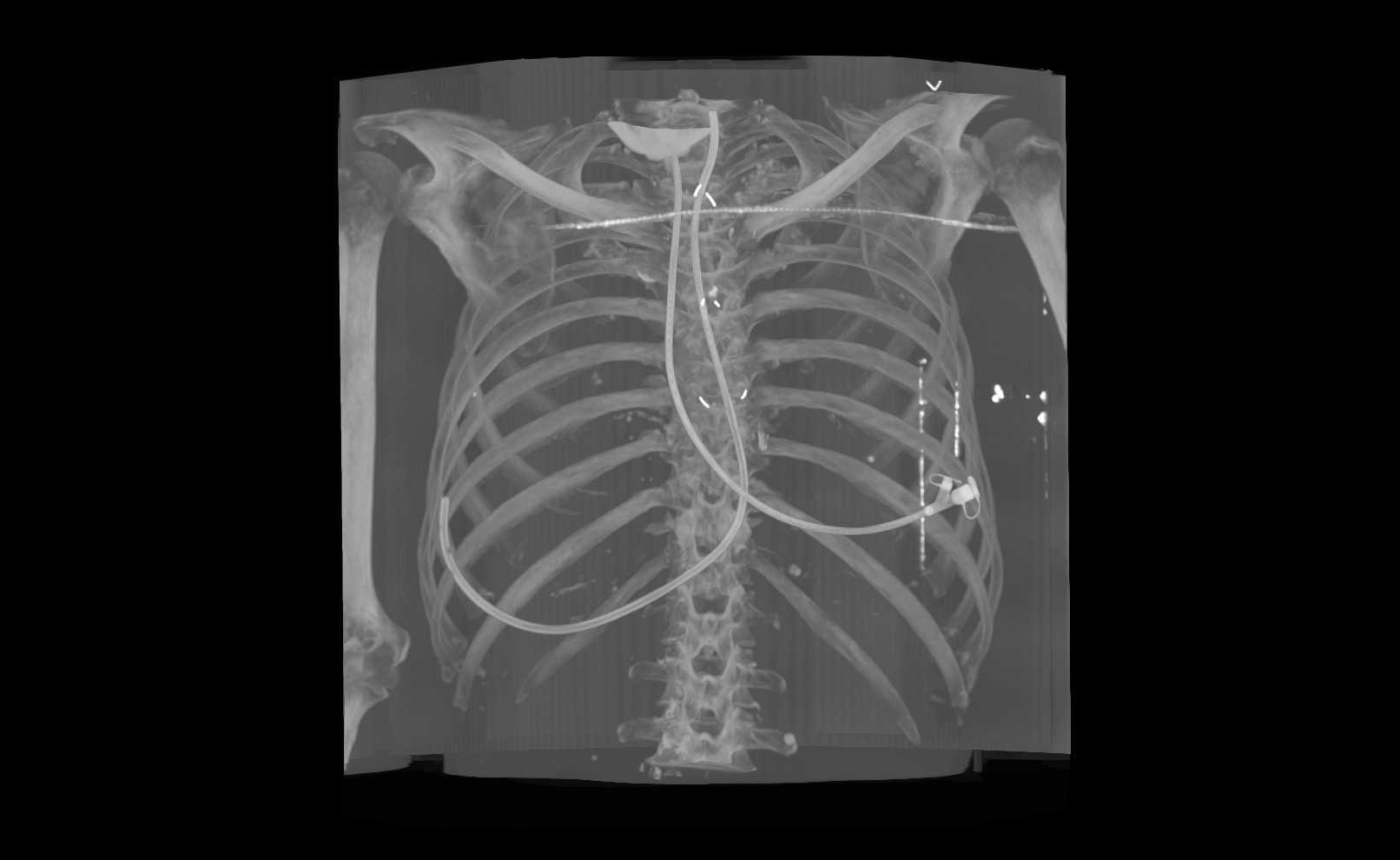}
        \hfill
        \includegraphics[width=0.48\linewidth]{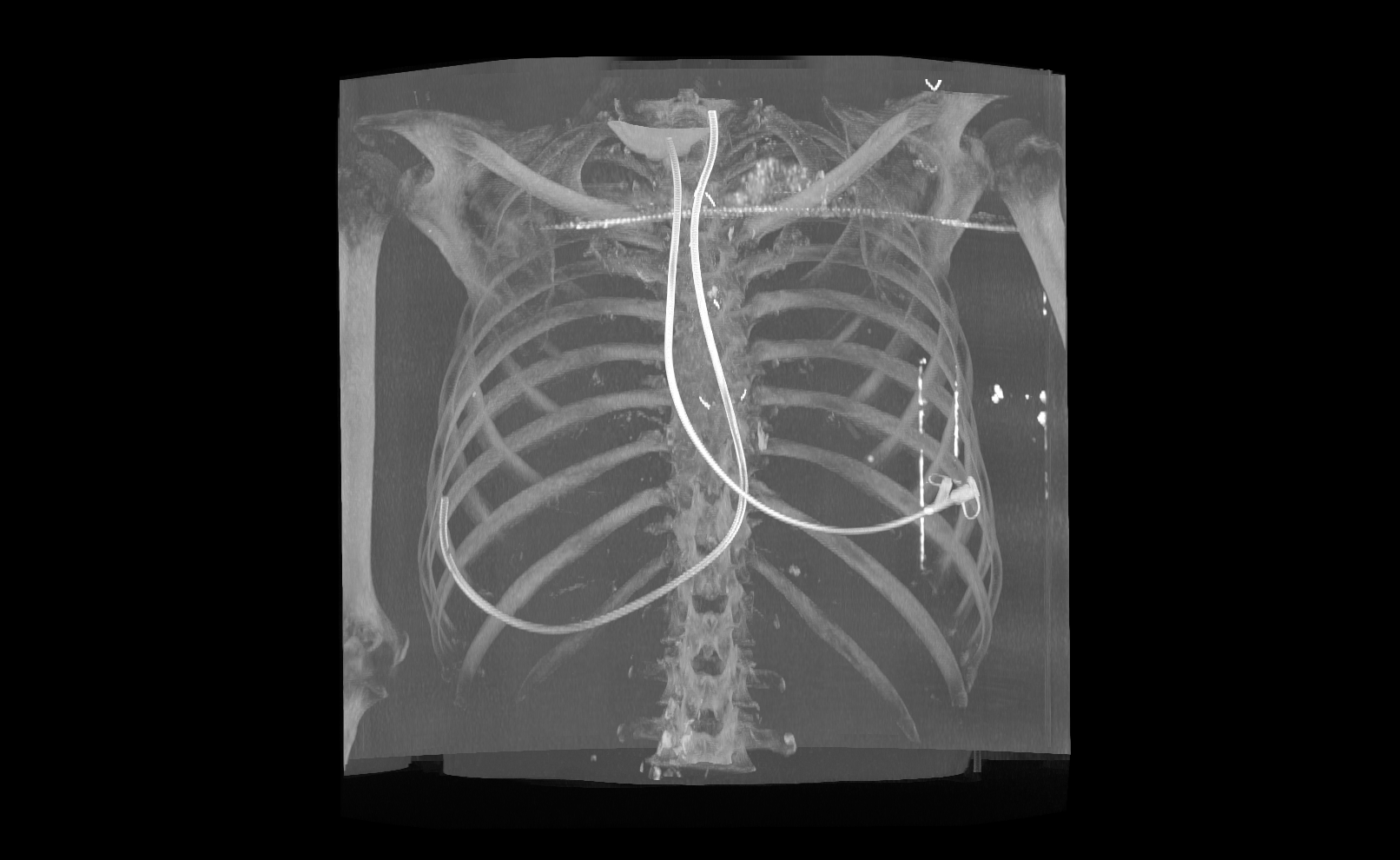}
        \caption{Hash distance: 424}
    \end{subfigure}

    \caption{Examples of duplicates identified via manual validation with hash distances larger than 100}
    \label{fig:fn_examples}
\end{figure}

\begin{figure}[ht]
    \centering
    \includegraphics[width=.45\textwidth]{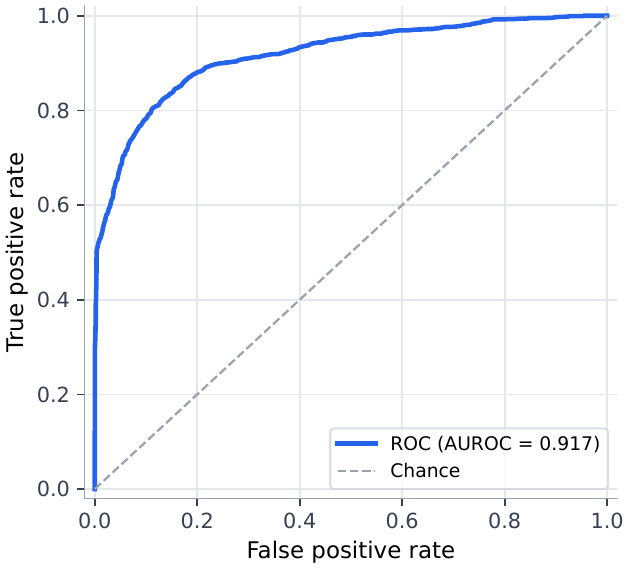}
    \caption{\textbf{ROC curve of the manual validation} (5{,}000 image pairs, AUROC 0.917), shown for completeness. Since the pairs were sampled separately below and above the threshold of 100, TPR and FPR are not proper estimates for the full collection, and the ROC curve is not a suitable metric for this stratified sample; PPV and NPV (\cref{fig:ppvnpv}) are better suited.}
    \label{fig:roc}
\end{figure}

\cref{fig:ppvnpv} shows PPV and NPV of the duplicate detection over the distance threshold. At the threshold of 100, the PPV is 0.99 and the NPV is 0.76: flagged pairs are almost always true duplicates, while about a quarter of the sampled pairs above the threshold are duplicates as well. Below the threshold, the PPV increases towards smaller distances (1.00 for identical hashes); above it, the NPV increases with the distance (0.86 at 200, 0.92 at 300), i.e., most missed duplicates lie at distances not far above the threshold. A further analysis of the false negative duplicates shows that they predominantly fall into three major categories:
\begin{enumerate}
    \item \textbf{Transformed data:} This category mainly includes cases, where data was modified using intensity clipping or (ROI-) cropping before inclduing it in a collection. We investigate the effects of such transformations further in \cref{apx:hashlimits}.
    \item \textbf{Different contrast phases:} Data from different contrast phases show visual similarity while shifting intensity statistics. While these are not strictly duplicates, they carry similar risks to exact duplicates, when it comes to training and evaluation. Consequently, we flagged them as duplicates during manual validation.
    \item \textbf{Different reconstructions:} This category mainly includes CT data, which was reconstructed with different reconstruction kernels. Similar to the previous category, we flagged them as duplicates during manual validation.
\end{enumerate}
Examples of false negatives with hash distance larger than 100 are visualized in \cref{fig:fn_examples}. For completeness, \cref{fig:roc} shows the ROC curve of the manual validation (AUROC 0.917).

\FloatBarrier

\section{The effect of transformations on hash similarity}
\label{apx:hashlimits}

\begin{figure*}[ht]
    \centering
    \setlength{\tabcolsep}{4pt}
    \renewcommand{\arraystretch}{1.15}

    \begin{tabular}{>{\centering\arraybackslash}m{0.22\textwidth}
                    >{\centering\arraybackslash}m{0.22\textwidth}
                    >{\centering\arraybackslash}m{0.22\textwidth}
                    >{\centering\arraybackslash}m{0.22\textwidth}}
        \toprule
        Original
        & Intensity clipping $[-1000,1000]$ HU
        & Cropping by 2 voxels
        & Left/right flipping \\
        \midrule

        \multicolumn{4}{c}{\textbf{AMOS}} \\
        \includegraphics[width=0.20\textwidth]{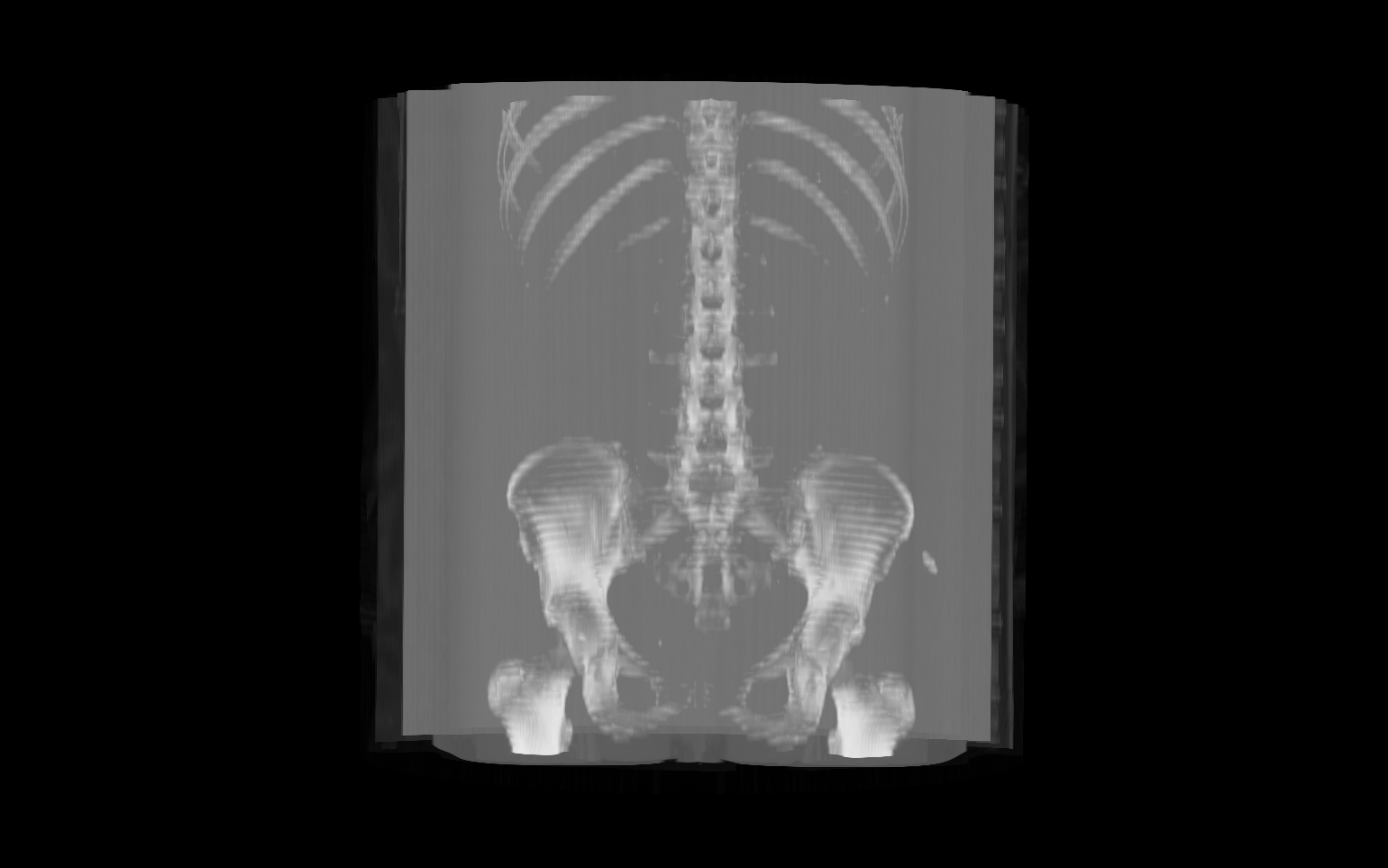}
        \newline
        Original Image
        &
        \includegraphics[width=0.20\textwidth]{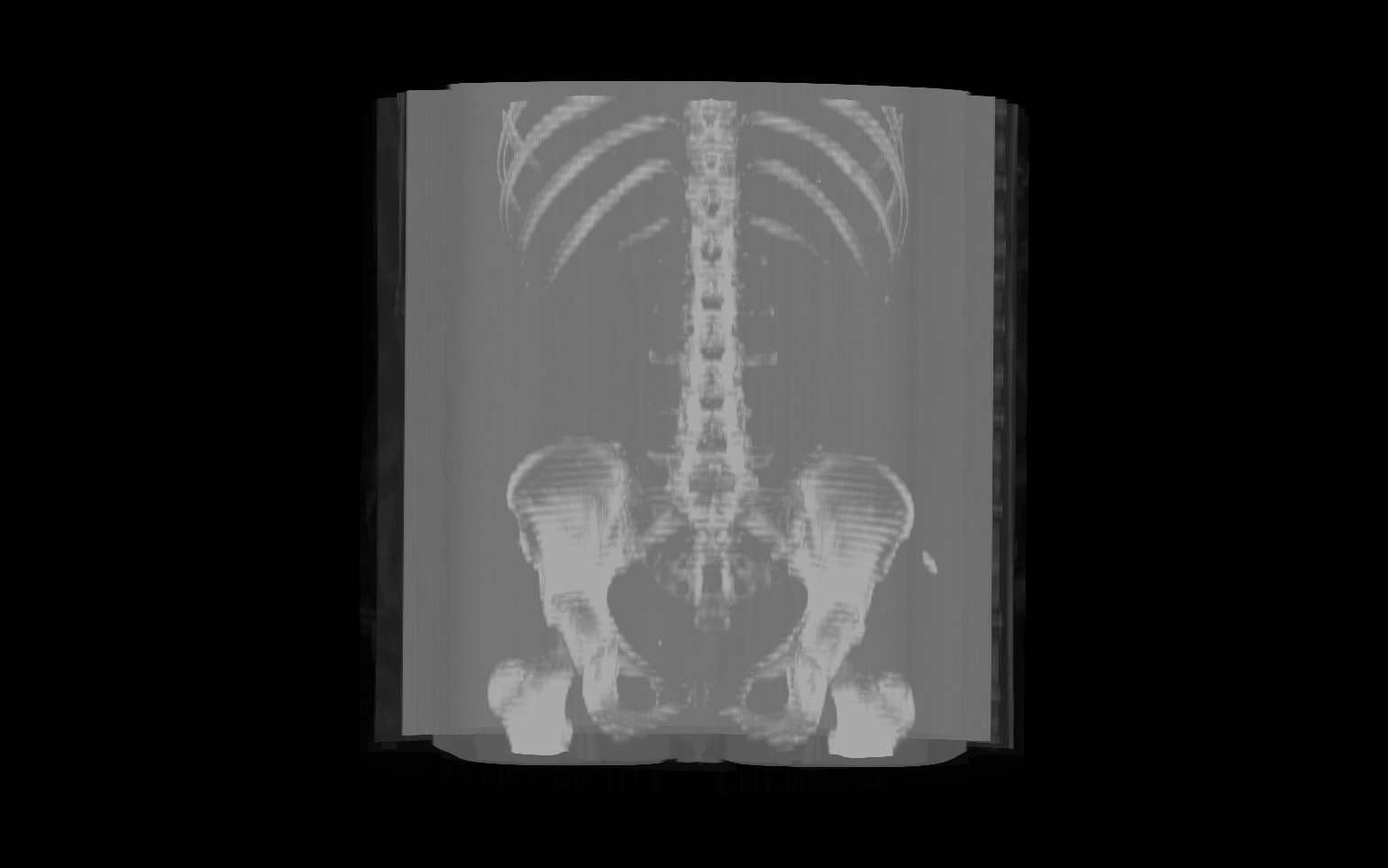}
        \newline
        Hash distance: 96
        &
        \includegraphics[width=0.20\textwidth]{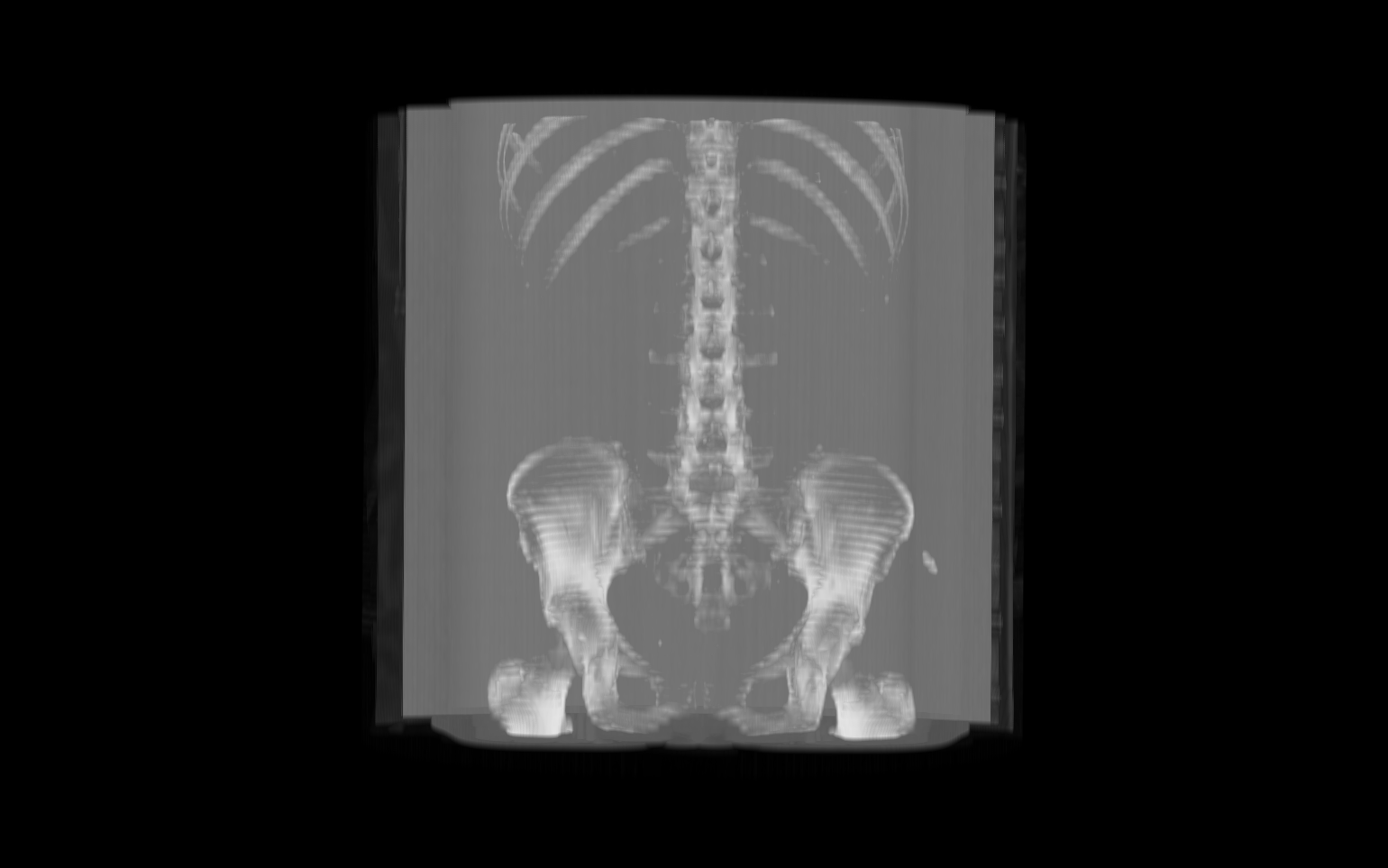}
        \newline
        Hash distance: 472
        &
        \includegraphics[width=0.20\textwidth]{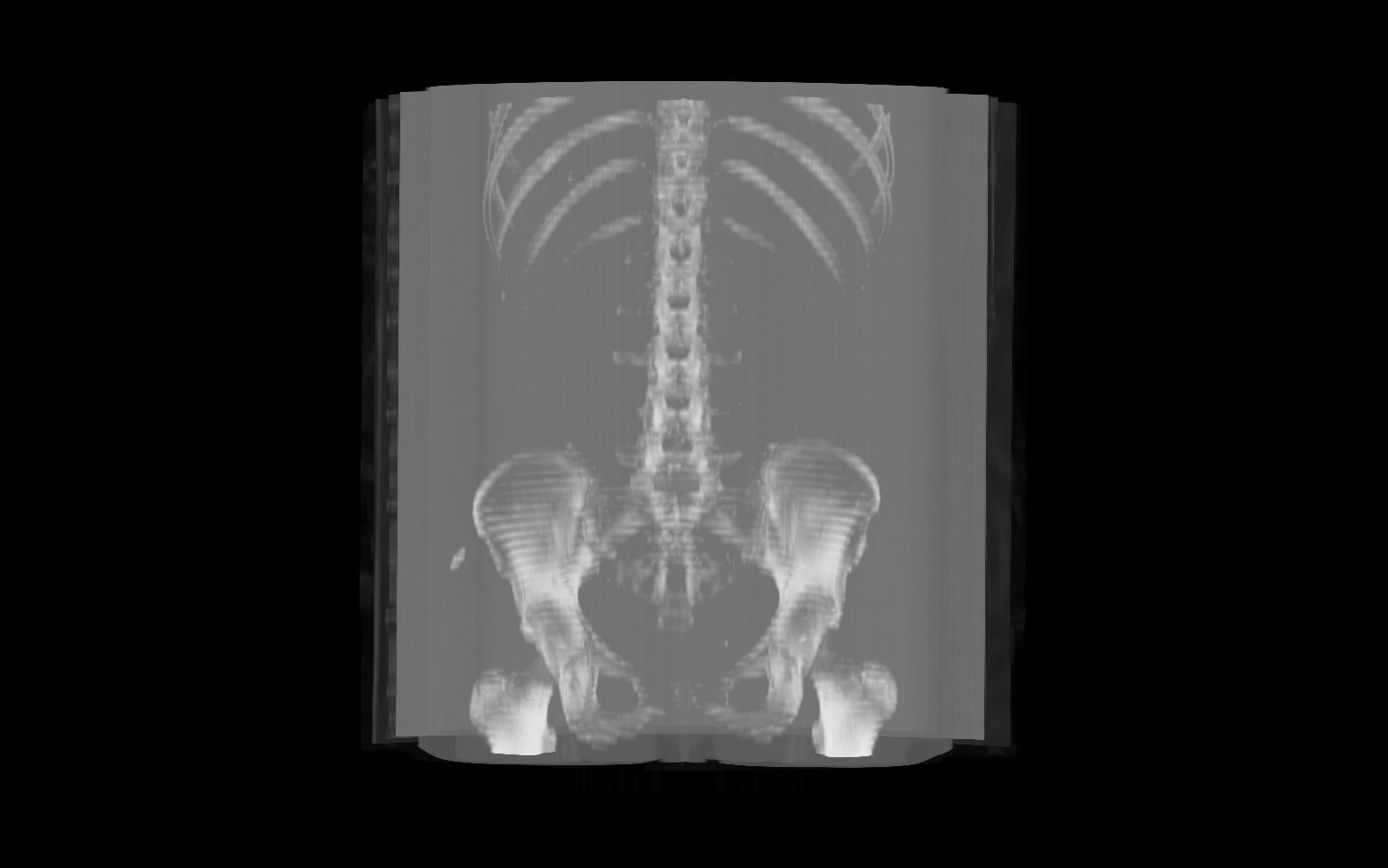}
        \newline
        Hash distance: 556
        \\

        \multicolumn{4}{c}{\textbf{AutoPET}} \\
        \includegraphics[width=0.20\textwidth]{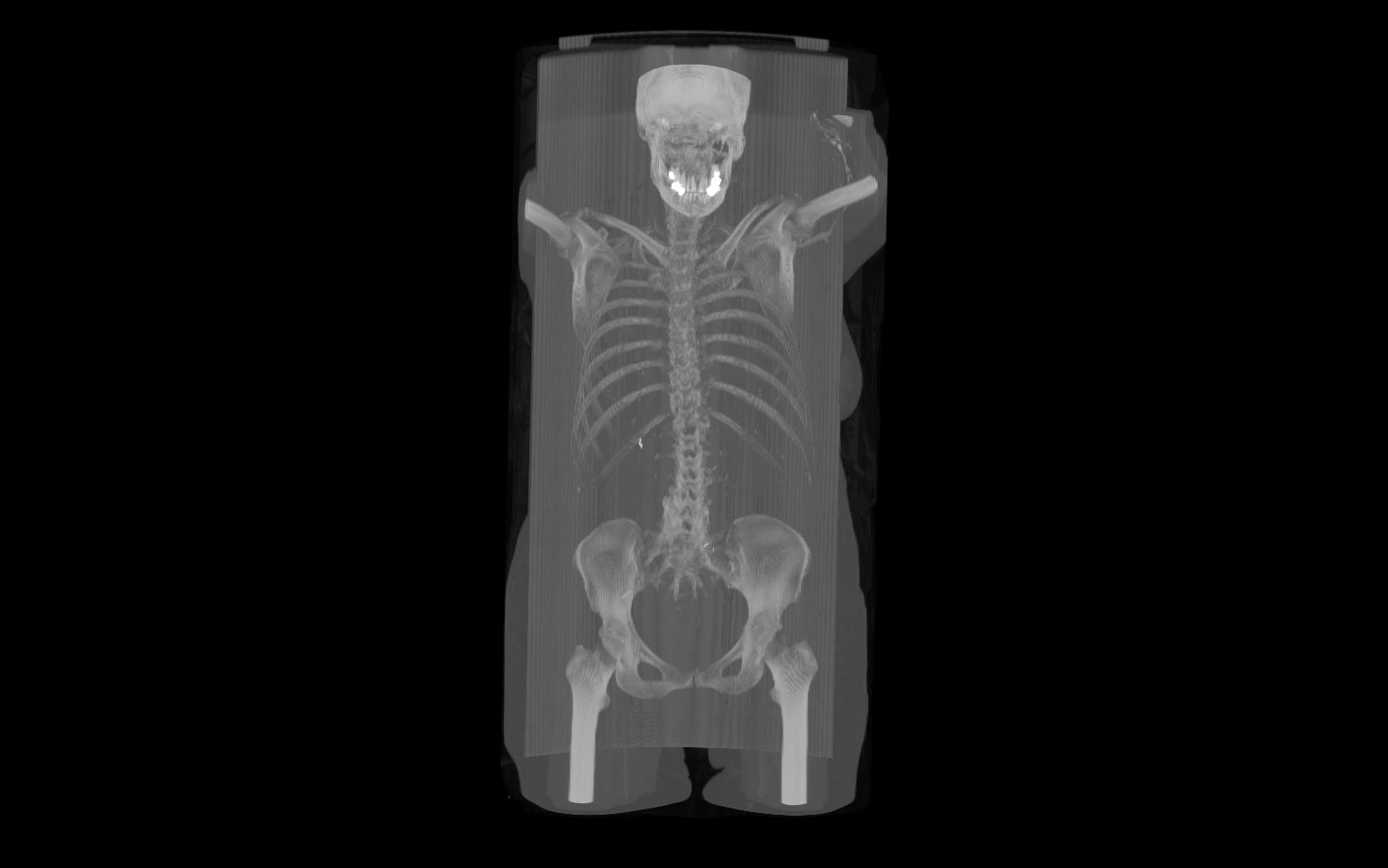}
        \newline
        Original Image
        &
        \includegraphics[width=0.20\textwidth]{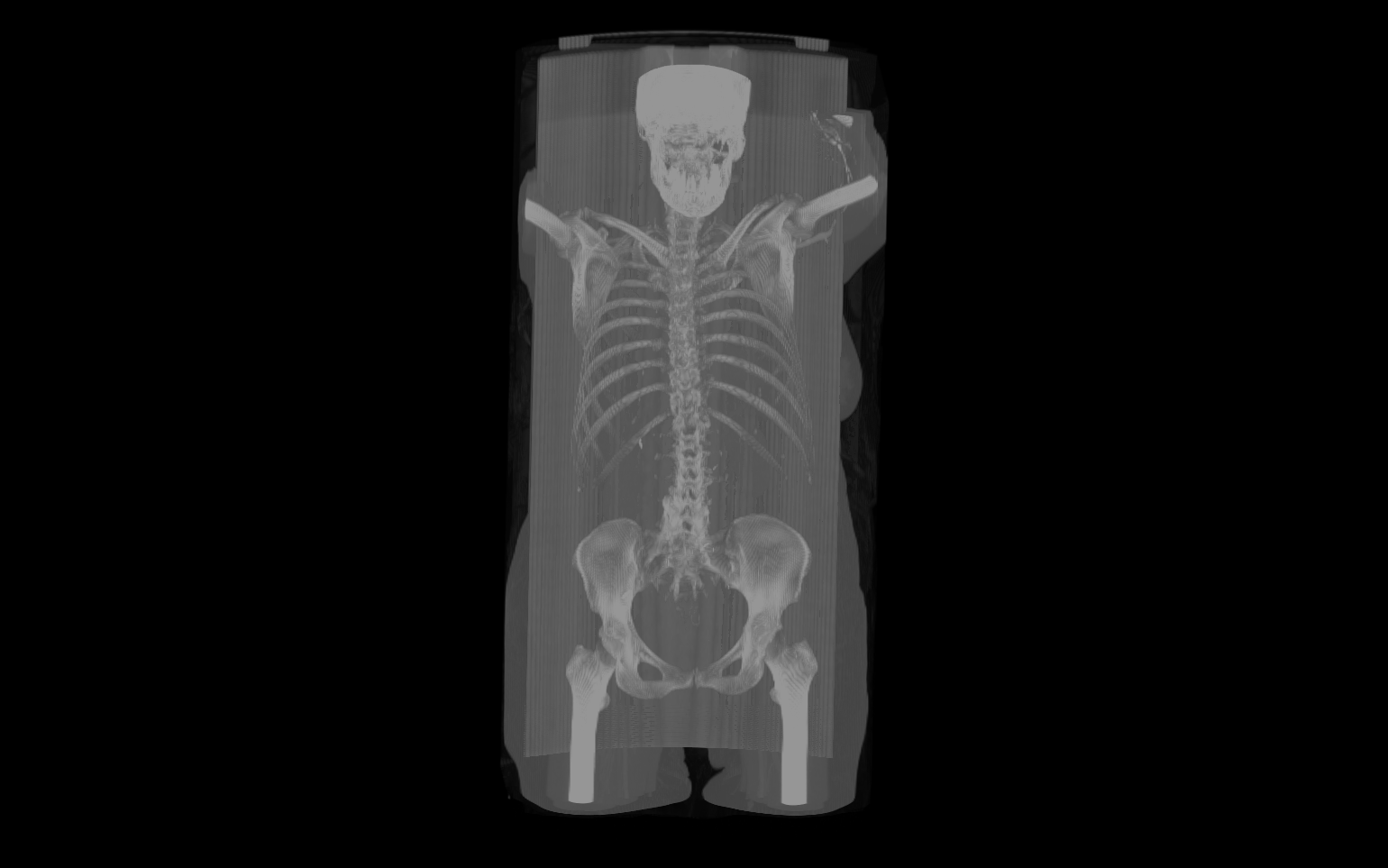}
        \newline
        Hash distance: 113
        &
        \includegraphics[width=0.20\textwidth]{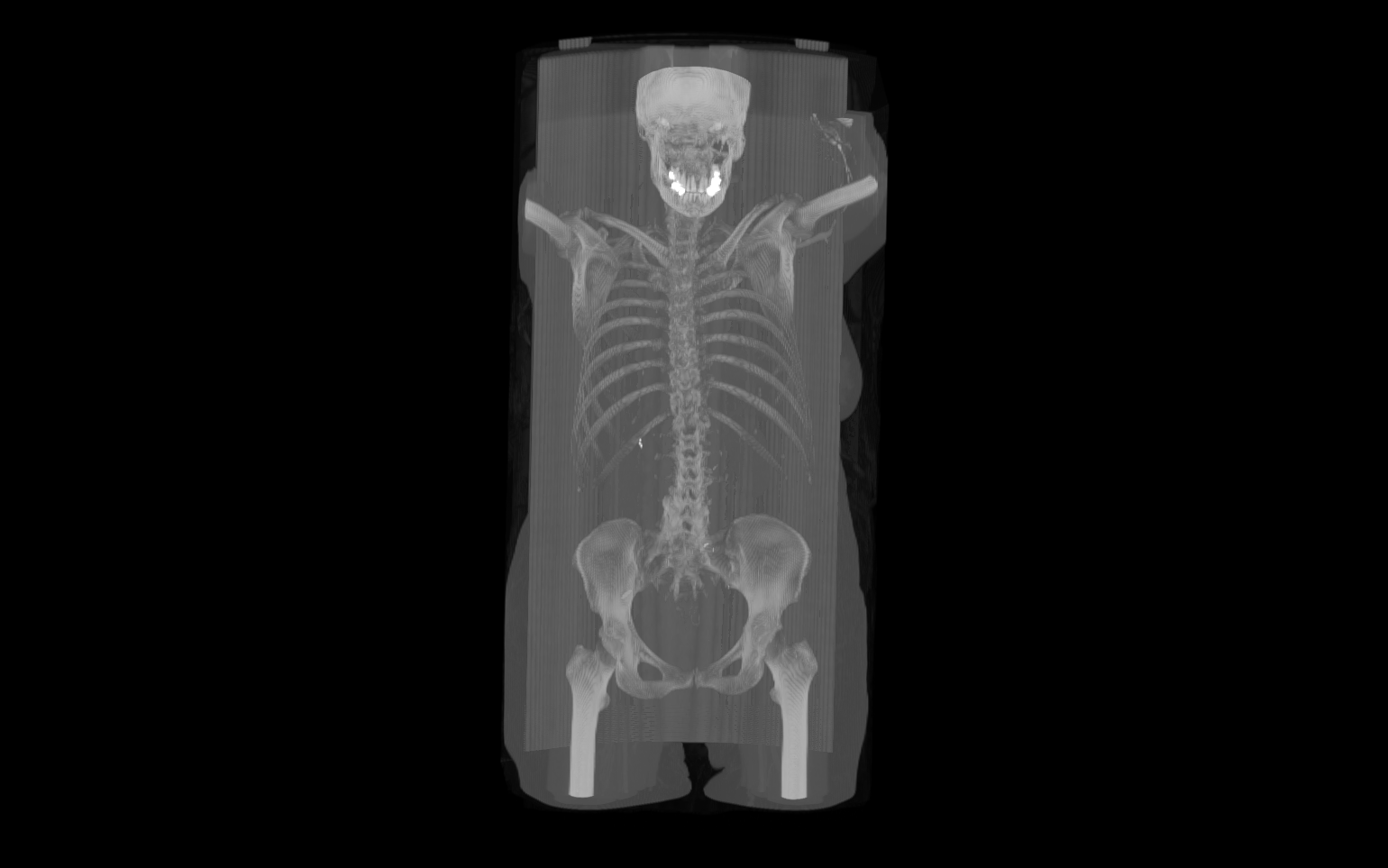}
        \newline
        Hash distance: 359
        &
        \includegraphics[width=0.20\textwidth]{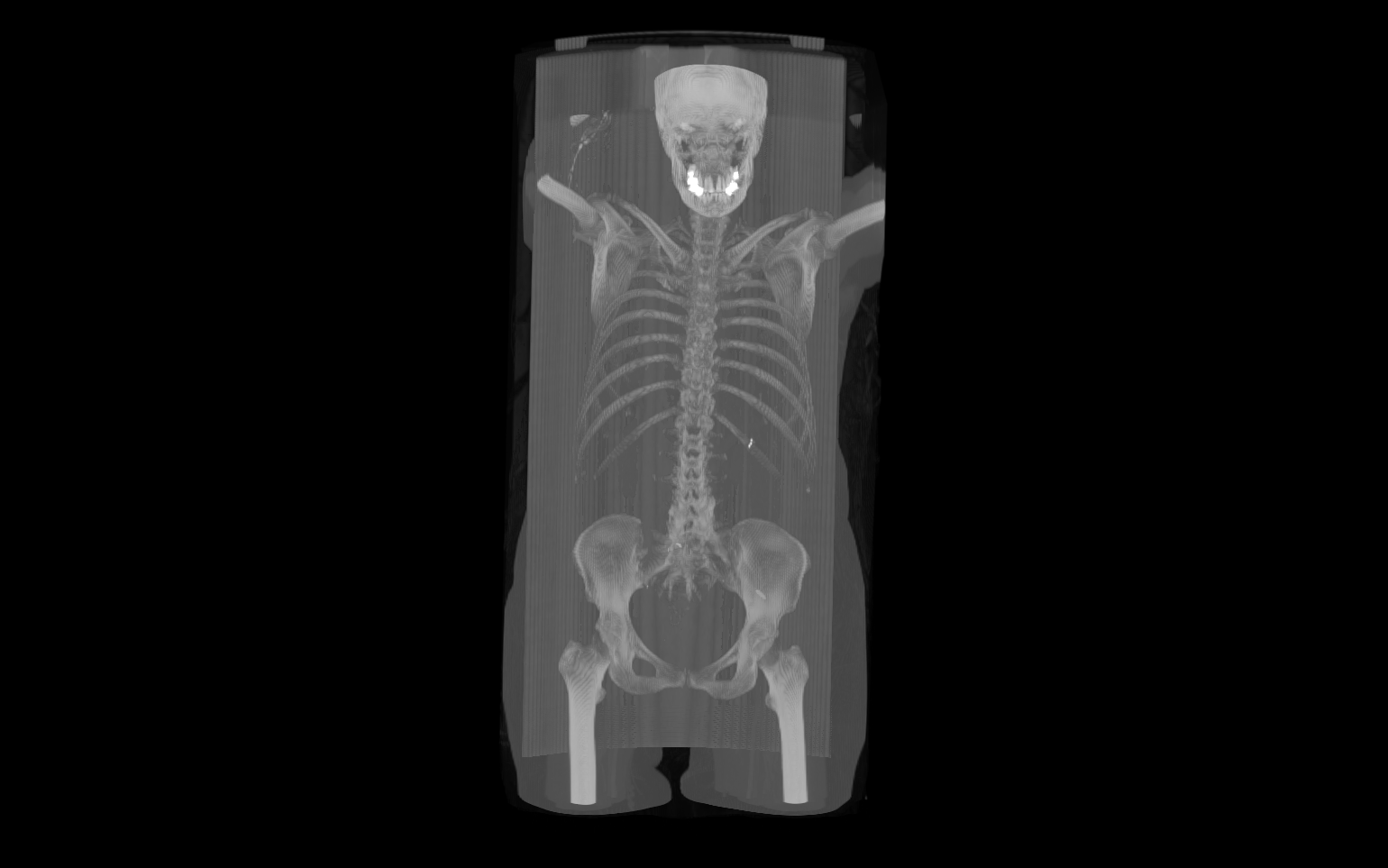}
        \newline
        Hash distance: 28
        \\

        \multicolumn{4}{c}{\textbf{KiTS}} \\
        \includegraphics[width=0.20\textwidth]{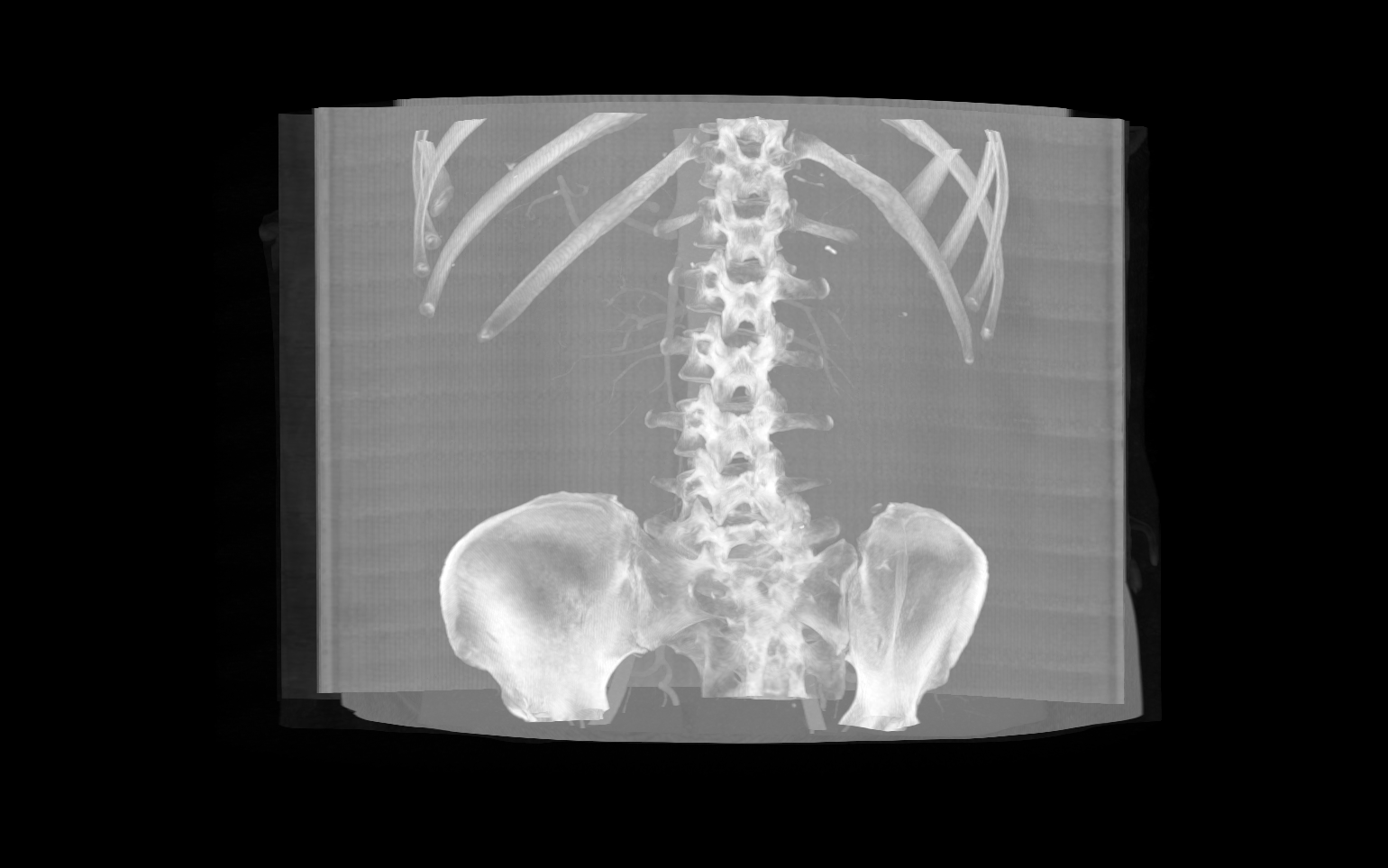}
        \newline
        Original Image
        &
        \includegraphics[width=0.20\textwidth]{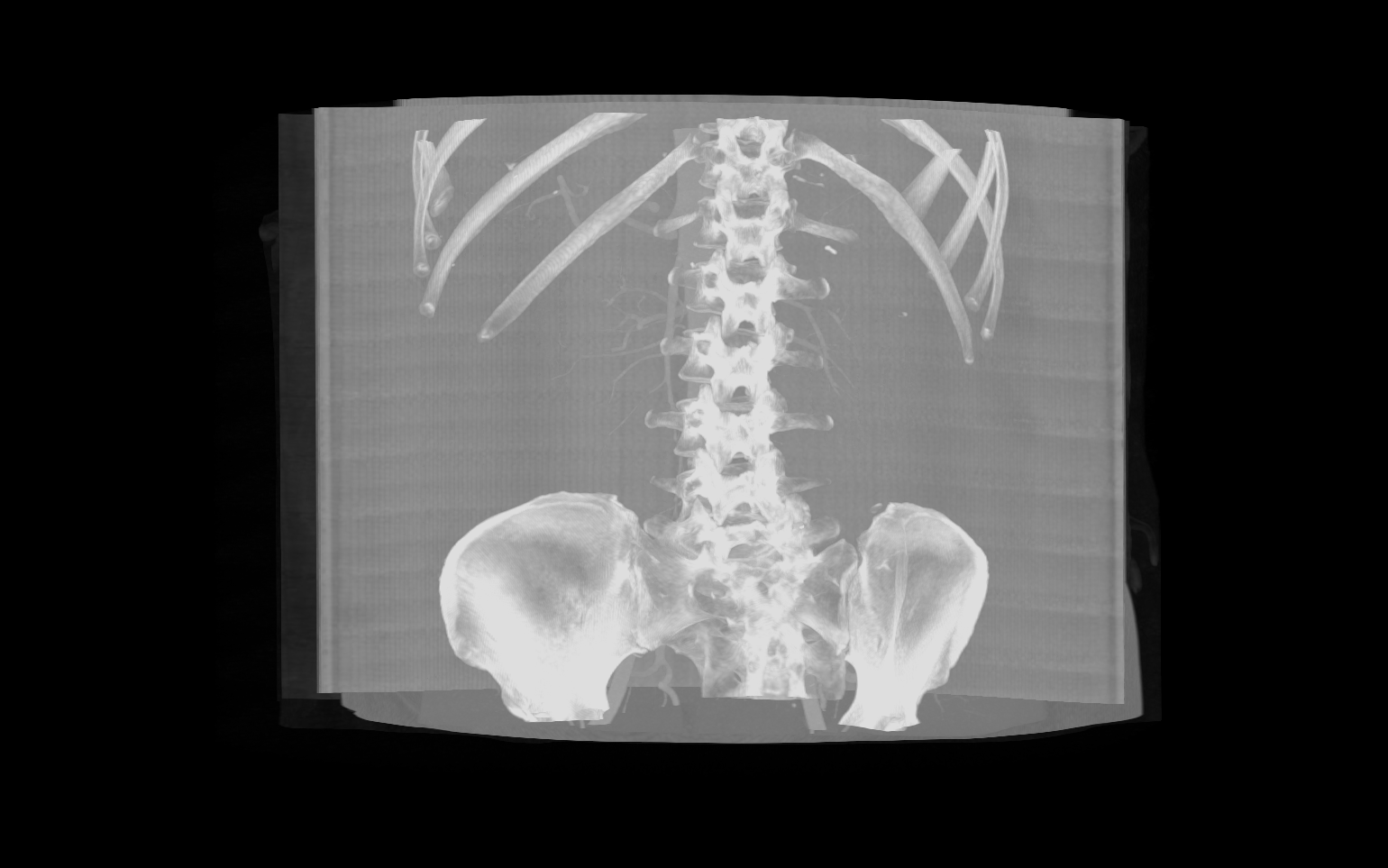}
        \newline
        Hash distance: 22
        &
        \includegraphics[width=0.20\textwidth]{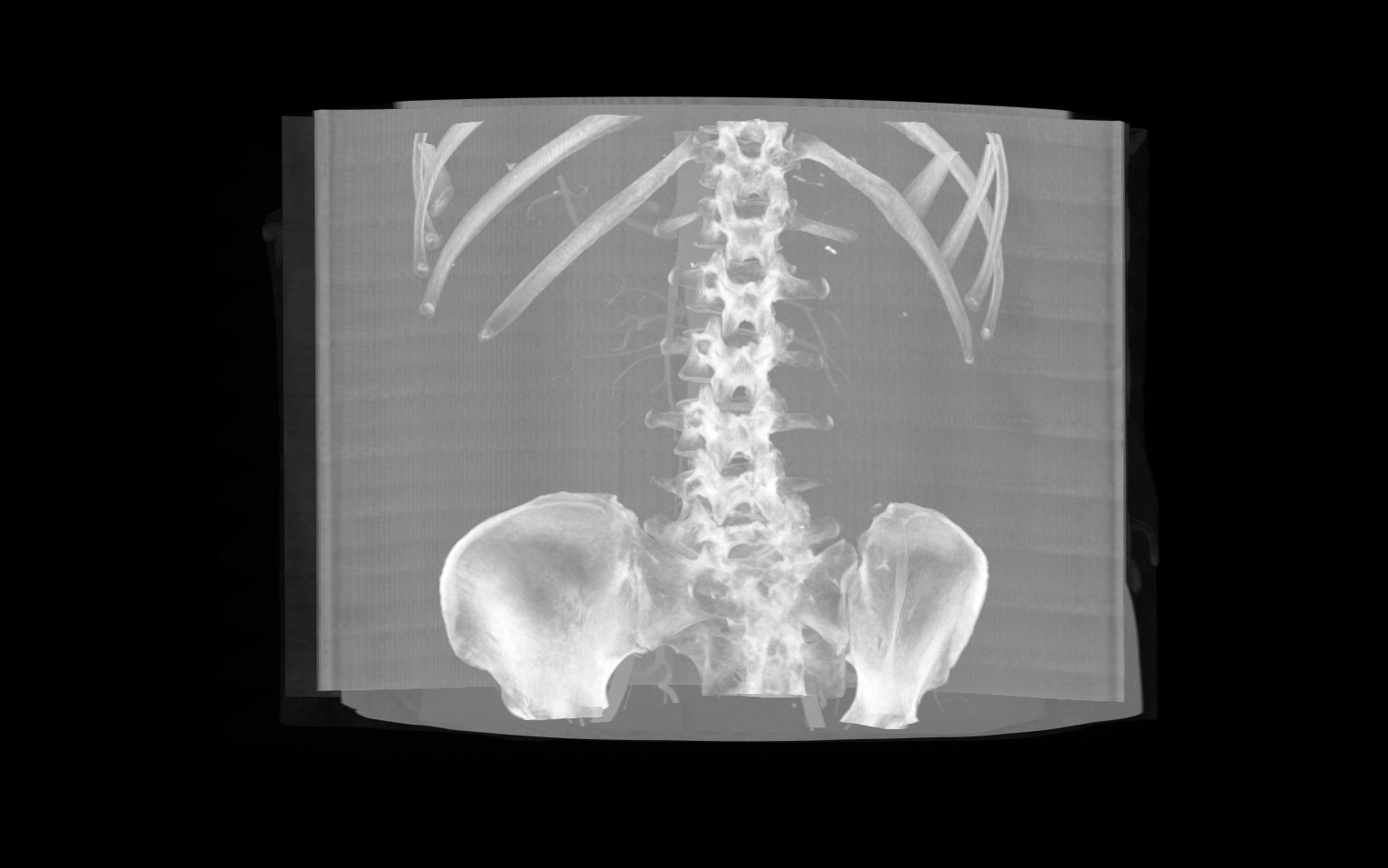}
        \newline
        Hash distance: 144
        &
        \includegraphics[width=0.20\textwidth]{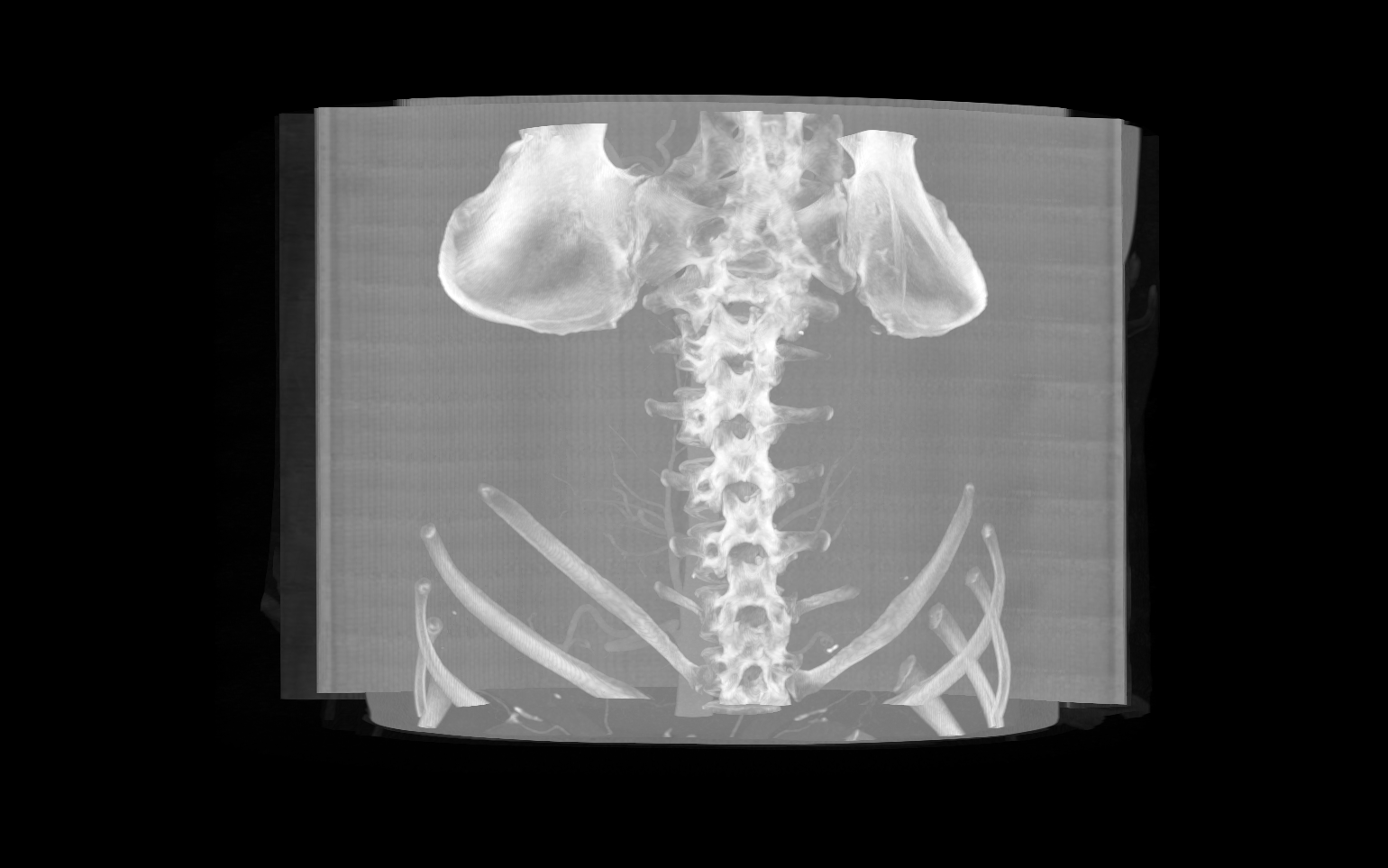}
        \newline
        Hash distance: 252
        \\
        \bottomrule
    \end{tabular}

    \caption{
        Examples of image transformations and corresponding perceptual hash distances from the original image.
        The transformations are intensity clipping to $[-1000,1000]$ Hounsfield Units, cropping by 2 voxels along each axis, and left/right flipping.
    }
    \label{fig:hash_distance_transformations}
\end{figure*}

\cref{fig:hash_distance_transformations} shows the effect of different realistic transformations, including intensity clipping, minor cropping and unintentional left/right flips, on the perceptual hashes of cases from the \texttt{AMOS}, \texttt{AutoPET} and \texttt{KiTS} datasets.

\textbf{Controlled evaluation.}
To quantify this effect beyond individual examples, we apply controlled transformations to 883 CT scans from four datasets with comparable fields of view (\texttt{AMOS}: 200, \texttt{KiTS}: 489, \texttt{LiTS}: 131, \texttt{MSD Lung}: 63) and compare the hash of each original scan with the hashes of all transformed scans. A scan and its own transformed version form a positive pair (883 pairs); a scan and the transformed version of any other scan form a negative pair (778,806 pairs). We evaluate intensity clipping to $[-1000,1000]$ HU, symmetric cropping by 2 voxels on each side along every axis, symmetric padding by 2 voxels on each side, and clipping combined with cropping. \cref{tab:transform_eval} reports TPR and FPR at the distance threshold of 100 used throughout this work, as well as the number of top-1 identification errors, i.e., scans for which the transformed version of a different scan has a smaller hash distance than their own transformed version. The hashes are largely robust to intensity clipping (TPR 0.962), whereas spatial transformations of only a few voxels strongly increase hash distances, so that fewer than 9\% of the transformed scans are still detected as duplicates. The FPR remains below $10^{-5}$ for all transformations, and top-1 identification errors remain below 5\%.

\begin{table}[ht]
    \centering
    \caption{Controlled transformation experiment on 883 CT scans (\texttt{AMOS}, \texttt{KiTS}, \texttt{LiTS}, \texttt{MSD Lung}) at a hash distance threshold of 100. TPR: share of the 883 scans whose transformed version lies within the threshold. FPR: share of the 778,806 pairs of different scans that lie within the threshold. Top-1 errors: scans for which a different transformed scan is closer than their own transformed version.}
    \label{tab:transform_eval}
    \begin{tabular}{lccc}
        \toprule
        Transformation & TPR & FPR & Top-1 errors \\
        \midrule
        Intensity clipping $[-1000,1000]$ HU & 0.962 & 0 & 3/883 \\
        Cropping (2 voxels per side) & 0.087 & 0 & 8/883 \\
        Padding (2 voxels per side) & 0.078 & 0 & 38/883 \\
        Clipping + cropping & 0.085 & 0 & 7/883 \\
        \bottomrule
    \end{tabular}
\end{table}

\FloatBarrier

\section{Detailed MAP Profile}
\label{apx:MAP}
A detailed field-specific view of the MAP Profile is made available in \cref{tab:metadata_fields}. We also discuss several nuances of the profile in the following section.

\begin{table}[t]
\centering
\caption{\textbf{Metadata profile fields, descriptions, and accepted inputs.}}
\label{tab:metadata_fields}
\scriptsize
\setlength{\tabcolsep}{4pt}
\renewcommand{\arraystretch}{1.13}
\begin{adjustbox}{width=0.975\textwidth}
\begin{tabular}{p{2.3cm} p{3.7cm} p{5.9cm} p{4.1cm}}
\toprule
\textbf{Section} & \textbf{Field} & \textbf{Description} & \textbf{Accepted input} \\
\midrule

\textbf{Basic information}
& \field{resource\_type} & What the record describes & \emph{dataset} \quad \emph{collection} \\
\cdashline{2-4}
& \field{title} & Official name of the dataset or collection & Text \\
\cdashline{2-4}
& \field{identifier} & Persistent identifier (DOI, URL, or dataset ID) & Text / URL / DOI \\
\cdashline{2-4}
& \field{version} & Version or release label & Text \\
\cdashline{2-4}
& \field{license} & License or terms of use & Controlled list (e.g., CC BY 4.0, MIT, Apache-2.0, Custom, Unknown) \\
\cdashline{2-4}
& \field{license\_other} & Custom license details, if applicable & Text \\
\cdashline{2-4}
& \field{item\_count} & Number of items in the dataset or collection & Integer $\geq 0$ \\
\cdashline{2-4}
& \field{item\_count\_unit} & Unit of the item count & \emph{subject} \quad \emph{study} \quad \emph{series} \quad \emph{image} \\
\cdashline{2-4}
& \field{modality} & Imaging modality & Multi-select: CT, MR, PET, PET-CT, Ultrasound, X-ray, etc. \\
\cdashline{2-4}
& \field{modality\_other} & Other modality, if not listed & Text \\
\cdashline{2-4}
& \field{body\_region} & Body region & Multi-select: Abdomen, Chest, Brain, etc. \\
\cdashline{2-4}
& \field{body\_region\_other} & Other body region, if not listed & Text \\

\midrule

\textbf{Source datasets}
& \field{source\_name} & Name of the reused source dataset & Text \\
\cdashline{2-4}
& \field{source\_identifier} & Identifier of the source dataset & Text / URL / DOI \\
\cdashline{2-4}
& \field{reuse\_extent} & How much and in what form the source dataset was reused & \emph{full} \quad \emph{subset} \quad \emph{partial} \quad \emph{unknown} \\
\cdashline{2-4}
& \field{relationship\_to\_source} & Conceptual relationship to the source dataset & \emph{is\_subset\_of} \quad \emph{is\_version\_of} \quad \emph{is\_derived\_from} \quad \emph{is\_aggregation\_of} \quad \emph{unknown} \\
\cdashline{2-4}
& \field{notes} & Optional explanation specific to this source dataset & Text \\

\midrule

\textbf{Duplicate handling}
& \field{duplicates\_removed} & Whether duplicates were checked and removed & \emph{yes} \quad \emph{no} \quad \emph{unknown} \\
\cdashline{2-4}
& \field{duplicate\_check\_level} & Level at which duplicates were checked & \emph{subject} \quad \emph{study} \quad \emph{series} \quad \emph{image} \\
\cdashline{2-4}
& \field{duplicate\_method} & Method used to detect duplicates & Text \\
\cdashline{2-4}
& \field{duplicate\_method\_reference} & Reference for the method & Text / URL / citation \\

\midrule

\textbf{Overlap assessment}
& \field{between\_datasets} & Datasets involved in overlap assessment & List of text values (min. 2) \\
\cdashline{2-4}
& \field{overlap\_level} & Level at which overlap was assessed & \emph{subject} \quad \emph{study} \quad \emph{series} \quad \emph{image} \\
\cdashline{2-4}
& \field{overlap\_result} & Outcome of overlap assessment & \emph{none} \quad \emph{suspected} \quad \emph{confirmed} \quad \emph{unknown} \\
\cdashline{2-4}
& \field{method} & Method used to assess overlap & Text \\

\midrule

\textbf{Processing steps}
& \field{step\_label} & Short name of the processing step & Text \\
\cdashline{2-4}
& \field{step\_type} & Category of processing step & \emph{filtering}, \emph{normalization},  \emph{aggregation}, \emph{deduplication}, \emph{conversion}, \emph{annotation}, \quad \emph{other} \\
\cdashline{2-4}
& \field{description} & Description of what was done & Text \\
\cdashline{2-4}
& \field{tool\_used} & Software, script, or pipeline used & Text \\
\cdashline{2-4}
& \field{date} & Date of the step & Date (YYYY-MM-DD) \\
\cdashline{2-4}
& \field{purpose} & Intended purpose of the step & \emph{benchmark} \quad \emph{training} \quad \emph{validation} \quad \emph{challenge} \quad \emph{other} \\
\cdashline{2-4}
& \field{input\_datasets} & Input datasets & List of text values \\
\cdashline{2-4}
& \field{output\_datasets} & Output datasets & List of text values \\

\midrule

\textbf{Contributors}
& \field{contributor\_name} & Name of the contributor (person, organization, or software/pipeline) & Text \\
\cdashline{2-4}
& \field{name\_type} & Type of contributor & \emph{Personal}, \emph{Organizational}, \emph{Software} \\
\cdashline{2-4}
& \field{name\_identifier} & Persistent identifier for the contributor & Text / URL \\
\cdashline{2-4}
& \field{name\_identifier\_scheme} & Identifier scheme & \emph{ORCID}, \emph{ROR}, \emph{ISNI}, \emph{GRID}, \emph{GitHub}, \emph{URL}, \emph{Other} \\
\cdashline{2-4}
& \field{affiliation} & Institutional affiliation & Text \\
\cdashline{2-4}
& \field{contributor\_type} & DataCite-aligned contribution role & \emph{ContactPerson}, \emph{DataCurator}, \emph{DataManager}, \emph{Researcher}, \emph{ProjectLeader}, \emph{HostingInstitution}, \emph{Software}, \emph{Other} \\
\cdashline{2-4}
& \field{contribution\_description} & Short description of the contribution & Text \\

\midrule

\textbf{Additional notes}
& \field{additional\_notes} & Notes about the dataset or collection & Free text \\

\bottomrule
\end{tabular}
\end{adjustbox}
\end{table}

\subsection{Sections of the MAP Profile}
The MAP Profile is structured in sections to segregate different elements on tracing provenance and processing in a dataset. They are described as follows:

\begin{enumerate}[leftmargin=*]
    \item \textbf{\textit{Basic Information:}} Captures the core identity of the dataset or collection, including title, identifier, version, license, size, modality, and body region.
    This section provides the minimal contextual metadata needed to distinguish and reference a dataset across repositories and publications.
    The structure is intentionally lightweight while remaining compatible with established metadata standards and repository ecosystems.
    \item \textbf{\textit{Source Datasets:}} Documents which datasets were reused and how they contributed to the new dataset or collection.
    The section distinguishes between the extent of reuse (e.g., full, subset, partial) and the semantic relationship to the source dataset (e.g., derived from, aggregation of).
    This supports explicit provenance tracking and helps clarify lineage in reused or republished collections.
    \item \textbf{\textit{Duplicate Handling:}} Records whether duplicates were checked or removed, at which level (e.g., subject, series, image), and by which method.
    The section is intended to make duplicate management practices transparent and reproducible.
    It also supports interpretation of downstream analyses and benchmark results where hidden duplication may introduce bias.
    \item \textbf{\textit{Overlap Assessment:}} Describes known or suspected overlap between datasets, supporting transparency about shared cases or images.
    The goal is to make potential data leakage and hidden reuse more visible in benchmarking and machine learning settings.
    \item \textbf{\textit{Preprocessing Steps:}} Summarizes the main curation and transformation steps used to create the dataset or collection.
    This includes operations such as filtering, normalization, aggregation, conversion, annotation, or deduplication.
    Capturing preprocessing activities supports workflow traceability and provides context for how the published dataset differs from its source materials.
    \item \textbf{\textit{Contributors:}}
Identifies the people, organizations, or software pipelines involved in creating, curating, or managing the dataset. The structure is designed to align with established metadata frameworks, enabling straightforward integration and interoperability.

    \item \textbf{\textit{Additional Notes:}}
Provides space for uncertainties, caveats, or important context that does not fit elsewhere.
\end{enumerate}

\subsection{What is a MAP Profile not supposed to be?}
There are two important highlights of what MAP is not designed to be. \textit{First}, MAP is a disclosure mechanism, not a duplicate detection method. It does not identify duplicates automatically or reconstruct missing lineage. Instead, it makes reuse, derivation, overlap risk, and deduplication decisions explicit so that they can be systematically inspected and interpreted. MAP is also not intended as a comprehensive dataset metadata standard. In contrast to descriptive frameworks such as Datasheets, \cite{gebru2021datasheets}, VIDs \cite{muthu2026vids}, ROADMAP \cite{suri2026roadmap}, which aim to capture full dataset characteristics, MAP focuses narrowly on collection-specific characteristics. Therefore, it captures only the minimal information required to describe how datasets are combined, transformed, and potentially overlap, enabling analysis of dependencies across dataset collections without attempting to fully characterize the underlying data. However, as we stated earlier, they may be incorporated into existing metadata profiles.

\subsection{Why Not Have Sample level identifiers?}
Reliable duplicate control could be, in principle, best achieved through stable image-level identifiers, robust to transformations (\cref{sec:data_dedup}). In clinical imaging workflows, the DICOM standard provides such identifiers via the \texttt{SOP Instance UID}, which uniquely identifies each image and enables exact matching across datasets when preserved.

In practice, however, this ideal rarely holds for public dataset collections. Many datasets are distributed in non-DICOM formats (e.g., NIfTI or PNG), where such identifiers are absent, and identifiers are often removed or altered during de-identification or preprocessing, breaking links between images across datasets. Moreover, identifiers only support detection of exact duplicates and do not capture transformed or derived data (e.g., resampling, cropping, or relabeling), which are common in dataset reuse.

As a result, image-level identifiers, while a gold standard when available, cannot be relied upon for post hoc duplicate control in large-scale collections. We therefore recommend preserving original identifiers whenever possible and retaining them as auxiliary metadata during format conversion. At the same time, identifiers alone are insufficient and must be complemented by explicit provenance information. MAP addresses this gap by providing a structured representation of dataset-level reuse, transformation, and overlap. MAP currently records provenance at the dataset and collection level, but it can be extended to support case-level identifiers where these are available (e.g., DICOM \texttt{SOP Instance UID}s), enabling applications such as the GDPR right to erasure.

\subsection{Operationalizing Provenance in Dataset Collections}
\label{apx:prospect}
To support adoption in practice, we implement MAP as part of a lightweight, web-based tool for analyzing and documenting dataset collections. The platform enables users to explore existing public datasets and collections, inspect overlap through similarity search against a reference database of perceptual hashes, and generate structured provenance metadata via a guided interface. Importantly, MAP is designed to operate both \textit{prospectively} and \textit{retrospectively}: it can be applied at the time of dataset release to document reuse and deduplication decisions, but also to existing datasets and collections where provenance is incomplete or missing. In the latter case, similarity-based inspection (e.g., via perceptual hashes computed locally and compared against indexed datasets) provides partial signals of overlap, which can then be formalized through the MAP profile. MAP is not intended to be fully automated: when a source dataset is created, its profile requires information that only the dataset creators can provide. When a collection is built from datasets that already have MAP profiles, however, potential overlap can be flagged directly from the profiles, since shared sources become visible before the datasets are combined. The resulting metadata can be exported in standard formats (e.g., JSON) and published alongside datasets, enabling consistent documentation across workflows. By linking dataset discovery, overlap inspection, and structured provenance reporting, the tool provides a practical pathway for making dataset reuse explicit and supporting community-driven governance of dataset collections.

\section{What does \textit{Limited} Provenance Mean in \cref{tab:dataset_collections}?}
Provenance extends beyond the identification of a source dataset and instead encompasses the origin, modification, and usage of the data. We use the term \textit{limited provenance} to indicate that the original publication does not provide sufficient information for detailed provenance reconstruction, such as that enabled by \textit{MAP}. However, in most cases, the authors do report basic information regarding the data source, and we do not intend to suggest otherwise.

\section{Datasets in our Collection}
\label{apx:collection}
This appendix lists all \nDatasetsAll{} datasets of the collection with their number of images. The \nDatasetsOpenNeuro{} datasets from \texttt{OpenNeuro} are listed individually; in the dataset-level analysis (\cref{sec:dedup_results}), they are treated as a single dataset, which gives \nDatasetsMerged{} datasets.

\begingroup
\small
\setlength{\tabcolsep}{3pt}
\begin{longtable}{>{\raggedright\arraybackslash}p{0.55\linewidth}r>{\raggedright\arraybackslash}p{0.20\linewidth}}
\toprule
Name & Num. images & Link \\
\midrule
\endfirsthead
\toprule
Name & Num. images & Link \\
\midrule
\endhead
\midrule
\multicolumn{3}{r}{Continued on next page} \\
\endfoot
\bottomrule
\endlastfoot
A Combined Computational Modeling and Functional Imaging Study - Stage 2 (ds004529) \cite{openneuro_ds004529_modeling_face_recognition_in_the_pr} & 34 & \href{https://openneuro.org/datasets/ds004529}{openneuro.org} \\
AbdomenAtlas 1.1 \cite{li2024abdomenatlas,li2024well,bassi2024touchstone,qu2023abdomenatlas} & 9262 & \href{https://huggingface.co/datasets/AbdomenAtlas/_AbdomenAtlas1.1Mini}{huggingface.co} \\
AbdomenCT-1K \cite{ma2021abdomenct,simpson2019large,antonelli2022medical,heller2021state,heller2019kits19,roth2016data} & 996 & \href{https://github.com/JunMa11/AbdomenCT-1K}{github.com} \\
ACDC \cite{bernard2018deep} & 300 & \href{https://www.creatis.insa-lyon.fr/Challenge/acdc/}{www.creatis.insa-lyon.fr} \\
ACRIN-HNSCC-FDG-PET/CT \cite{kinahan2019data_hnscc,rozenfeld2023annotations_2} & 134 & \href{https://www.cancerimagingarchive.net/collection/acrin-hnscc-fdg-pet-ct/}{www.cancerimagingarchive.net} \\
AMOS 2022 \cite{ji2022amos} & 600 & \href{https://zenodo.org/records/7262581}{zenodo.org} \\
AortaSeg24 \cite{imran2026multiclass_2,imran2026multiclass} & 50 & \href{https://aortaseg24.grand-challenge.org/}{aortaseg24.grand-challenge.org} \\
ATLAS (Tumour and Liver Automatic Segmentation) \cite{quinton2023tumour} & 60 & \href{https://atlas-challenge.u-bourgogne.fr}{atlas-challenge.u-bourgogne.fr} \\
\end{longtable}
\endgroup

\end{document}